%% file: main.tex
\documentclass{aa} %

\usepackage{graphicx}
\usepackage{txfonts}
\usepackage[breaklinks=true]{hyperref}
\makeatletter
\renewcommand*\aa@pageof{, page \thepage{} of \pageref*{LastPage}}
\usepackage{amsmath}    %
 \usepackage{amssymb}   %
\usepackage{natbib}
\usepackage{placeins}
\usepackage{afterpage}
\usepackage{colortbl}
\usepackage{siunitx}
\graphicspath{{./}{Figures/}}

\newcommand{\HII}{\textrm{H~{\textsc{ii}}}}
\newcommand{\HI}{\textrm{H~{\textsc{i}}}}
\newcommand{\jwst}{\textit{JWST}}
\newcommand{\spitzer}{\textit{Spitzer}}
\newcommand{\wise}{\textit{WISE}}
\newcommand{\iso}{\textit{ISO}}
\newcommand{\akari}{\textit{AKARI}}

\newcommand{\mum}{$\mu$m\xspace}
\newcommand{\micron}{$\mu$m}
\newcommand{\molh}{H$_2$\xspace}

\usepackage{color}

\definecolor{lightgrey}{rgb}{0.84, 0.84, 0.84}

\newcommand{\calsymb}[1]{A_\mathrm{#1}}	%
\newcommand{\fcontrib}[1]{\mathcal{C}_{#1}}	%

\newcommand{\mjysr}{\,MJy\,sr$^{-1}$} %
\newcommand{\FeII}{\textrm{[Fe~{\textsc{ii}}]}}
\newcommand{\ArII}{\textrm{[Ar~{\textsc{ii}}]}}
\newcommand{\ArIII}{\textrm{[Ar~{\textsc{iii}}]}}
\newcommand{\NeIII}{\textrm{[Ne~{\textsc{iii}}]}}
\newcommand{\SIV}{\textrm{[S~{\textsc{iv}}]}}
\newcommand{\hi}{\textrm{H~{\textsc{i}}}}
\newcommand{\hei}{\textrm{He~{\textsc{i}}}}
\newcommand{\hii}{\textrm{H~{\textsc{ii}}}}
\newcommand{\um}{\,$\mu$m}

\input{affiliations}

\makeatletter
\renewcommand*\maketitle{%
  \thispagestyle{firstpage}
\begingroup
    \if@wideboxfn
    \setlength\bibindent{1.4\parindent}
    \else
    \setlength\bibindent{\parindent}
    \fi
    \renewcommand*\thefootnote{\@fnsymbol\c@footnote}%
    \renewcommand\@makefntext[1]{%
    \ifaa@longfn\hsize\textwidth\fi
    \noindent
    \hb@xt@\bibindent{\hss\@makefnmark\enspace}##1}
  \ifaa@twocolumn
  \begin{aa@strip}
    \aa@maketitle
  \end{aa@strip}
  \@thanks %
  \else
    \begingroup
      \let\thanks\footnote
      \aa@maketitle
    \endgroup
  \fi
\endgroup
  \setcounter{footnote}{0}%
}
\makeatother

\begin{document} 

\title{PDRs4All\\ XIII. Empirical prescriptions for the interpretation of \jwst\ imaging observations of star-forming regions}
\titlerunning{\textit{JWST} Imaging/IFU Calibrations in PDRs}
\author{Ryan Chown \inst{\uwo,\wspace,\osu} \and
Yoko Okada\inst{\koln} \and
Els Peeters\inst{\uwo, \wspace, \seti} \and 
Ameek Sidhu\inst{\uwo, \wspace} \and 
Baria Khan\inst{\uwo, \wspace} \and 
Bethany Schefter\inst{\uwo, \wspace} \and \\
Boris Trahin \inst{\paris, \stsci} \and
Am\'elie Canin \inst{\toulouse} \and
Dries Van De Putte \inst{\stsci, \uwo, \wspace} \and
Felipe Alarc\'on \inst{\umich} \and
Ilane Schroetter \inst{\toulouse} \and
Olga Kannavou \inst{\paris} \and
{E}milie Habart\inst{\paris} \and
Olivier Bern\'{e} \inst{\toulouse} \and
Christiaan Boersma \inst{\ames} \and
Jan Cami\inst{\uwo, \wspace, \seti} \and 
Emmanuel Dartois\inst{\cnrs} \and
Javier Goicoechea\inst{\madrid} \and
Karl Gordon\inst{\stsci} \and
Takashi Onaka\inst{\tokyo} 
}
\institute{
\uwoname; \email{rchown53@gmail.com} \and
\wspacename \and
\osuname \and
\kolnname \and
\setiname \and
\parisname \and
\stsciname \and
\toulousename \and
\umichname \and
\amesname \and
\cnrsname \and
\madridname \and %
\tokyoname
}

   \date{Received DD Month 2024; accepted DD Month 2024}

  \abstract
   {\jwst\ continues to deliver incredibly detailed infrared (IR) images of star forming regions in the Milky Way and beyond. IR emission from star-forming regions is very spectrally rich due to emission from gas-phase atoms, ions, and polycyclic aromatic hydrocarbons (PAHs). Physically interpreting IR images of these regions relies on assumptions about the underlying spectral energy distribution in the imaging bandpasses. }
   {%
   We aim to provide empirical prescriptions to derive line, PAH, and continuum intensities from \jwst\ images. These prescriptions will facilitate the interpretation of  images in a wide variety of astrophysical contexts. We also measure the level of agreement between \jwst\ imaging and integral field spectroscopy.
   }
   {We use \jwst\ PDRs4All Near-Infrared Camera (NIRCam) and Mid-Infrared Instrument (MIRI) imaging and Near-Infrared Spectrograph (NIRSpec) integral field unit (IFU) and MIRI Medium Resolution Spectrograph (MRS) spectroscopic observations of the Orion Bar, the proto-typical photodissociation region (PDR), %
   to directly compare and cross-calibrate imaging and IFU data at $\sim$100~AU resolution over a region where the radiation field and ISM environment evolves from the hot ionized gas to the warm neutral gas followed by the cold molecular gas.
   We study the relative contributions of line, PAH, and continuum emission to the NIRCam and MIRI filters as functions of local physical conditions, and investigate filter combinations which represent selected line and PAH emission.}
   {%
   We provide empirical prescriptions based on NIRCam and MIRI images that may be used to derive intensities of strong emission lines and PAH features. Within the range of the environments probed in this study, these prescriptions accurately predict Pa~$\alpha$, Br$\alpha$, PAH 3.3\um\ and 11.2\um\ intensities, while those for \FeII\ 1.644\um, H$_2$ 1--0 S(1) 2.12\um\ and 0--0 S(9) 4.69\um, and PAH 7.7\um\ show more complicated environmental dependencies. }
   {Linear combinations of \jwst\ NIRCam and MIRI images provide effective tracers of ionized gas, \molh, and PAH emission in PDRs. We expect these recipes to be useful for both the Galactic and extragalactic communities. The flux calibration between imaging and spectroscopy is found to agree within 1--20\% for NIRCam and NIRSpec, and  2--7\% for MIRI Imager and MRS. %
   }
   \keywords{astrochemistry  -- infrared: ISM -- ISM: molecules -- ISM: individual objects: Orion Bar -- ISM: photon-dominated region (PDR) --
                techniques: spectroscopic }

   \maketitle
   
\section{Introduction}
\label{sec:intro}

Near-infrared (NIR) and mid-infrared (MIR) observations give crucial information on the physical and chemical structure of the interstellar medium (ISM). Emission in this wavelength range ($\sim$1--28~\um) includes a large number of emission lines from gas-phase atoms, ions, and molecules, broad features from astronomical polycyclic aromatic hydrocarbons \citep[PAHs;][]{tielens2008}, and continuum  from stochastically-heated small dust grains \citep[e.g.][]{galliano2018}. PAHs play critical roles in both the thermal and charge balance in the ISM. PAH emission is found in photodissociation regions \citep[PDRs;][]{tielens1985, hollenbach1997, tielens2008, wolfire2022}, which are the neutral regions of the ISM where stellar FUV photons dominate the heating. Most of the \ion{H}{i} in galaxies lies in PDRs. Data from space-based NIR and MIR missions such as \spitzer, \akari, and the \textit{Infrared Space Observatory} (\iso) have been used to investigate the physical properties of the ISM in the Milky Way and beyond \citep{rosenthal2000,rubin2007,tielens2008,dale2009,lai2020, li2020a}. \jwst\ NIR and MIR spectra of the ISM have shown an unprecedented amount of detail, opening opportunities for high spatial resolution (e.g., $\sim$100~AU in the Orion Nebula) modeling of the ISM and PDRs \citep{chown2024,peeters2024,van-de-putte2024}.

Given its strong wavelength dependence, NIR and MIR emission is best characterized using infrared spectroscopy. However, IR integral field unit (IFU) spectroscopy is very resource-intensive and covers much smaller fields of view than imaging. This imaging/spectroscopy difference is a common challenge with space, ground, and airborne observatories. For the \jwst\ Mid-Infrared Instrument \citep[MIRI; ][]{rieke2015, bouchet2015, dicken2024}, the imaging field-of-view (FOV) is $74''\times 113''$, while the accompanying Medium-Resolution Spectrograph \citep[MRS;][]{wells2015, argyriou2023} FOV is much smaller at $3''\times 3''$ at the shortest wavelengths to $7''\times 7''$ at the longest wavelengths. Similarly, \jwst's Near Infrared Camera \citep[NIRCam;][]{rieke2023} has a $132''\times132''$ FOV for each of it's two modules (A and B), while NIRSpec \citep{boker2022} has $3''\times3''$ FOV. Imaging therefore has the advantage of being able to map the entirety of much larger extended objects than can be mapped with IFU observations.

With IR imaging (or imaging, in general) it is not possible to directly estimate the contributions of different spectral components to a given filter. For example, in nearby galaxies, the Rayleigh-Jeans tail of stellar continuum emission can contribute up to $65$\% of the flux in the  \jwst\ NIRCam F335M filter, which is mainly used for tracing the 3.3~\um\ PAH feature \citep{sandstrom2023}. Emission from small, hot dust grains can contribute significantly to the longer-wavelength PAH-dominated imaging filters, e.g., \jwst/MIRI F770W, F1130W, \spitzer/IRAC 8~\um, and \wise\ 12~\um\ \citep[e.g.,][]{whitcomb2023a}. In most cases it is not known beforehand what the PAH-to-dust ratios are in a given source, and so, interpretations of MIR images rely on some assumption about the underlying spectral energy distribution (SED).
With \spitzer, a variety of methods were used to isolate spectral components from multi-band images using SED modeling \citep[e.g.][]{helou2004, crocker2013, zhang2023}. Another approach is to decompose the observed spectra into a sum of line, continuum, and PAH feature spectra, and then estimate contributions of emission components to filters using synthetic photometry of each component \citep[e.g.][]{lai2020, whitcomb2023a}, and then apply this knowledge to the entire imaging field of view.

In this work we aim to quantify the contributions of line, continuum, and PAH emission to \jwst\ NIRCam and MIRI Imager filters. We utilize \jwst\ Near-Infrared Spectrograph \citep[NIRSpec][]{boker2022}, MIRI MRS, and imaging (NIRCam and MIRI Imager) observations obtained as part of the PDRs4All \jwst\ Early Release Science program \citep[program ID: 1288;][]{berne2022, habart2024, peeters2024, chown2024, van-de-putte2024}. These observations are centered on the nearly edge-on PDR, the Orion Bar \citep{tielens1993}. The IFU observations cover the \HII\ region, which includes emission from the face-on PDR of the background molecular cloud, OMC-1; the ionization front; the atomic PDR; and three dissociation fronts.\footnote{For a sketch of a classical example of a PDR and the Orion Bar PDR, see \citet{tielens1985, wolfire2022} and \citet{habart2024} respectively.} Our goal is to leverage the high-quality in spectral-spatial resolution and sensitivity of the \jwst\ IFU data to quantify the relative contributions of emission from the different ISM components to NIR and MIR \jwst\ imaging filters. We consider gaseous emission lines, PAH features and continuum in this study. Given the widespread use of \jwst\ images as tracers of the ISM from the Milky Way to nearby and distant galaxies, we expect this work to help interpret observations in a wide variety of contexts. 

In Sect.~\ref{sec:data} we describe the data that are used and how they are processed. 
In Sect.~\ref{sec:xcal}, we use the existing imaging data and spectra to cross calibrate the fluxes of JWST's IFUs (MIRI/MRS and NIRSpec) from their corresponding imagers (MIRI Imager and NIRCam). In Sect.~\ref{subsec:continuum_fraction}, we generate synthetic images for a 
set of imaging filters of interest using the IFU spectra and explore the fidelity of
utilizing pure imaging to estimate line and feature intensities. We compare our results with previous work in Sect.~\ref{sec:comparison_previous_work}.

\section{Observations and Data Reduction}\label{sec:data}

Between 10 September 2022 and 30 January 2023, \jwst\ observed the Orion Bar PDR as part of the PDRs4All Early Release Science program \citep[ID: 1288\footnote{MAST DOI: 10.17909/pg4c-1737}, ][]{berne2022}. The MIRI and NIRCam imaging observations in this program targeted the Orion Nebula (M42) as well as part of M43 \citep[NGC~1982;][]{habart2024}. NIRspec and MIRI/MRS IFU mosaics ($9\times 1$ pointings each) across the Orion Bar were obtained for a patch (roughly $3''\times 25''$) of the imaging FOV.

This paper is based on the following data:
\begin{enumerate}
    \item NIRSpec mosaics of all nine pointings combined, keeping each grating/filter combination separate (g140h-f100lp, g235h-f170lp, and g395h-f290lp). The individual pointings were reduced using context \texttt{jwst\_1084.pmap} of the Calibration References Data System (CRDS). The nine-pointing single-segment cubes were used to measure imaging/IFU cross-calibration factors (Sect.~\ref{sec:crosscal_NIR}). These calibration factors were used to produce the spatially and spectrally stitched, flux-calibrated NIRSpec cube presented in \citet{peeters2024}.
    \item The MIRI/MRS data were reduced using CRDS context \texttt{jwst\_1154.pmap}. These observations were first presented in \citet{chown2024} and \citet{van-de-putte2024} using earlier contexts. To create a stitched cube, individual sub-bands were multiplicatively scaled to match in flux where they overlap, using Channel 2 Long as the reference.
    \item NIRCam and MIRI Imager observations of the Orion Bar that were presented in \citet{habart2024}, were reduced using CRDS contexts \texttt{jwst\_0954.pmap} and \texttt{jwst\_1040.pmap}, respectively. The NIRCam filters are F140M, F164N, F162M, F182M, F187N, F210M, F212N, F277W, F300M, F323N, F335M, F405N, F470N, and F480M. The MIRI filters are F770W, F1130W, F1500W, and F2550W. We do not use the parallel imaging observations of M43, because they do not have overlapping IFU coverage.
\end{enumerate}

\begin{figure*}
\begin{center}
\includegraphics[width=\textwidth]
{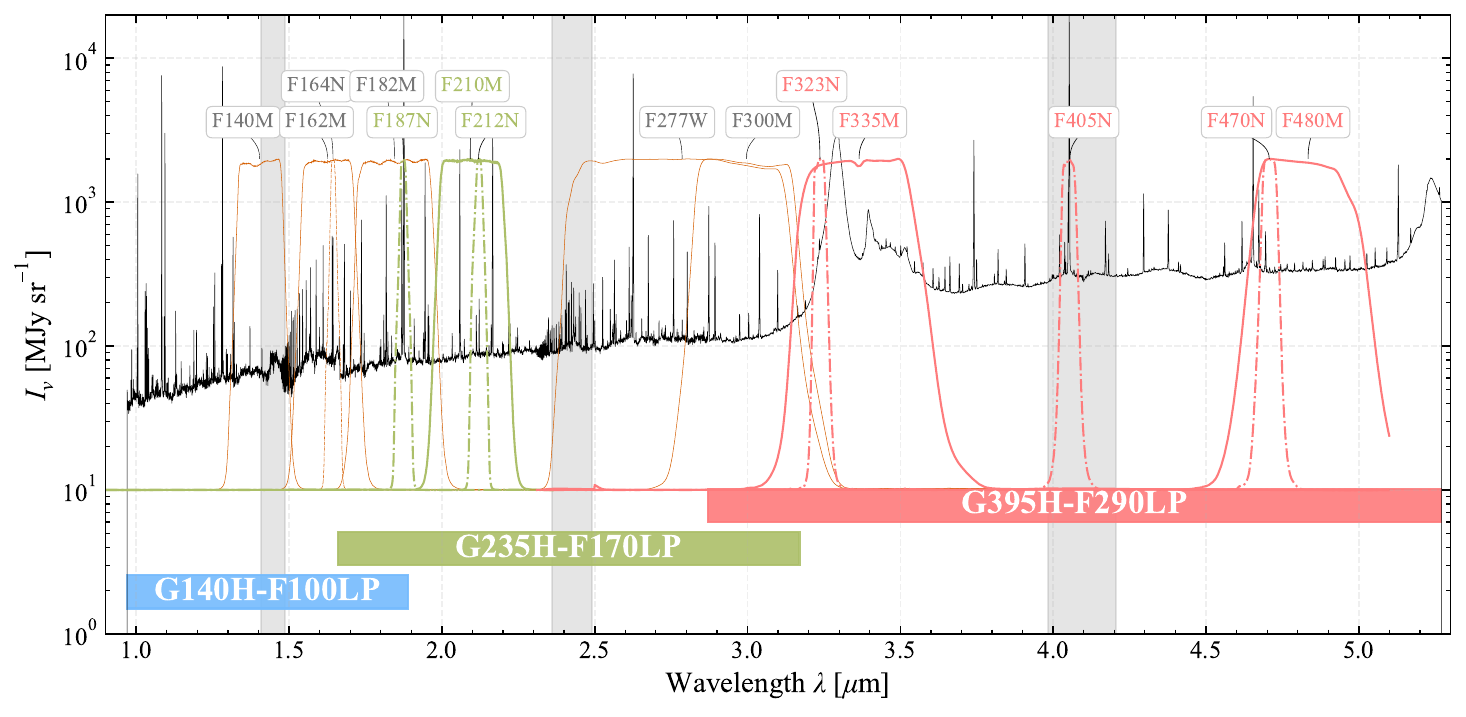}
\includegraphics[width=\textwidth]
{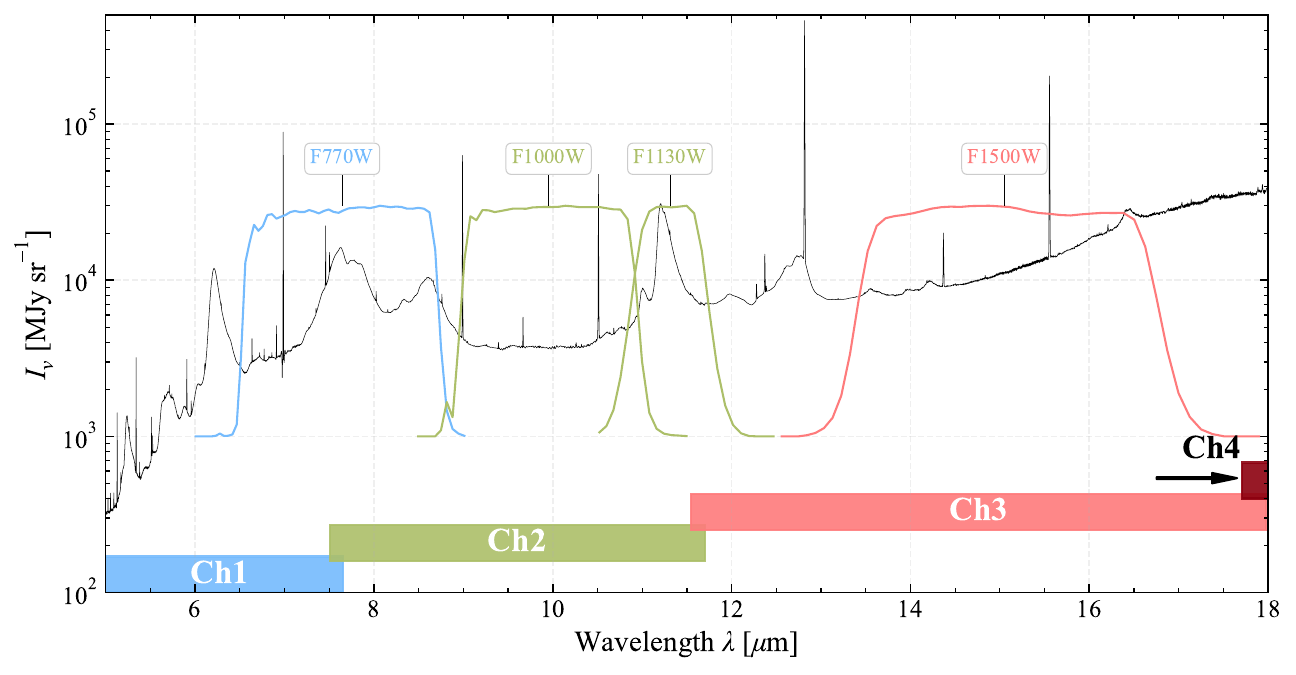}
\caption{\textit{Top:} NIRSpec template spectrum of the Orion Bar Atomic PDR \citep{peeters2024} with NIRCam filter throughputs, NIRSpec grating wavelength ranges, and NIRSpec wavelength gaps \citep[from Table 3 in][]{boker2022} shaded in grey. Flux calibration of the NIRSpec IFU data (\S\ref{sec:crosscal_NIR}) was carried out using filters coloured in green for G235H-F170LP and in red for G395H-F290LP. The calibration factors ($\calsymb{filter}$ in Eq.~\ref{eq:xcal}) are given in Table~\ref{tab:xcal_nirspec}. The other filters which were included in PDRs4All and observed the Orion Bar are shown in orange. \textit{Bottom:} MIRI MRS template spectrum (thick black lines) of the Orion Bar Atomic PDR from \citet{chown2024} and \citet{van-de-putte2024} together with the MIRI filter response functions (thin orange lines) employed in this study (F770W, F1000W, F1130W, and F1500W). We have MIRI images for all of these filters except for F1000W, which is included in the spectroscopy-only part of our analysis (\S\ref{subsec:continuum_fraction}). Flux cross-calibration with F770W and F1130W can only be performed on the stitched MRS cube because these filters span multiple MRS channels. The MRS/MIRI Imager cross-calibration results are shown in Table~\ref{tab:xcal_miri}.
}
\label{fig:template_with_gratings}
\end{center}
\end{figure*}

We check the absolute astrometric alignment for the spectroscopic observations by adjusting the image center such that the observed position of the proplyd d203-504 \citep[see Fig.~1 of][]{chown2024} matches the true position \citep{ricci2008}. We checked the astrometric alignment of the NIRCam and MIRI images by comparing with \textit{Gaia}-DR3 sources and do not find that any shifts are necessary. The IFU observations have a significantly smaller spatial footprint than the imaging observations \citep[see their positions in Fig.~2 of][]{habart2024}. We thus reproject the NIRCam and MIRI Imager observations onto the NIRSpec and MIRI MRS spatial grids (using \texttt{reproject.reproject\_exact}), respectively, when comparing the imaging and IFU observations.

We note that some artifacts are present in the data:
\begin{itemize}
    \item Horizontal stripes (relative to the detector orientation) in the NIRSpec spectral maps are due to residual $1/f$ noise. These stripes are visible in the synthetic images for the F470N filter in Fig.~\ref{fig:continuum_line_fractions_H2_469} and the F182M filter in Fig.~\ref{fig:xcal_nirspec3}.
    \item Incomplete flagging of glitches that are difficult to distinguish from strong emission lines appear as individual bright pixels in the wavelength ranges of the synthetic images for F187N (Fig.~\ref{fig:continuum_line_fractions_Pa_a}) and F210M (Fig.~\ref{fig:continuum_line_fractions_H2_212}).
    \item High-resolution NIRSpec spectra have a wavelength gap which varies with position due to the physical gap between detectors \citep{boker2022}. For our analysis this gap only affects the Br$\alpha$ line (4.05~\um) and the NIRSpec synthetic image of NIRCam F405N. %
    
\end{itemize}

\input{table_xcal_nirspec}

\section{Comparing imaging and IFU observations}\label{sec:xcal}
\subsection{Synthetic images}\label{subsec:synthetic_images}

To compare imaging and IFU surface brightnesses, and to assess the contribution of different spectral components in the imaging filters, we compute ``synthetic images'' by integrating each IFU spaxel over a \jwst\ imaging filter's spectral response curve.
The integration is given by Equation 5 of \citet{Gordon2022}, which also defines the units of NIRCam and MIRI images. Since \jwst\ is a photon-counting instrument, the imaging units are defined in photon units (i.e. counts of photons over a bandpass), and so the units are a weighted sum of the number of photons (i.e. $F_\nu(\lambda)/h\nu=F_\nu(\lambda) \lambda /hc$) in a given wavelength bin.
For a given IFU spaxel $F_\nu(\lambda)$ (in units of \mjysr), the synthetic flux in an imaging filter is given by Equation 5 of \citet{Gordon2022}:
\begin{align}\label{eq:synth}
    F^\mathrm{synth.}_{\mathrm{filter}}~[\mathrm{MJy~sr^{-1}}] &\equiv \mathcal{S}(F_\nu(\lambda); t_{\lambda,~\mathrm{filter}}) \\ &= \frac{\int_{\lambda_0}^{\lambda_1}(F_\nu(\lambda)~\lambda / hc)~t_{\lambda,~\mathrm{filter}}~d\lambda}{\int_{\lambda_0}^{\lambda_1}(\lambda/hc)~t_{\lambda,~\mathrm{filter}}~d\lambda},    
\end{align}
where $\mathcal{S}$ is the synthetic image operator (to make notation easier in the rest of the paper), and $t_{\lambda,~\mathrm{filter}}$ is the total throughput for that imaging filter. The integration is performed over the wavelength range, $\lambda_0 \leq \lambda \leq \lambda_1$, where $t_{\lambda,~\mathrm{filter}} \geq 10^{-3}\,\mathrm{max}(t_{\lambda,~\mathrm{filter}})$. We compute the uncertainty on $F^\mathrm{synth.}_{\mathrm{filter}}$ by propagating measurement uncertainties on $F_\lambda$ through Equation~\ref{eq:synth}.

We obtained throughputs $t_{\lambda,~\mathrm{filter}}$ from the \jwst\ user documentation\footnote{\url{https://jwst-docs.stsci.edu/jwst-near-infrared-camera/nircam-instrumentation/nircam-filters}}, version 5.0, produced in November 2022 for NIRCam, and using the \texttt{Pandeia} engine\footnote{\url{https://pypi.org/project/pandeia.engine/}} for MIRI. 
Fig.~\ref{fig:template_with_gratings} shows the NIRCam bandpass functions with a template spectrum, together with NIRSpec grating wavelength ranges and NIRSpec wavelength gaps. The MIRI bandpass functions are shown in the lower panel of Fig.~\ref{fig:template_with_gratings}. We do not consider F2550W because the MRS spectra do not cover the full wavelength range of that filter.

Wavelength gaps in the NIRSpec data (\S\ref{sec:data}), pipeline-generated data quality flags, and manually-identified spikes in surface brightness along a spectrum, all lead to gaps of missing data in a spaxel. 
We do not measure $F^\mathrm{synth.}_{\mathrm{filter}}$ in a spaxel if more than 5\% of the surface brightness measurements in the wavelength range of integration are missing, except for the following cases. The threshold of 5\% was chosen to avoid discarding too many pixels, and to avoid increasing uncertainties due to interpolation. There is a larger fraction of missing data in the F140M and F277W filters, and to be able to include these bands in our analysis for completeness we raise the 5\% threshold to 20\%, with the cost of less reliable measurements. We tested higher and lower fractions, and 20\% led to a larger number of pixels without including extreme outliers. 
For F405N, a large wavelength gap hits the line of interest, \mbox{Br~$\alpha$} (at 4.052\,$\mu$m) for some parts of the map, and a single threshold does not properly mask the pixels where \mbox{Br~$\alpha$} falls inside this gap. Thus, we allowed missing data between 4.06\,$\mu$m and the upper edge of the F405N filter in order to capture the \mbox{Br~$\alpha$} line. 

\subsection{Cross-calibrating IFU and imaging data}\label{subsec:cross-cal}

We examine the level of consistency between the IFU (NIRSpec and MIRI MRS) and imaging data (NIRCam and MIRI Imager) used in this paper.  We compute synthetic images from the IFU data (Sec.~\ref{subsec:synthetic_images}) and compare them pixel-by-pixel with imaging data for each filter over the IFU FOVs. Our procedure is identical for the two IFU/imager combinations. 

We assume a linear relationship between the surface brightness in a NIRCam or MIRI Imager pixel ($F_{\mathrm{filter}} \equiv \langle F(\lambda) \rangle $ in \mjysr) and that in the corresponding pixel of a synthetic image
of the filter ($F^\mathrm{synth.}_{\mathrm{filter}}$ in \mjysr, defined in Eq.~\ref{eq:synth}):
\begin{equation}\label{eq:xcal}
    F_{\mathrm{filter}} = \calsymb{0, filter} + \calsymb{filter}F^\mathrm{synth.}_{\mathrm{filter}},
\end{equation}
where the free parameters are a DC offset $\calsymb{0, filter}$ (in \mjysr) and gain, or relative calibration factor, $\calsymb{filter}$ (unitless). 

We compute synthetic images for all NIRCam filters and MIRI Imager filters that are included in the PDRs4All observations. Then we mask the bright protoplanetary disk (d203-504) and plot $F_{\mathrm{filter}}$ against $F^\mathrm{synth.}_{\mathrm{filter}}$. Next, we 
perform Bayesian linear regression on $F_{\mathrm{filter}}$ against $F^\mathrm{synth.}_{\mathrm{filter}}$ using \texttt{linmix} \citep{kelly2007}, which incorporates uncertainties in the independent and dependent variables. There are no upper limits in our fits.

\subsubsection{Tying NIRSpec flux calibration to that of NIRCam}\label{sec:crosscal_NIR}

\begin{figure*}
\begin{center}
\includegraphics[width=0.79\textwidth]{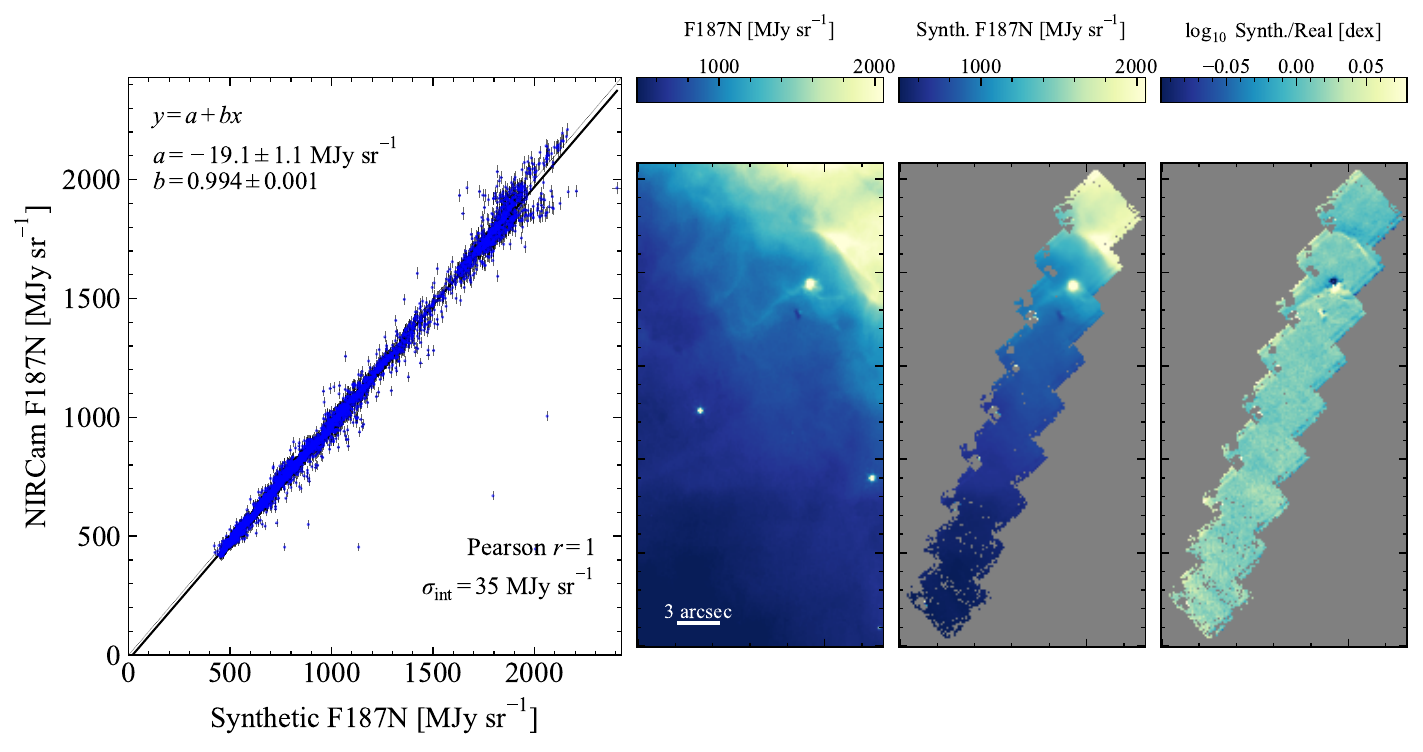}
\includegraphics[width=0.79\textwidth]{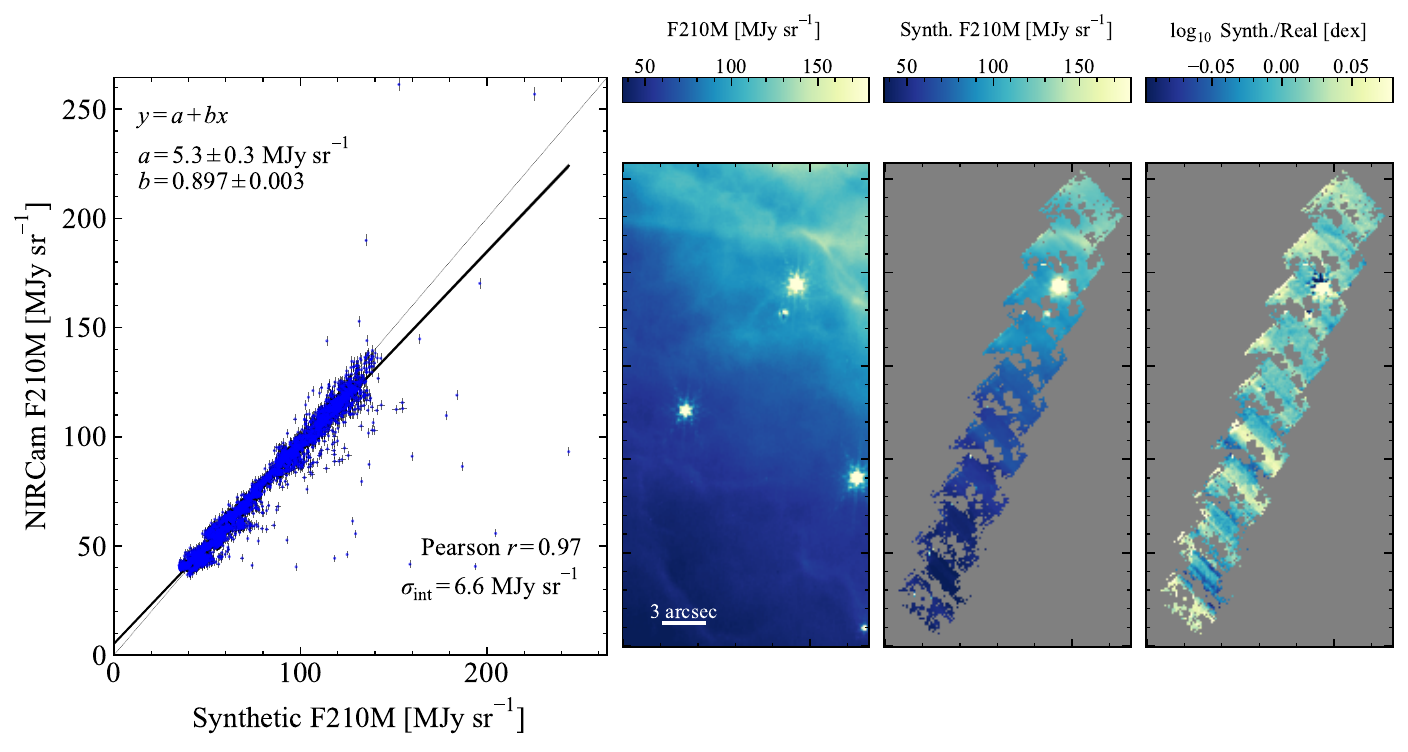}
\includegraphics[width=0.79\textwidth]{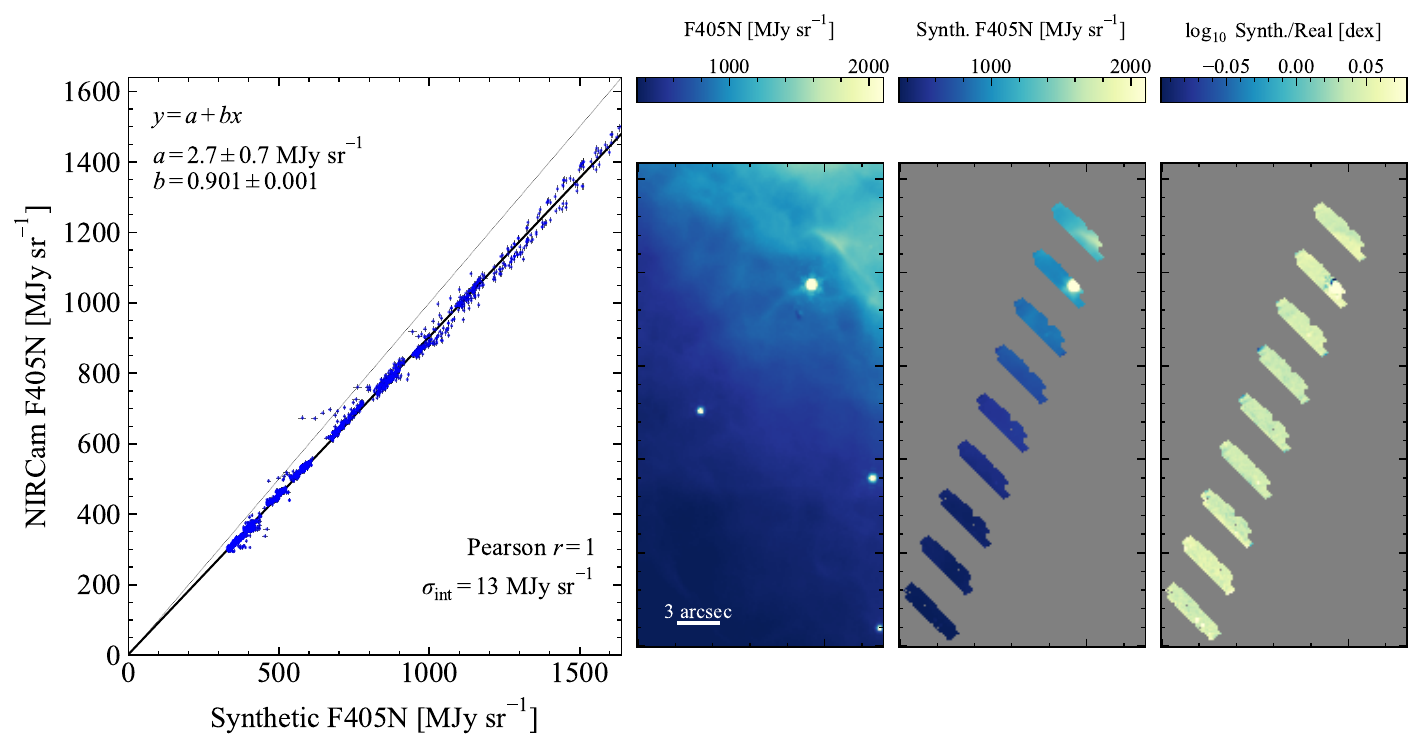}
\caption{NIRSpec/NIRCam cross-calibration results for selected NIRCam filters: F187N (top row), F210M (middle row), and F405N (bottom row). From left to right: linear fit of fluxes from individual pixels from the real NIRCam image (y axis) and synthetic NIRCam image (x axis); NIRCam image zoomed in on the NIRSpec FOV; synthetic image measured from the un-stitched, uncalibrated NIRSpec cube; ratio of the synthetic image to the real image. The best-fit parameters are shown in the top left panel of each scatter plot and are given in Table~\ref{tab:xcal_nirspec}. The thin dotted line is a 1:1 relationship. See Section~\ref{sec:crosscal_NIR} for further details. Versions of this figure for F335M and F212N are shown in Appendix~\ref{sec:xcal_additional}. Some pixels are missing from the synthetic images due to the physical gap between detectors (affecting F405N), and artifacts stemming from the NIRSpec data reduction pipeline. }
\label{fig:xcal_nirspec}
\end{center}
\end{figure*}

We use the NIRCam/NIRSpec cross-calibration results to improve the flux calibration of our three individual grating/filter combination NIRSpec cubes before stitching them together. Fig.~\ref{fig:xcal_nirspec} shows the comparison between NIRCam and synthetic NIRCam images obtained from single grating/filter NIRSpec cubes \textit{without any post-pipeline flux scaling applied}, as well as the best-fit line relating surface brightness of the real and synthetic images. The best-fit slope in each row is our estimate of the scaling factor that must be applied to NIRSpec grating/filter cube in order to match the NIRCam surface brightness. The fit results are summarized in Table~\ref{tab:xcal_nirspec}.

Since background subtraction was performed during the reduction of the NIRSpec data but not during that of the NIRCam data, the y-intercept $\calsymb{0, filter}$ absorbs any residual background flux (adding a constant background surface brightness to $F_{\mathrm{filter}}$ simply shifts the best-fit value of $\calsymb{0, filter}$). The best-fit slopes with background-subtracted NIRCam data for those few filters for which we have off-source observations \citep{habart2024} are consistent with the non-background-subtracted results, and so subtracting a background from NIRCam does not affect the relative calibration factor $\calsymb{filter}$.

\begin{figure*}
\begin{center}
\includegraphics[width=0.65\textwidth]{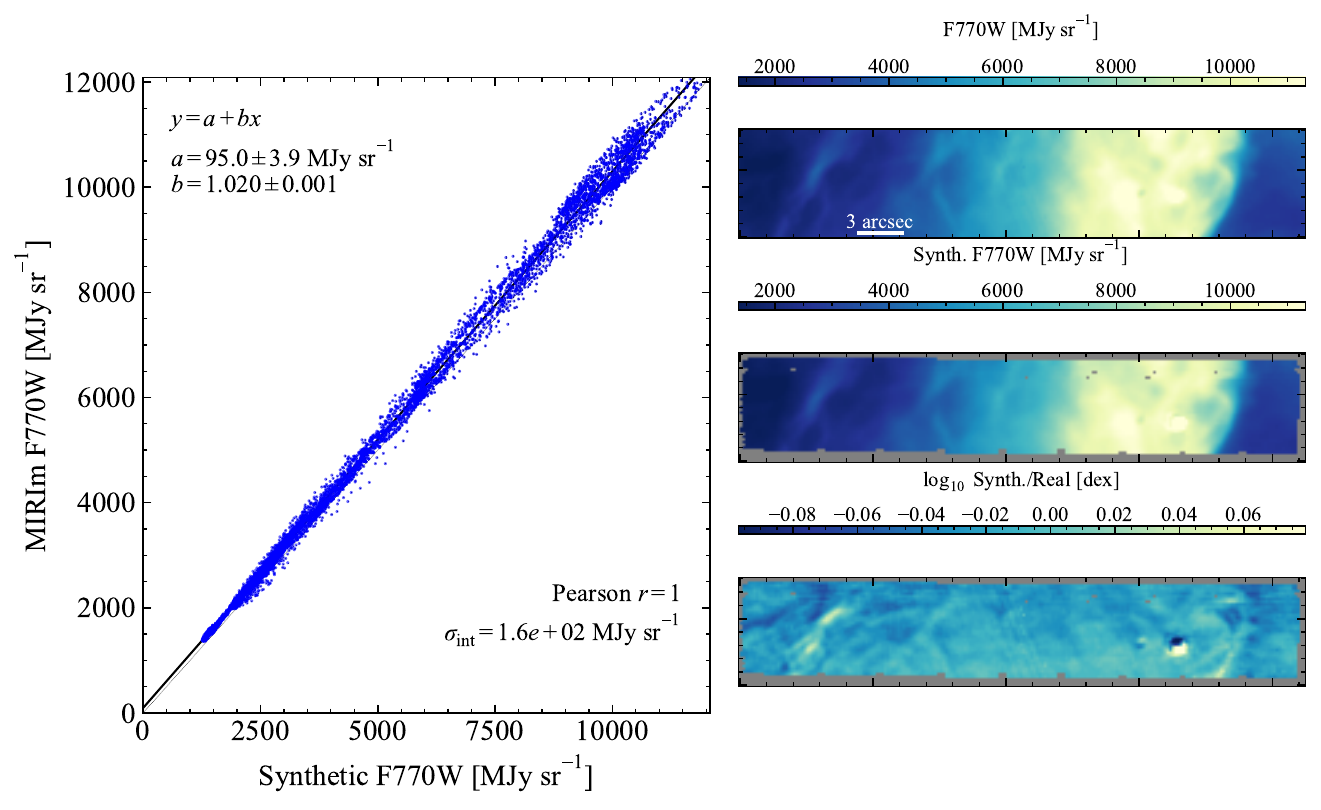}
\includegraphics[width=0.65\textwidth]{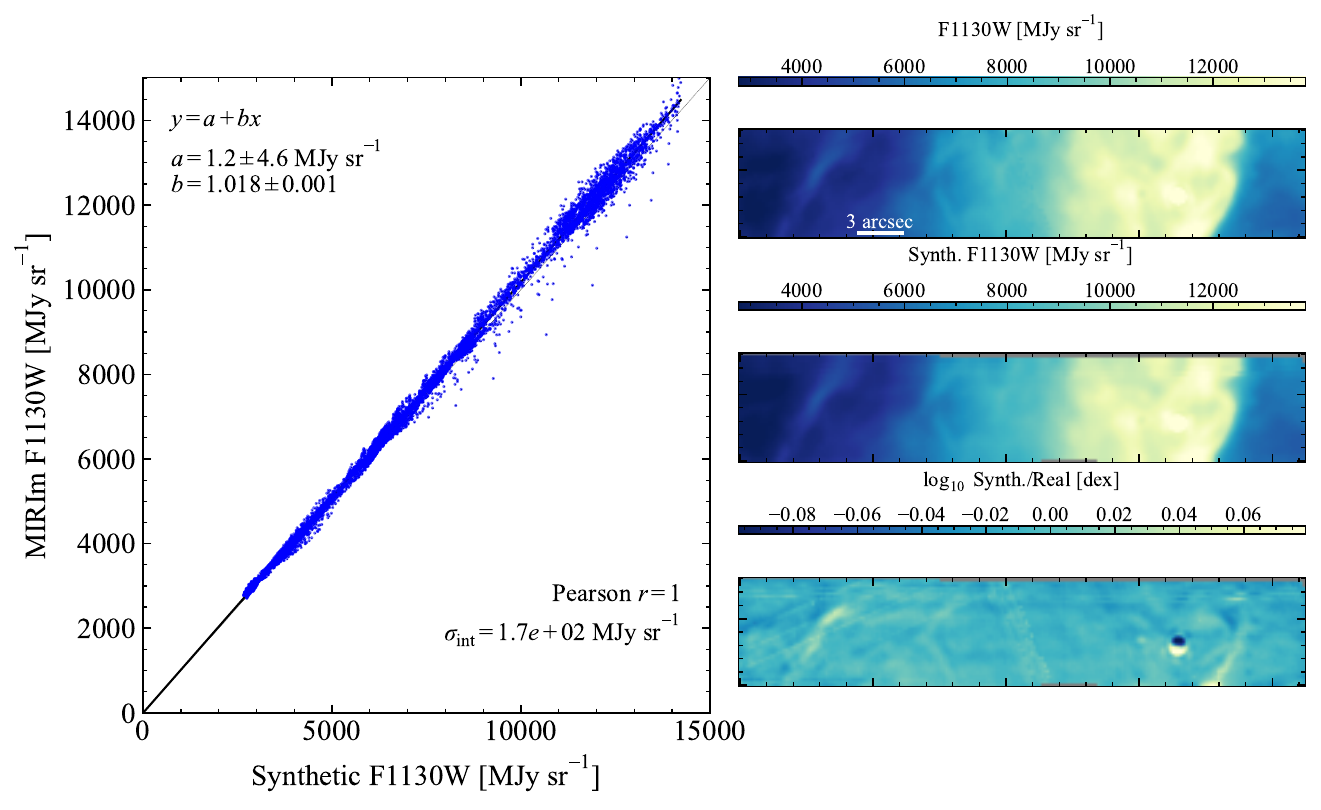}
\includegraphics[width=0.65\textwidth]{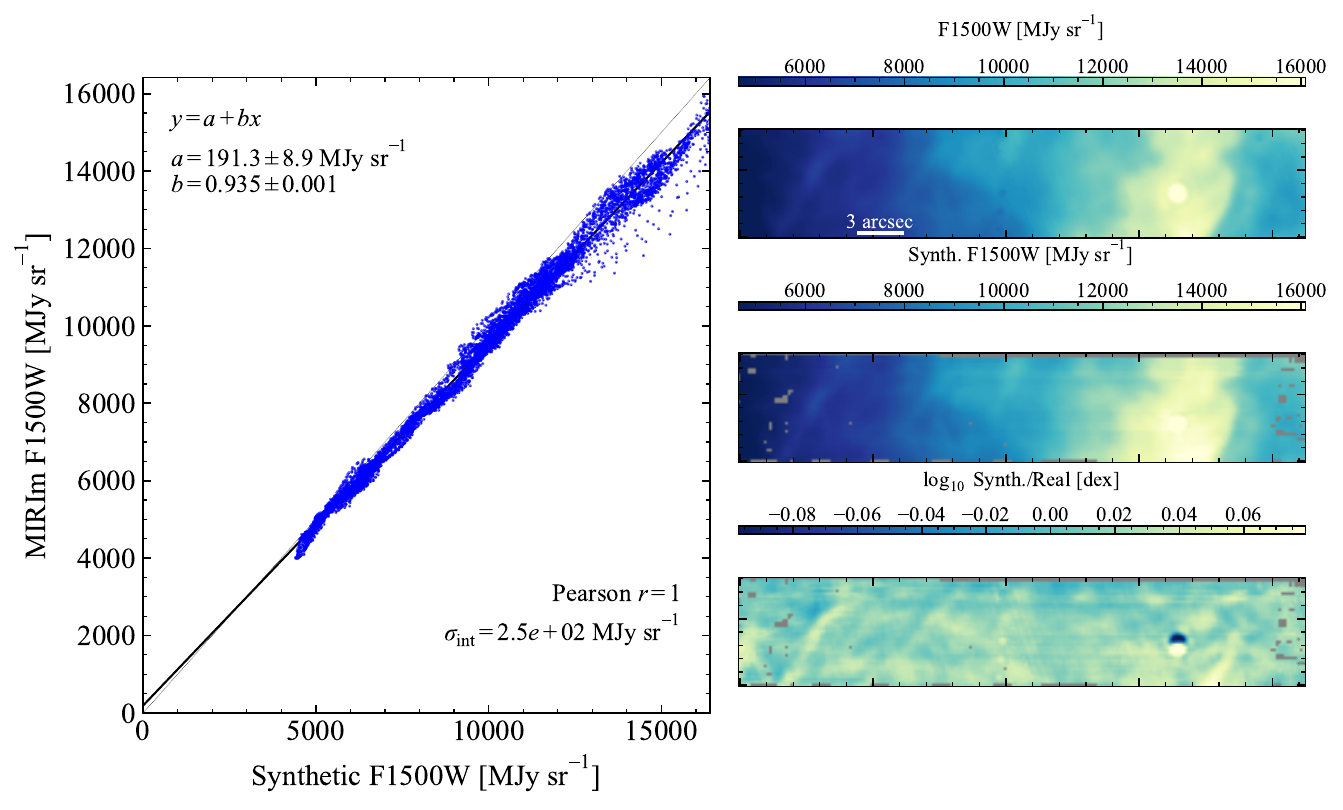}
\caption{MIRI MRS/MIRI Imager cross-calibration results for MIRI filters: F770W (top), F1130W (middle), and F1500W (bottom). For each filter, the three images on the right (top to bottom) are MIRI image zoomed in on the MRS field of view, synthetic image measured from the stitched, scaled MRS cube, and the ratio of the synthetic image to the real image. The left panels show the linear fit of fluxes from individual pixels from the real MIRI image (y axis) and synthetic MIRI image (x axis). The best-fit parameters are given in Table~\ref{tab:xcal_miri}. The thin dotted line is a 1:1 relationship. The shadow-like features in the ratio maps are largely due to residual (but expected) astrometric offsets between the imaging and IFU observations.
}
\label{fig:xcal_miri}
\end{center}
\end{figure*}

Within each grating/filter wavelength range, the spread in best-fit calibration factors $\calsymb{filter}$ is significantly larger than the statistical uncertainties of $\calsymb{filter}$. Additionally, the intrinsic scatter of each fit tends to be large (of order 10~\mjysr). Both of these findings suggest that the relative flux calibration between NIRCam and NIRSpec is affected by wavelength-dependent systematic errors, likely due to the not-yet-finalized NIRSpec flux calibration. Future calibration reference files should bring the relative flux calibration of NIRCam and NIRSpec into better agreement.

As described in \citet{peeters2024}, we multiply the g235h-f170lp cube by the inverse-variance weighted mean of $\calsymb{filter}$ for F187N, F210M, and F212N reported in Table~\ref{tab:xcal_nirspec}), and we propagate the uncertainty in $\calsymb{filter}$ into the uncertainties of this cube. These filters were chosen because they are completely contained within the wavelength range spanned by the g235h-f170lp grating/filter combination  (F277W and F300M both extend beyond the wavelength range of this cube). Likewise, we multiply the g395h-f290lp cube by the inverse-variance weighted mean of $\calsymb{filter}$ for F323N, F335M, F405N, F470N, and F480M. For the filters that overlap with g140h-f100lp, the NIRSpec wavelength gap hampers our ability to measure synthetic F140M flux over most of the observed area, while the synthetic F164N and F162M filters are affected by artifacts (see Fig.~\ref{fig:xcal_nirspec2}). Instead of multiplying the g140h-f100lp cube by a calibration factor derived using F140M, F164N and F162M, we apply the g235h-f170lp calibration factor because it leads to better consistency between the two segments where they overlap, likely as a result of the effects just mentioned. 

\input{table_xcal_miri}

\subsubsection{Flux calibration consistency between MIRI MRS and MIRI Imager}\label{sec:crosscal_MIR}

The best-fit parameters of Eq.~\ref{eq:xcal}, measured for MIRI filters are shown in Table~\ref{tab:xcal_miri}. The comparison between the real MIRI images and synthetic images, as well as the fits for all three filters that we use (F770W, F1130W, and F1500W) are shown in Fig.~\ref{fig:xcal_miri}. As shown in the bottom panel of Fig.~\ref{fig:template_with_gratings}, the F770W and F1130W filters extend well beyond Channel 1 and Channel 2 respectively, and so for these filters it is not possible to cross-calibrate fluxes with imaging prior to stitching. F1500W is contained within Channel 3 however, so cross-calibration is possible there. Although we have MIRI Imager F2500W observations, this filter extends too far beyond the MRS wavelength limit, and so we do not use this filter for cross-calibration. 

The best-fit values of 
$\calsymb{filter}$ applied to the stitched MRS data cube are significantly closer to 1.0 than for NIRCam/NIRSpec (Sect.~\ref{sec:crosscal_NIR}). The ratio maps in Fig.~\ref{fig:xcal_miri} show some structure especially near the ionization front. Some of this structure is due to slight imperfections in the IFU astrometric alignment (which is sensitive to small-scale structures like point sources and sharp edges). The MIRI Imager/MRS calibration factors are within 2\% of unity for F770W and F1130W, and about 6\% for F1500W, which is close to the MRS absolute flux calibration accuracy \citep{law2025}. Given how close these calibration factors are to unity, and to the wavelength range complication mentioned above, we do not find it necessary to apply any flux calibration corrections to the MRS IFU data.

\section{Deriving line and PAH intensities from NIRCam and MIRI images}\label{subsec:continuum_fraction}

We employ three methods to estimate line and PAH intensities: 1) measuring the flux within synthetic filters containing the target features (\S\ref{subsec:frac_cont}), 2) calculating the difference between feature-containing filters and appropriate continuum filters (\S\ref{subsec:lincomb}), and 3) applying a general linear combination of these filters (also \S\ref{subsec:lincomb}). Detailed descriptions of each approach follow, with the prescriptions themselves presented in Sect.~\ref{subsec:prescriptions} and best-fit parameters shown in Table~\ref{table:line_prescriptions}.

\subsection{Relative contributions of emission components to NIRCam and MIRI images}\label{subsec:frac_cont}

Here we present the relative contributions of bright lines, PAHs, and continuum emission to NIRCam and MIRI imaging filters (and combinations of filters) to facilitate the interpretation of \jwst\ images.
The fractional contribution of emission type $i$ (either line, PAH band, or continuum emission) to filter $j$ for a spectrum $F_\lambda$ is given by
\begin{equation}\label{eq:fcont}
    \fcontrib{i,j} \equiv \frac{\mathcal{S}(F_\nu(\lambda)^i; t_{\lambda,~j})}{\mathcal{S}(F_\nu(\lambda); t_{\lambda,~j})},
\end{equation}
where the synthetic image operator $\mathcal{S}$ is defined by Eq.~\ref{eq:synth}
and $F_\nu(\lambda)^i$ is the spectrum of component $i$ obtained by decomposing the total spectrum in its emission components (i.e. $F_\nu(\lambda) = \sum_i F_\nu(\lambda)^i$). 
The results in this section rely solely on spectroscopic data since $\fcontrib{i,j}$ is computed using only spectroscopic data. Hence, these results are not affected by any flux calibration differences between NIRSpec/NIRCam (Section~\ref{sec:crosscal_NIR}) or MIRI Imager/MRS (Section~\ref{sec:crosscal_MIR}). In addition to the filters listed in Sect.~\ref{sec:data}, we add the MIRI F1000W filter to this 
analysis for two main reasons. While F1000W is not part of the filter suite employed by PDRs4All, it is often used to subtract the continuum from neighboring PAH band images (F770W and F1130W). Second, it is important to quantify contributions of PAH, line, and continuum emission to F1000W in Orion for comparison with galaxies, where this filter can be strongly affected by the silicate absorption feature at $9.8~\mu$m \citep[e.g.][]{draine2007a}.

We apply a spectral decomposition to each spaxel from the NIRSpec and MRS cubes. We also decompose the spectra extracted from the five template regions presented in \citet{peeters2024, chown2024, van-de-putte2024}. The decomposition is performed as follows. We selected bright lines from the line list %
based on model calculations using the \texttt{Cloudy} \citep{ferland2017} and \texttt{Meudon} PDR codes \citep{le-petit2006} \citep[for more details about the line list, see Section~4.1 of][]{peeters2024}. 
We apply a lower limit cut-off to the modeled intensity of $10^{-12}$\,W\,$\mathrm{m}^{-2}$\,sr$^{-1}$ in either the \hii\ region or PDR regions. The threshold is very conservative: it is two orders of magnitude lower than the noise level of the spectra used in this analysis in order not to miss any lines that maybe much stronger than the model. 
A local linear baseline was fit around each line on each spatial pixel in the spectroscopic cube.
We used these baselines to decompose the spectra into a set of bright line spectra and continuum spectra. Integrated line intensities were measured from the line spectra.
To extract PAH spectra, we used the Gaussian decomposition and underlying continuum presented in \citet{peeters2024} for NIRSpec (see their Fig. 16). For MIRI MRS, we determined the continuum as presented in \citet{chown2024} (see their Fig. C1), and extract the PAH spectra by subtracting the continuum and emission lines from the observed spectra. Here the plateau component is included in the PAH emission.

The decomposed spectra for bright line emission, continuum emission, and PAH emission were fed into Equation~\ref{eq:fcont} to yield maps of the fractional contributions of these emission components $\fcontrib{i,j}$ (e.g. $\fcontrib{\mathrm{Pa~\alpha},~\mathrm{F187N}}$) over the entire NIRSpec and MIRI MRS footprints. Tables~\ref{table:cont_frac_temp_all}  and~\ref{table:line_frac_temp_all} show these fractions in the five template spectra.
Figs.~\ref{fig:continuum_line_fractions_FeII}--\ref{fig:continuum_line_fractions_H2_469} and Figs.~\ref{fig:continuum_PAH_fractions_F770W}--\ref{fig:continuum_PAH_fractions_F1130W} show their results for the entire map (for NIRCam and MIRI filters, respectively) as well as their cumulative histograms.

\subsection{Line and PAH intensities from linear combinations of images}\label{subsec:lincomb}

\begin{figure*}
\begin{center}
\includegraphics[width=0.83\textwidth]{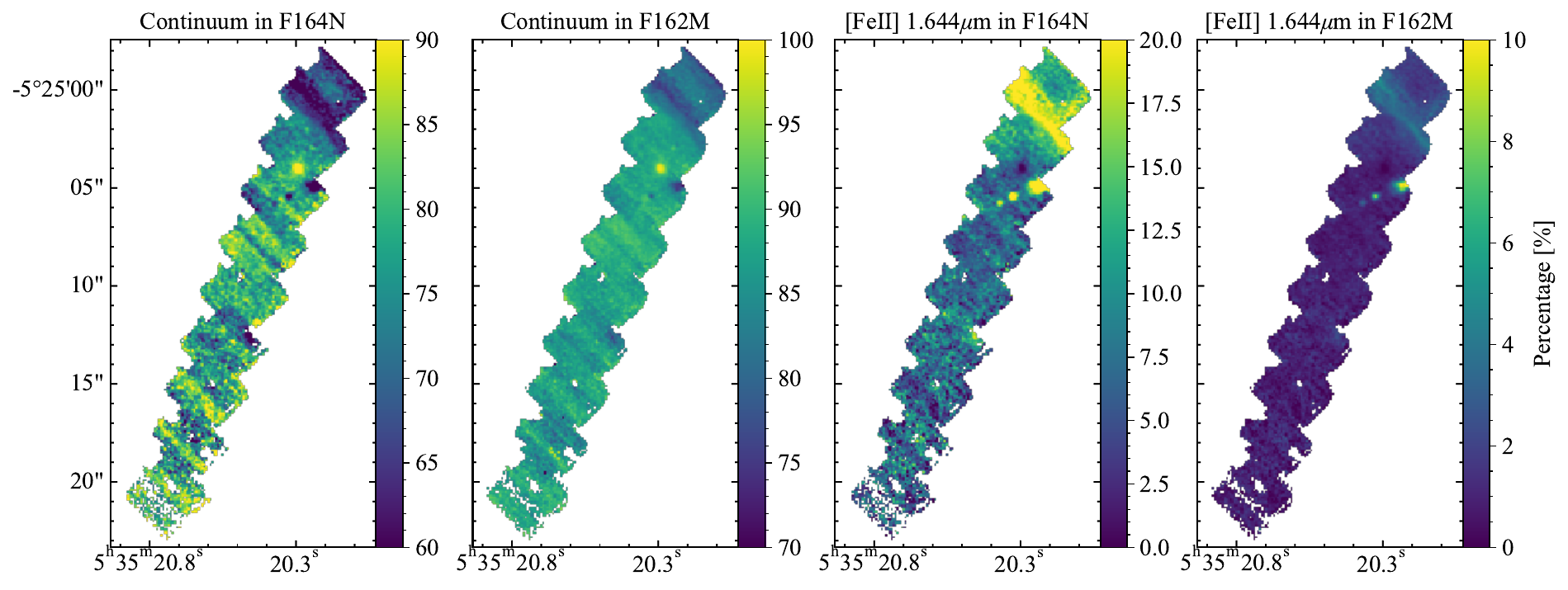}
\includegraphics[width=0.4\textwidth]
{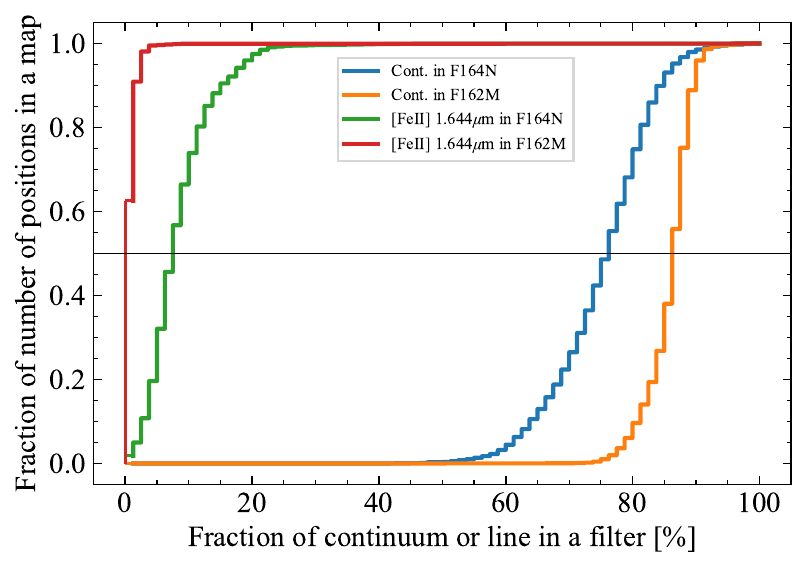}
\includegraphics[width=0.4\textwidth]{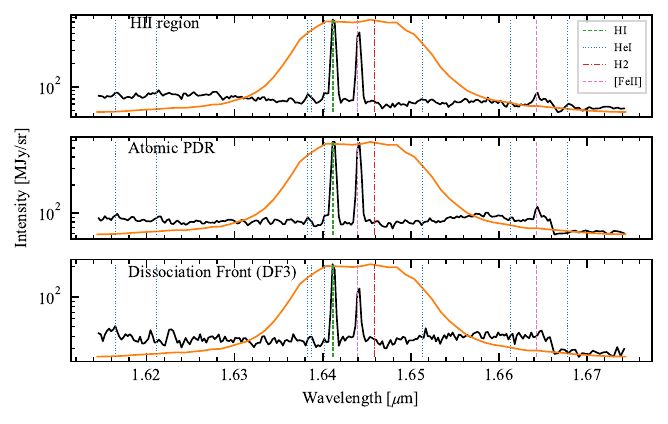}
\includegraphics[width=0.83\textwidth]
{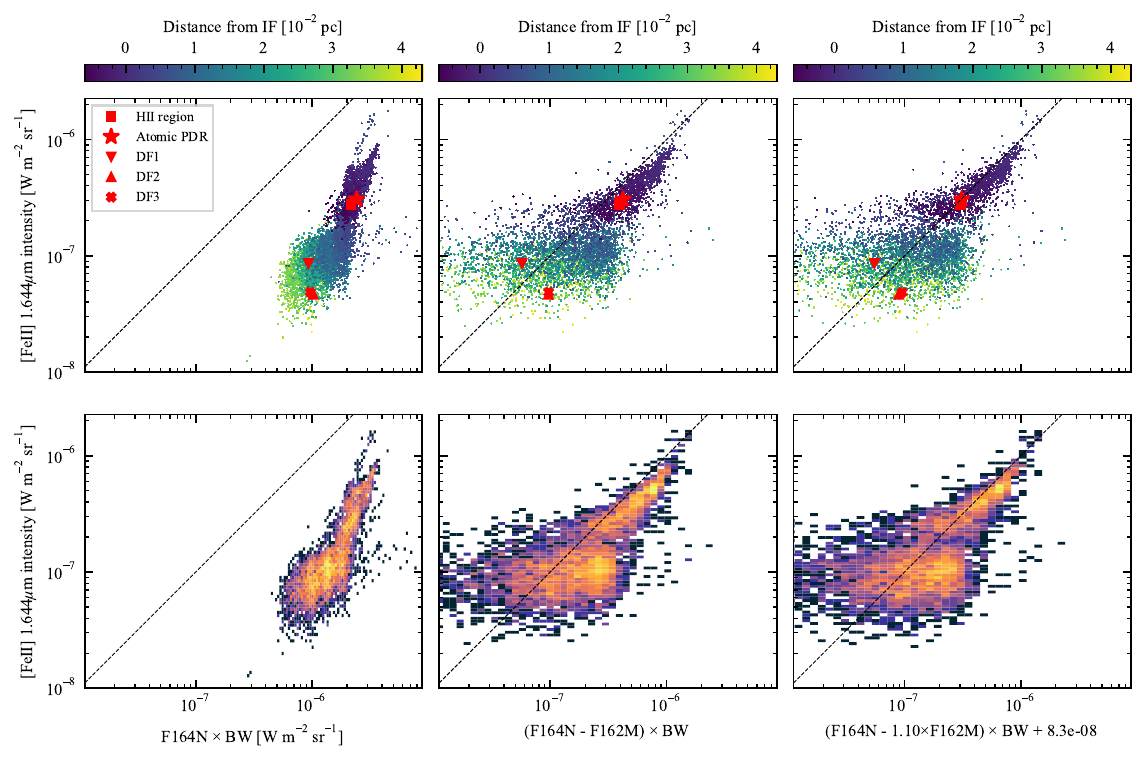}
\caption{Analysis of the continuum and \FeII\ 1.64~$\mu$m emission in F164N and F162M calculated from the NIRSpec spectra. \textbf{Top:} Maps of the continuum fraction in F164N and F162M (left two panels) and the \FeII\ 1.64~$\mu$m fraction (right two panels). \textbf{Middle left:} Cumulative histograms of the continuum and line fractions in the top panels. The black line indicates 50\%, i.e. the median value. \textbf{Middle right:} Three template spectra in the range covered by F164N. Response functions of the filters are shown with thin orange lines. Expected positions of \hi, \hei, and \molh, and \FeII\ lines are marked. \textbf{Bottom:} Correlation between the synthetic images (left: F164N, middle: F164N$-$F162M, right: the best fit with Eq.~\ref{eq:correlation}) and the measured \FeII\ 1.64~$\mu$m line intensities. x-axis is the surface brightness of the synthetic images multiplied by the band width (BW) of F164N (see main text). The second last row shows the correlations color coded by the distance of each pixel from the ionization front \citep[IF; see][]{peeters2024}, while the bottom row shows two-dimensional histograms color coded by the density of pixels. Only pixels detected at $3\sigma$ or higher are shown. 
The black line indicates the 1:1 relationship, i.e. perfect agreement between the x and y axes. Red points show measurements from the five template spectra. See Sec.~\ref{subsec:feii} for details.}
\label{fig:continuum_line_fractions_FeII}
\end{center}
\end{figure*}

\begin{figure*}
\begin{center}
\includegraphics[width=0.83\textwidth]{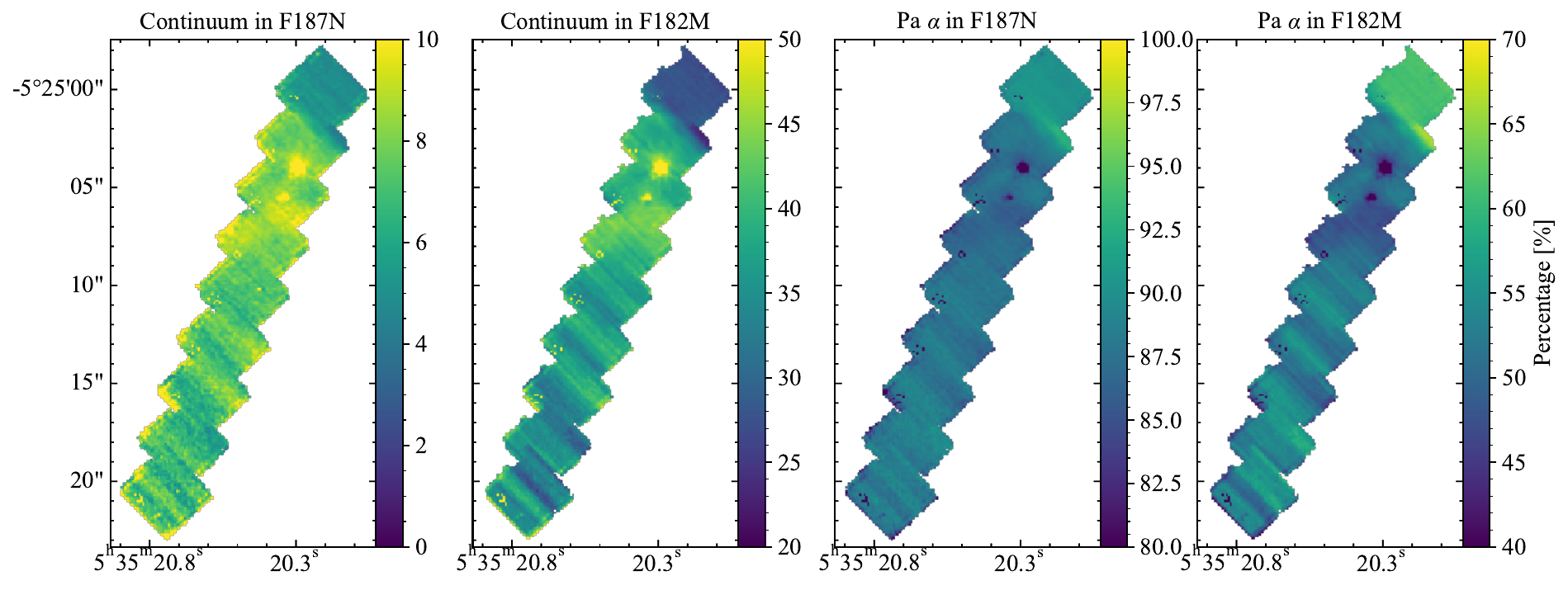}
\includegraphics[width=0.4\textwidth]{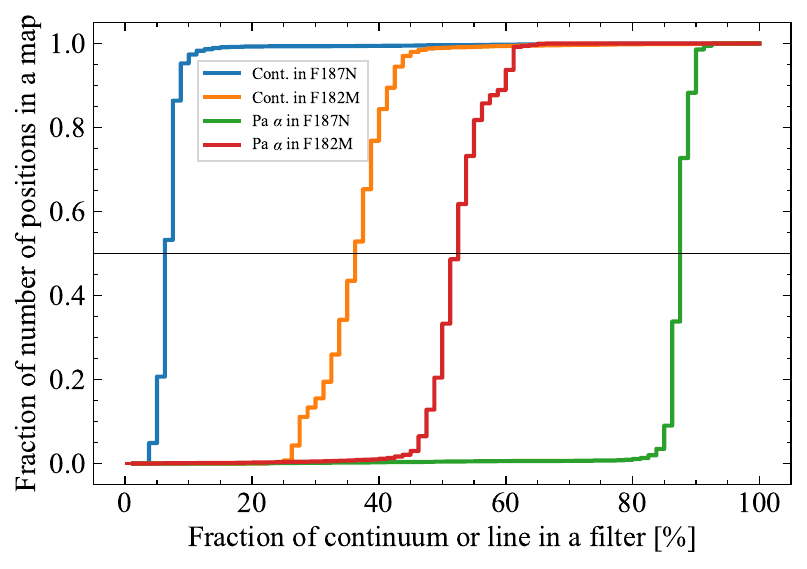}
\includegraphics[width=0.4\textwidth]{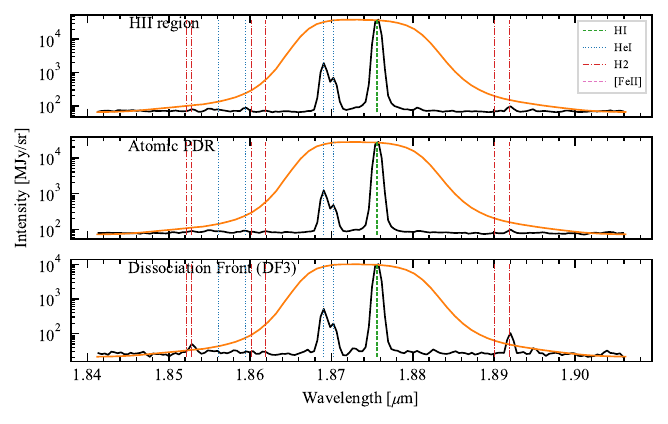}
\includegraphics[width=0.83\textwidth]{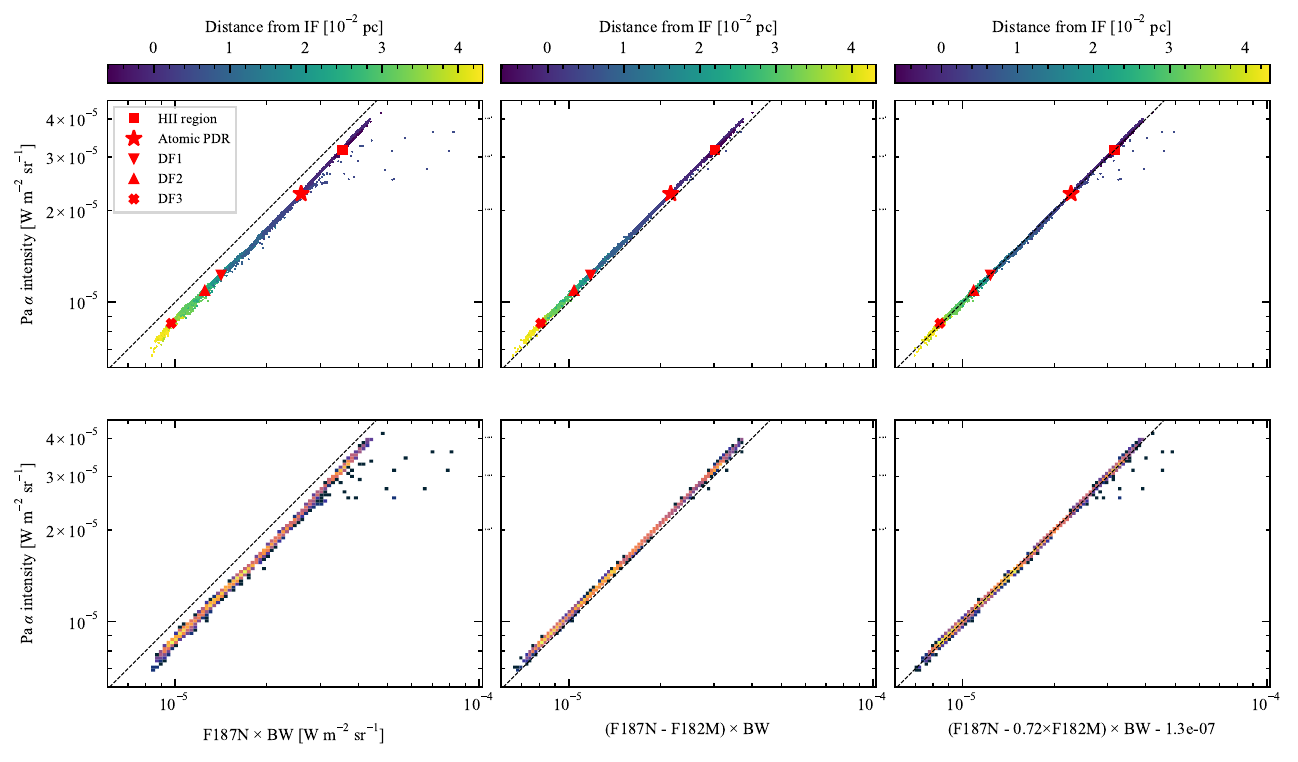}
\caption{Analysis of the continuum and Pa~$\alpha$ in F187N and F182M calculated from the NIRSpec spectra. \textbf{Top: } Maps of continuum fraction in F187N and F182M (left two panels) and Pa~$\alpha$ (right two panels). \textbf{Middle left: } Cumulative histograms of the continuum and line fractions in the maps above. \textbf{Middle right: } Template spectra in the range covered by F187N. \textbf{Bottom: } Correlations between the synthetic images (left: F187N, middle: F187N$-$F182M, right: the best fit with Eq.~\ref{eq:correlation}) and the measured Pa~$\alpha$ line intensities. Symbols, lines, and plot types are the same as in Fig.~\ref{fig:continuum_line_fractions_FeII}. See Sec.~\ref{subsec:paschen} for details.}
\label{fig:continuum_line_fractions_Pa_a} 
\end{center}
\end{figure*}

\begin{figure*}
\begin{center}
\includegraphics[width=0.83\textwidth]{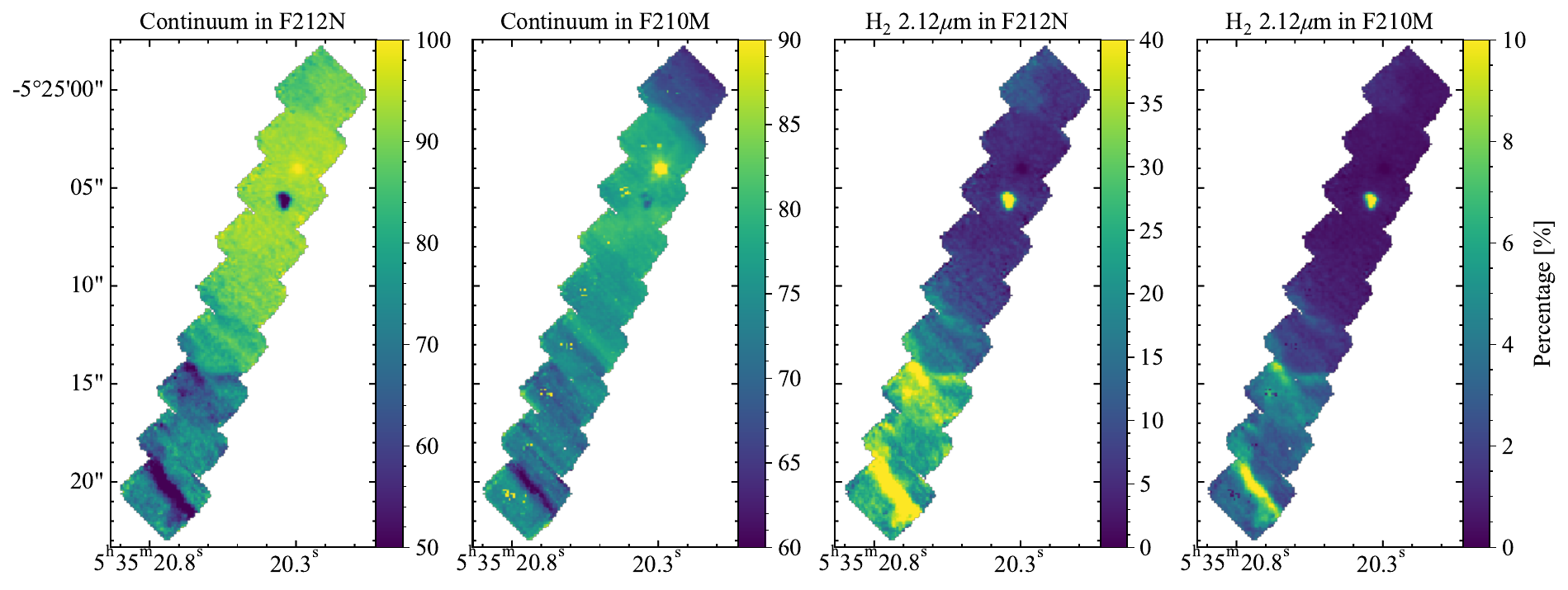}
\includegraphics[width=0.4\textwidth]{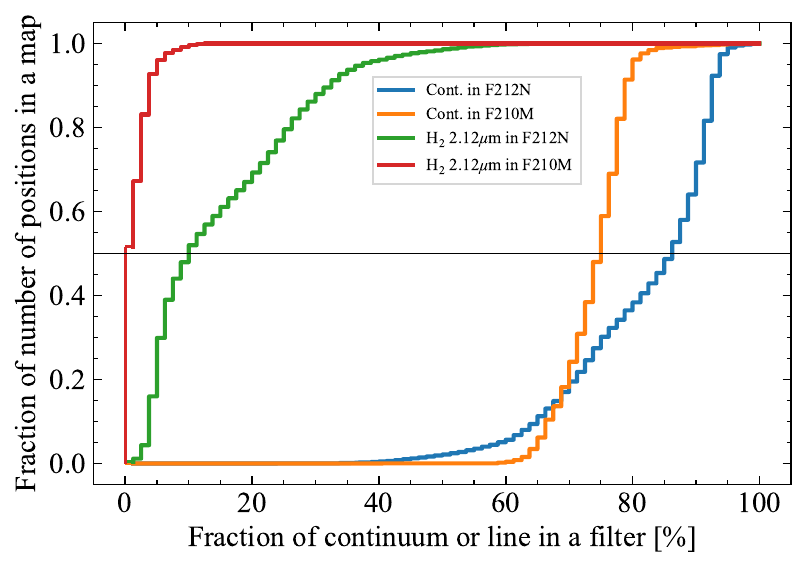}
\includegraphics[width=0.4\textwidth]{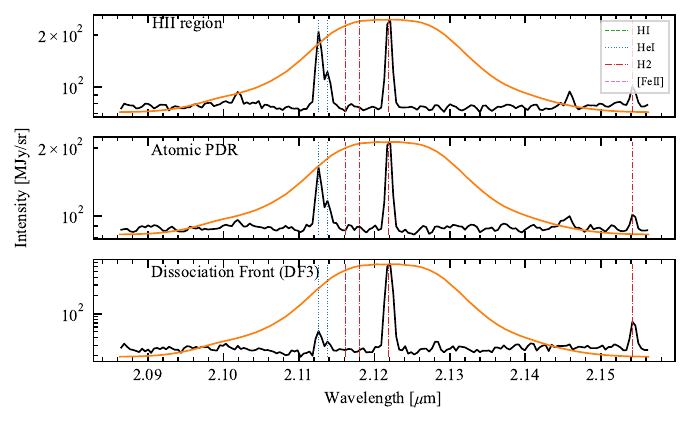}
\includegraphics[width=0.83\textwidth]{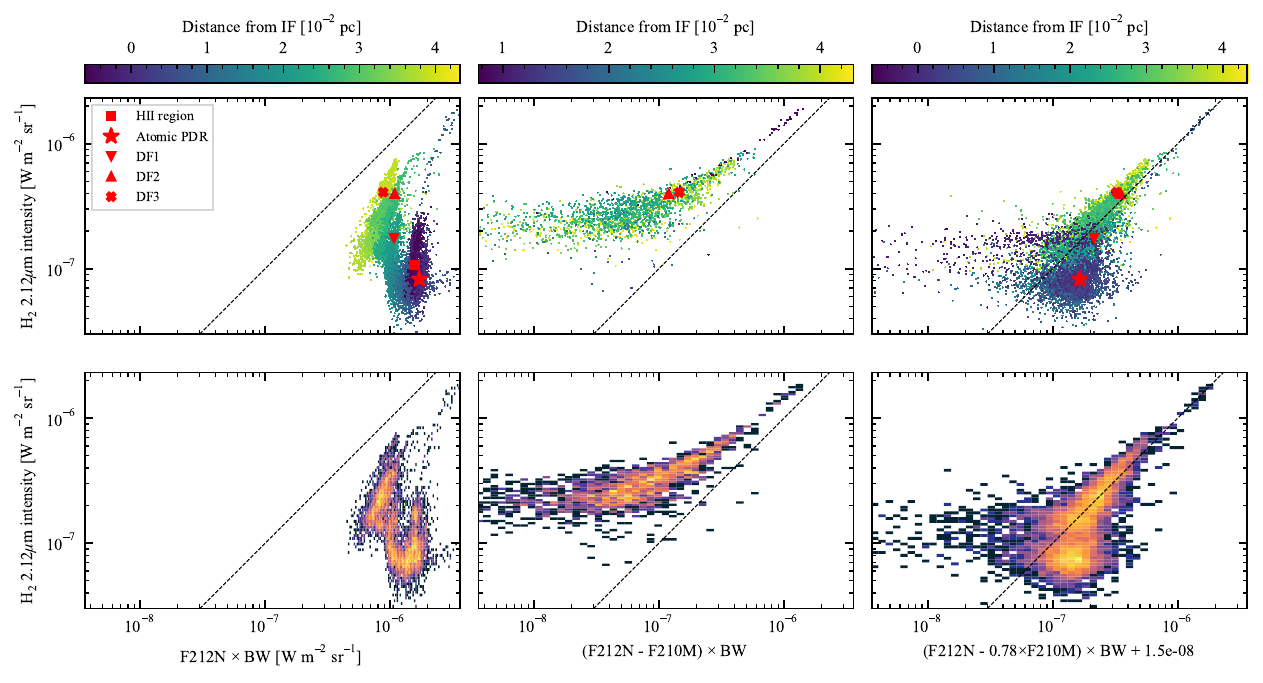}
\caption{Analysis of the continuum and H$_2$ 2.12~$\mu$m in F212N and F210M calculated from the NIRSpec spectra. \textbf{Top: } Maps of continuum fraction in F212N and F210M (left two panels) and H$_2$ 2.12~$\mu$m (right two panels). \textbf{Middle left: } Cumulative histograms of the continuum and line fractions in top panels. \textbf{Middle right: } Template spectra in the range covered by F212N. \textbf{Bottom: } Correlation between the synthetic images (left: F212N, middle: F212N$-$F210M, right: the best fit with Eq.~\ref{eq:correlation}) and the measured H$_2$ 2.12~$\mu$m line intensities.  In the bottom middle panels (F212N$-$F210M), positions closer to the IF (including the templates of \hii\ region, Atomic PDR and DR1) give negative values, therefore not shown. Symbols, lines, and plot types are the same as in Fig.~\ref{fig:continuum_line_fractions_FeII}. See Sec.~\ref{subsec:h2_2p12} for details.
}
\label{fig:continuum_line_fractions_H2_212}
\end{center}
\end{figure*}

In order to derive prescriptions for the emission line and PAH intensities based on NIRCam and MIRI images, we first plot the correlations between the line or PAH intensity (on the y-axis) and the synthetic image of the filter that captures the 
emission of interest (on the x-axis, shown in the bottom left panels of Fig.~\ref{fig:continuum_line_fractions_FeII}--\ref{fig:continuum_line_fractions_H2_469} for NIRCam, and Figs.~\ref{fig:continuum_PAH_fractions_F770W}--\ref{fig:continuum_PAH_fractions_F1130W} for MIRI). The PAH features are not confined within each MIRI filter, so we use the synthetic image of the PAH component in F770W and F1130W (namely the PAH component within each filter) as the y-axis quantity instead of the total integrated intensity. Because all of the MIRI filters are broad and the fractional contribution of emission lines to the total intensity is at most a few percent (Table~\ref{table:line_frac_temp_all}), we do not provide prescriptions to measure line intensities from MIRI filters, but simply show their correlation with the corresponding synthetic image in Fig.~\ref{fig:MIRI_line_fractions}.

\input{table_continuum_fraction_all}

\input{table_line_fraction_all}

The simplest prescription that takes into account the continuum contribution is $(F^\mathrm{synth.}_{\mathrm{A}} - F^\mathrm{synth.}_{\mathrm{B}})$, where filter ``A'' captures the emission of interest and filter ``B'' is used to subtract the continuum emission that contributes to filter A. The filter combinations are given in Table~\ref{table:N-W_fit}, and the correlations between $(F^\mathrm{synth.}_{\mathrm{A}} - F^\mathrm{synth.}_{\mathrm{B}})$ and the line or PAH intensities are shown in bottom middle panels of Figs.~\ref{fig:continuum_line_fractions_FeII}--\ref{fig:continuum_PAH_fractions_F1130W}.

\input{table_N-W_fit_all_pointings_paper}

Pixels are included in this and later correlation analysis if the following conditions are satisfied:
\begin{enumerate}
    \item the line (or PAH) intensity in the pixel is greater than three times the RMS noise in the underlying continuum, and
    \item $F^\mathrm{synth.}_{\mathrm{filter}} > 3\sigma^\mathrm{synth.}_{\mathrm{filter}}$ for both ``A'' and ``B'' bands, where $\sigma^\mathrm{synth.}_{\mathrm{filter}}$ is the uncertainty of the synthetic image (Section~\ref{subsec:synthetic_images}).
\end{enumerate} 
We then model the relationship between the intensity of an emission line or PAH feature $I_\mathrm{line~or~PAH}$ according to
\begin{equation}
    I_\mathrm{line~or~PAH} = \left(F^\mathrm{synth.}_{\mathrm{A}} - \alpha F^\mathrm{synth.}_{\mathrm{B}}\right)~\mathrm{BW_A} + \beta, \label{eq:correlation}
\end{equation}
where filter ``A'' and ``B'' are the same as above; $\mathrm{BW_A}$ is the bandwidth of filter A (in \micron), and $\alpha$, $\beta$ are free parameters. The factor $\mathrm{BW_A}$ is required in order to convert the units of the synthetic images ($F_\nu$ in MJy~sr$^{-1}$) to line intensity units (W~m$^{-2}$~sr$^{-1}$). This conversion was performed except for the PAH 7.7\micron\ and 11.2\micron, where the PAH features are more continuous and there is no unique way to define the intensity of each PAH feature in the y-axis (see Sect.~\ref{subsec:7p7_11p3_pah}).

We used \texttt{linmix} (as in Sec.~\ref{subsec:cross-cal})
to fit for $\alpha$ and $\beta$, incorporating uncertainties in all measured quantities. 
Before fitting, we rearranged Eq.~\ref{eq:correlation}. Specifically, we fit $y=\alpha x + \beta$, with
\begin{equation}
    y \equiv I_\mathrm{line~or~PAH} - F^\mathrm{synth.}_{\mathrm{A}}~\mathrm{BW_A}, \label{eq:correlation_rearranged}
\end{equation}
and
\begin{equation}
    x \equiv - F^\mathrm{synth.}_{\mathrm{B}}~\mathrm{BW_A}.
    \label{eq:correlation_rearranged2}
\end{equation}
Uncertainties for $y$ were computed by propagating formal uncertainties in $I_\mathrm{line~or~PAH}$ and $F^\mathrm{synth.}_{\mathrm{A}}$, ignoring covariance between these variables. Covariance is likely to be significant since both $I$ and $F$ were computed from the same NIRSpec spectrum, and as a result, the uncertainties are likely underestimated. 
To test the impact of covariance, Monte Carlo calculations of the uncertainty in $F^\mathrm{synth.}_{\mathrm{A}}-F^\mathrm{synth.}_{\mathrm{B}}$ shows that the true uncertainty (including covariance) is about 30\% larger than the individual uncertainties added in quadrature. 
We did not include these errors into the fit because deriving the precise errors for all pixels is computationally too expensive. 
On the x-axis of the bottom right columns of Figs.~\ref{fig:continuum_line_fractions_FeII}--\ref{fig:continuum_PAH_fractions_F1130W}, the right-hand side of Eq.~\ref{eq:correlation} is computed from our best-fit values of $\alpha$ and $\beta$, which are summarized in Table~\ref{table:N-W_fit}. We tested that these values are insensitive to the used signal-to-noise ratio (for example less than a few percent difference when changing the threshold from 3$\sigma$ to 5$\sigma$) because the lines discussed here have sufficient signal-to-noise at the majority of the pixels.

\input{table_line_prescriptions}
The goodness of the tested prescriptions is summarized in Table~\ref{table:line_prescriptions}, by providing the ratio between the expected line or PAH intensity using each prescription and the actual intensity for five template spectra. For the 11.2\um\ feature in the F1130W filter, we also provide a prescription combining the F1000W and F1500W filters (see \S\ref{subsec:7p7_11p3_pah} and footnote to Table~\ref{table:line_prescriptions}).
In the correlation plots of Figs.~\ref{fig:continuum_line_fractions_FeII}--\ref{fig:continuum_PAH_fractions_F1130W}, we illustrate deviations from a 1:1 relation which is shown as black dashed lines. These deviations can originate from contributions from other lines or PAHs, and/or inaccurate representation of the continuum contribution by an imaging filter ``B''  when the continuum level is not constant across the wavelength range spanned by the two filters. We calculate the average distance of the data to the 1:1 line (the expected relationship) in each correlation plot (in log space) as a measure of the overall scatter about the expected relationship. This scatter is reported in the last column of Table~\ref{table:line_prescriptions}. The rows in boldface in the Table indicate the recommended prescription (having the smallest scatter) to be used to trace the intensity of each emission line or PAH feature. In the following, we present our analysis of empirical prescriptions for emission line and PAH feature intensities based on linear combinations of \jwst\ images.
We note that the two proplyds, d203-504 and d203-506 \citep[see Fig.~1 of][]{chown2024} are not masked in our analysis, but their influence on the fitted quantities are insignificant. These proplyds will be separately discussed in detail in Vicente et al. (in prep.) and are not further discussed in this paper.

\subsection{Prescriptions}\label{subsec:prescriptions}

\subsubsection{F164N--F162M as a tracer of \FeII\ 1.64~\um\ line emission}\label{subsec:feii}

Figure~\ref{fig:continuum_line_fractions_FeII} shows the spatial distributions of the fraction of continuum and \FeII\ 1.64\um\ emission to the F164N and F162M filters (Eq.~\ref{eq:fcont}). The maps indicate that  $\fcontrib{\mathrm{cont.},~\mathrm{F164N}}$ and $\fcontrib{\mathrm{cont.},~\mathrm{F162M}}$ are distributed roughly into two regimes: the \ion{H}{ii} region, where the continuum fraction is lowest; regions inside the molecular cloud (south of the Atomic PDR), where the continuum fraction is highest. The second row (left) in the figure shows the cumulative distributions of the maps above, where the medians can be read off: F162M is $\approx 90$\% continuum, while F164N is $\approx 75$\% continuum. As shown in the template spectra (second row right in the figure), the rest of the emission in F164N originates from the \FeII\ 1.64\um\ and \hi\ (12-4) lines). 

The two bottom rows of Fig.~\ref{fig:continuum_line_fractions_FeII} show the relationships between \FeII\ 1.64~\um\ intensity and different linear combinations of F164N and F162M. The second last row shows how the trend depends on distance from the ionization front, while the bottom row is color-coded by the number density of pixels. One can see that as expected, F164N on its own does not trace the \FeII\ line intensity. The difference F164N--F162M provides a closer correspondence with \FeII\ line intensity, but there is significant scatter in this relationship in pixels toward the dissociation fronts (DFs). The rightmost panel -- the fit based on Eq.~\ref{eq:correlation}, with best-fit parameters provided in Table~\ref{table:N-W_fit} -- provides the better fit to the data in pixels closest to the ionizing source on one hand, but the scatter toward the DFs remains or is even slightly worse. This is likely due to the fact that the fit with Eq.~\ref{eq:correlation} is dominated by the pixels with stronger line intensities. Indeed the scatter about the 1:1 relationship, which is measured in the log-space, is slightly worse in the right columns (shown in Table~\ref{table:line_prescriptions}). Therefore we recommend to use the simple difference (F164N--F162M) with keeping the scatter in mind.

\subsubsection{F187N--F182M as a tracer of Pa~$\alpha$ line emission}\label{subsec:paschen}

The Pa~$\alpha$ line contributes $\approx 90$\% to the total emission in the F187N filter (Fig.~\ref{fig:continuum_line_fractions_Pa_a}) without subtracting the underlying continuum emission using the F182M band. The top row of the figure shows that the fraction of continuum emission and line emission follows a qualitatively similar spatial distribution as in the \FeII\ line (\S\ref{subsec:feii}), namely the fraction of the line contribution is stronger in the ionized region.

As shown in the second row (left) of Fig.~\ref{fig:continuum_line_fractions_Pa_a}), continuum emission contributes 30--50\% to the total surface brightness in F182M, while Pa~$\alpha$ contributes 40--60\%. This raises a caution to use the F182M filter as a continuum band. The bottom two rows in the Figure show very strong correlations ($r\approx 1.0$) between Pa~$\alpha$ intensity and combinations of F187N and F182M. This is the strongest correlation we find. The second lowest row shows that both x and y axes strongly correlate with distance from the ionization front. The intensities from each template spectrum follow this trend of increasing intensity with as distance from the IF decreases. 
The middle column in the bottom row shows that F187N--F182M intensity is slightly lower than the true Pa~$\alpha$ intensity (by about a few per cent), which is due to self-subtraction of the Pa~$\alpha$ line. The right column of the bottom rows show the fit based on Eq.~\ref{eq:correlation}, with best-fit parameters shown in Table~\ref{table:N-W_fit}, has the lowest scatter (Table~\ref{table:line_prescriptions}) and is therefore our recommended prescription for inferring Pa~$\alpha$ intensity. The dominant fit parameter is $\alpha$ (value of $0.72$), whereas the intercept $\beta$ is almost negligible, which may indicate that the underlying continuum between these filters is almost identical.

\subsubsection{F212N--F210M as a tracer of H$_2$ 2.12\um}
\label{subsec:h2_2p12}

The top row of Fig.~\ref{fig:continuum_line_fractions_H2_212} shows that the H$_2$ 2.12\um\ contributions to F212N and F210M vary strongly across the mapped region, ranging from $\approx 0$ to $50$\% (F212N). The highest fractions appear in the molecular PDRs (as expected), followed by a diffuse component in the southern half of the mapped region (where the gas is molecular). As expected, the F212N filter has a higher fraction of H$_2$ 2.12\um\ emission than F210M, which is seen in the maps as well as the second row. The cumulative distributions show median fractional continuum contributions of $\approx 85$\% (with a strong variation across the field) in F212N, while $\approx 75$\% in F210M. This is due to additional bright \hi\ and \hei\ emission lines that fall in the F210M filter but fall outside of the F212N filter. 

The bottom rows of the Figure show that  H$_2$ 2.12\um\ intensity is poorly correlated with F212N. Subtracting continuum from F212N using F210M provides a better correlation, but with significantly underestimated line intensity due to the presence of \hi\ and \hei\ lines in F210M that are not captured by F212N, and the self-subtraction of H$_2$ 2.12\um\ at DFs. This even leads to negative fluxes in a large fraction of the mosaic closer to the exciting sources. The fit of H$_2$ 2.12\um\ against a linear combination of F212N and F210M using  Eq.~\ref{eq:correlation} significantly improves this problem, although scatter remains closer to the exciting sources. In summary, we recommend using the prescription with $\alpha$ and $\beta$ as free parameters (Table~\ref{table:N-W_fit}), however we advise caution when applying this calibration to sightlines where strong \hi\ and \hei\ emission lines are expected.%

\begin{figure*}
\begin{center}
\includegraphics[width=0.7\textwidth]{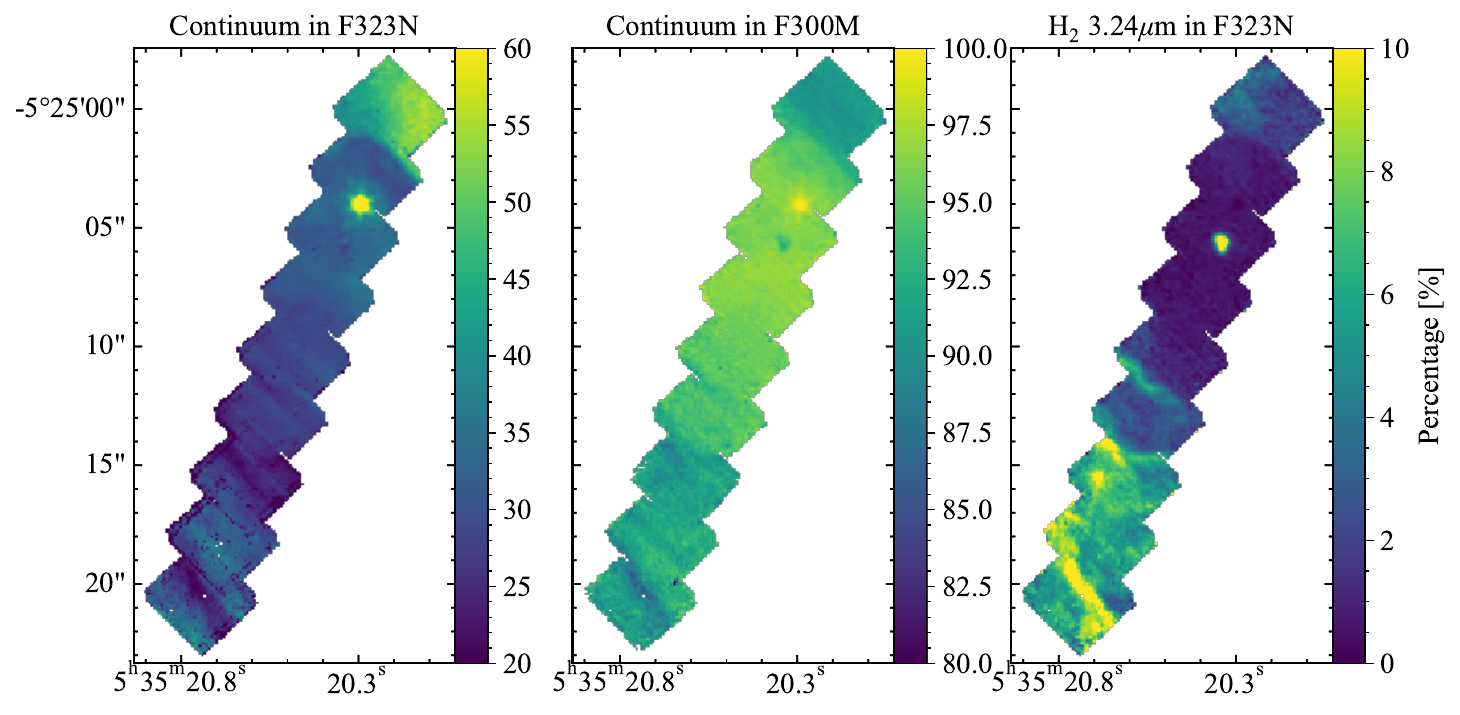}
\includegraphics[width=0.38\textwidth]{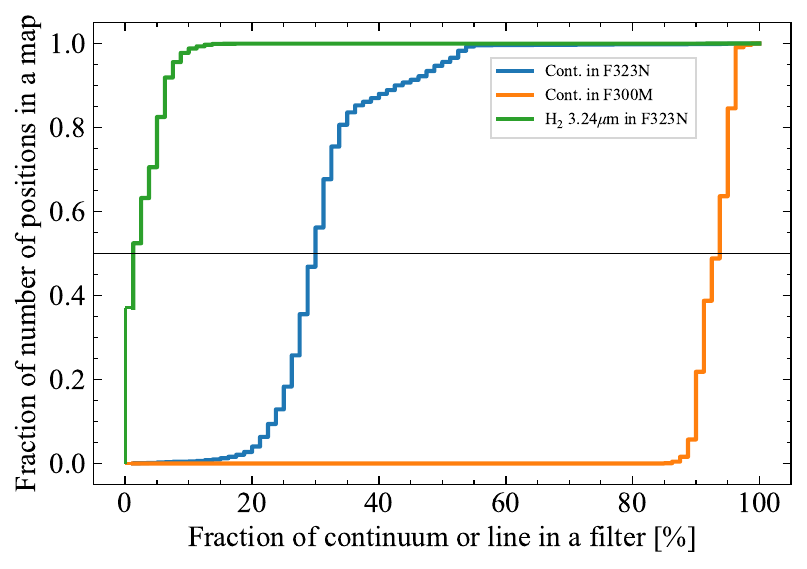}
\includegraphics[width=0.4\textwidth]{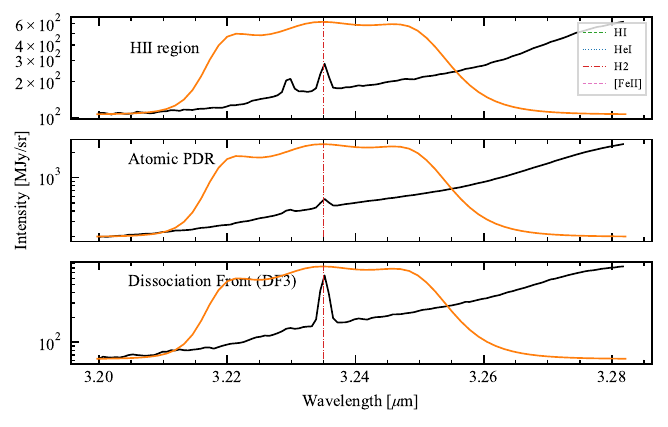}
\includegraphics[width=0.7\textwidth]{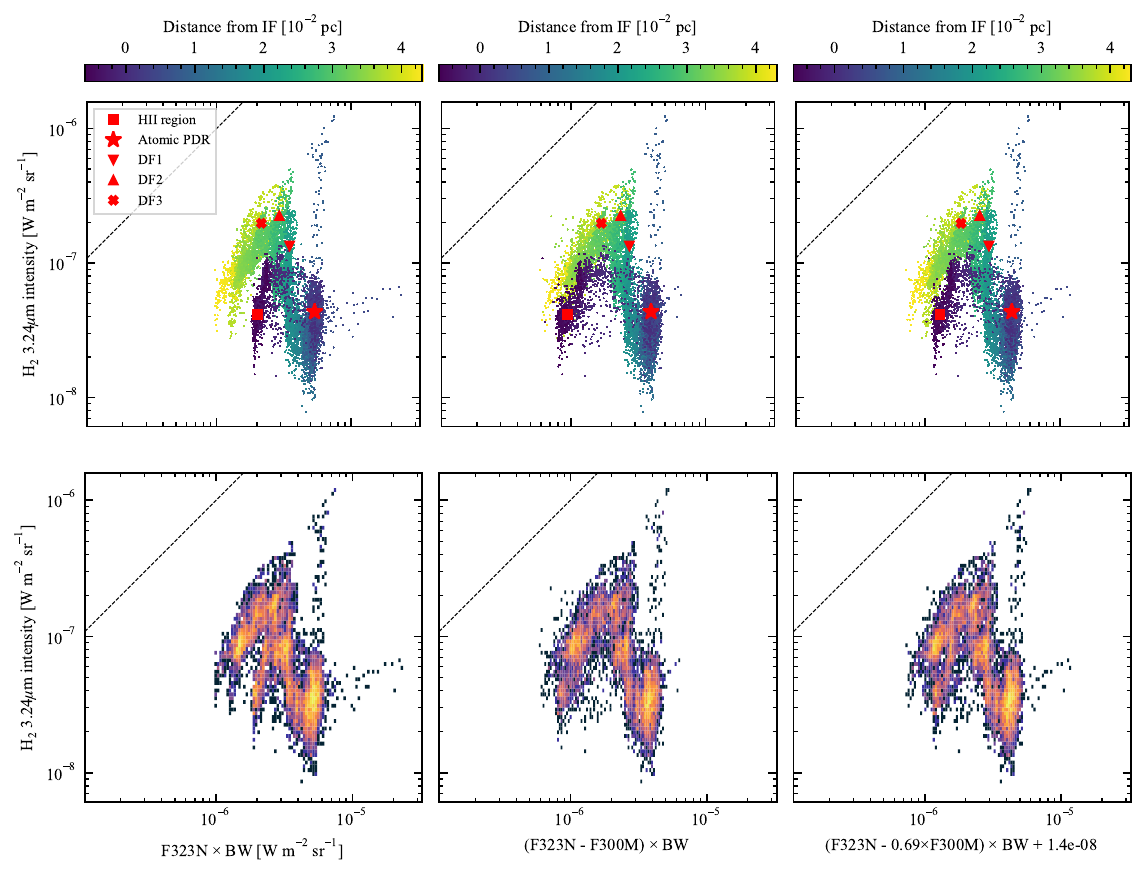}
\caption{Analysis of the continuum and H$_2$ 3.23~$\mu$m in F323N and F300M calculated from the NIRSpec spectra. \textbf{Top: } Maps of continuum fraction in F323N and F300M (left two panels) and H$_2$ 3.23~$\mu$m (right). H$_2$ 3.23~$\mu$m is outside of the filter F300M so its fraction in F300M is not shown. \textbf{Middle left: } Cumulative histograms of the continuum and line fractions in top panels. \textbf{Middle right: } Template spectra in the range covered by F323N. \textbf{Bottom: } Correlation between the synthetic images (left: F323N, middle: F323N$-$F300M, right: the best fit with Eq.~\ref{eq:correlation}) and the measured H$_2$ 3.23~$\mu$m line intensities. Symbols, lines, and plot types are the same as in  Fig.~\ref{fig:continuum_line_fractions_FeII}. See Sec.~\ref{subsec:h2_3p23} for details.}
\label{fig:continuum_line_fractions_H2_323}
\end{center}
\end{figure*}

\begin{figure*}
\begin{center}
\includegraphics[width=0.7\textwidth]{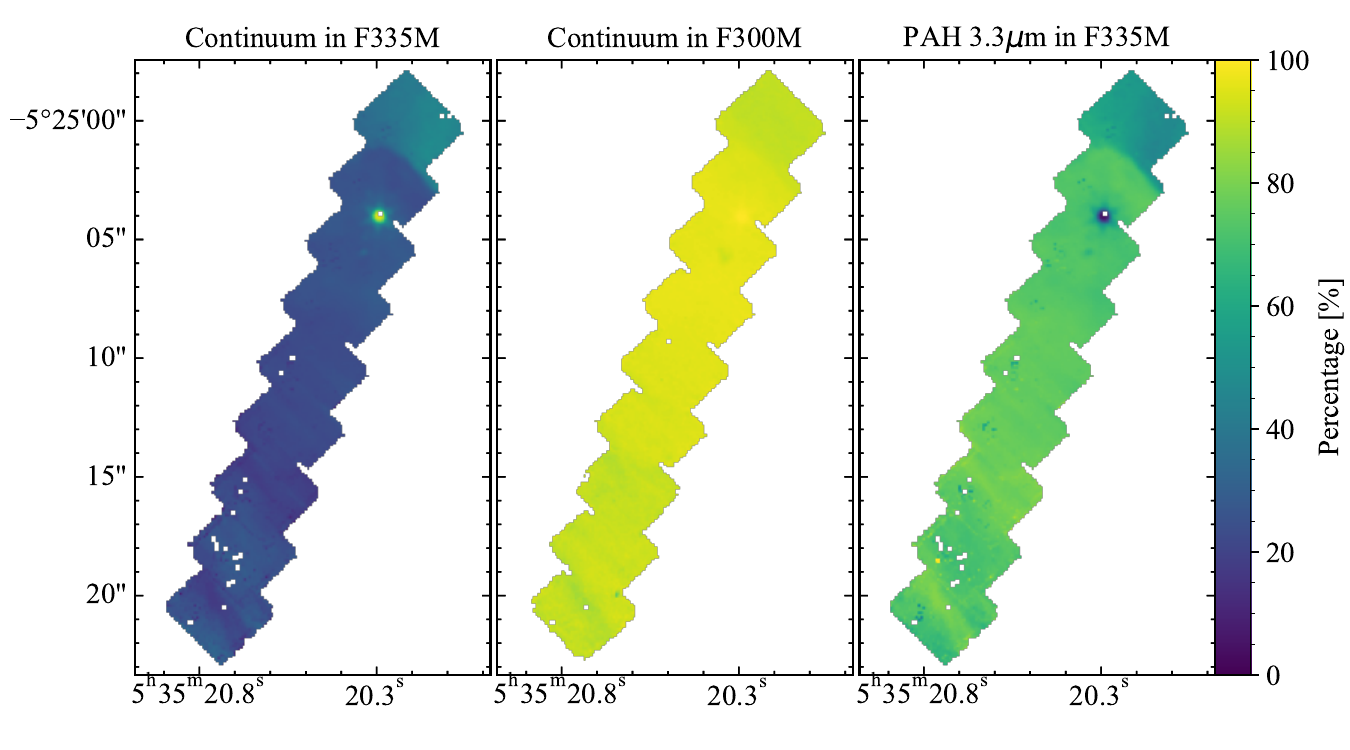}
\includegraphics[width=0.38\textwidth]{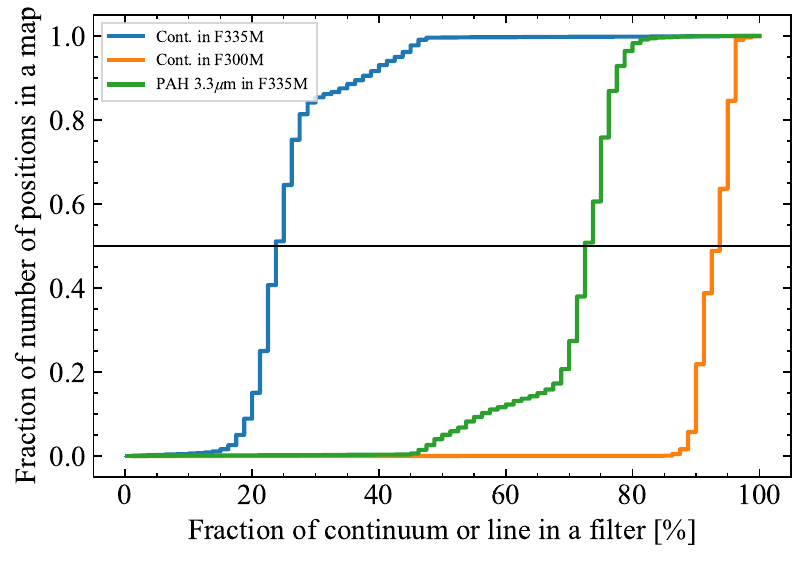}
\includegraphics[width=0.4\textwidth]{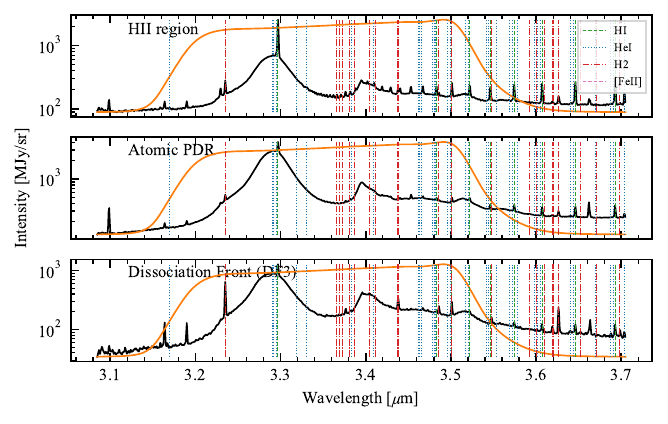}
\includegraphics[width=0.7\textwidth]{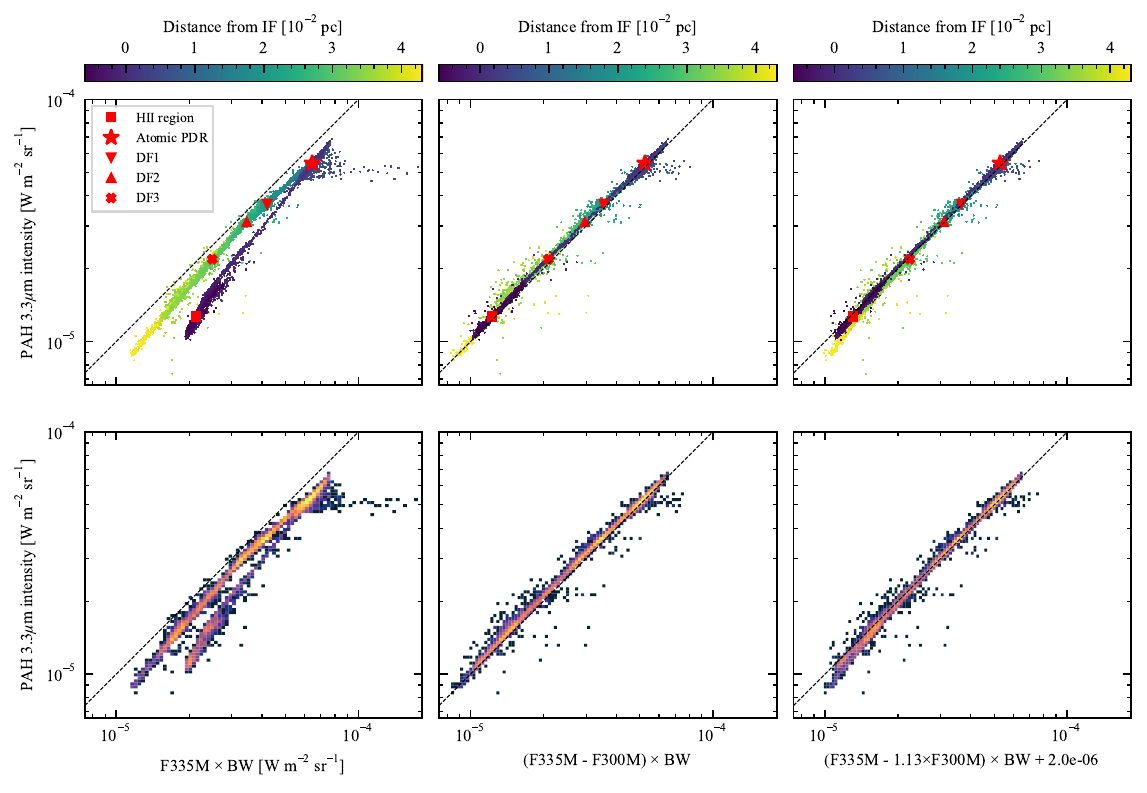}
\caption{Analysis of the continuum and PAH 3.3~$\mu$m emission in F335M and F300M calculated from the NIRSpec spectra. \textbf{Top: } Maps of the continuum fraction in F335M and F300M (left two panels) and the PAH 3.3~$\mu$m (right). The PAH 3.3~$\mu$m is outside of the filter F300M so its fraction in F300M is not shown. \textbf{Middle left: } Cumulative histograms of the continuum and line fractions in top panels. \textbf{Middle right: } Template spectra in the range covered by F335M.\textbf{Bottom: } Correlation between the synthetic images (left: F335M, middle: F335M$-$F300M, right: the best fit with Eq.~\ref{eq:correlation}) and the measured PAH intensities. Symbols, lines, and plot types are the same as in Fig.~\ref{fig:continuum_line_fractions_FeII}. See Sect.~\ref{subsec:3p3_pah} for details. }
\label{fig:continuum_line_fractions_AIB}
\end{center}
\end{figure*}

\begin{figure*}
\begin{center}
\includegraphics[width=0.7\textwidth]{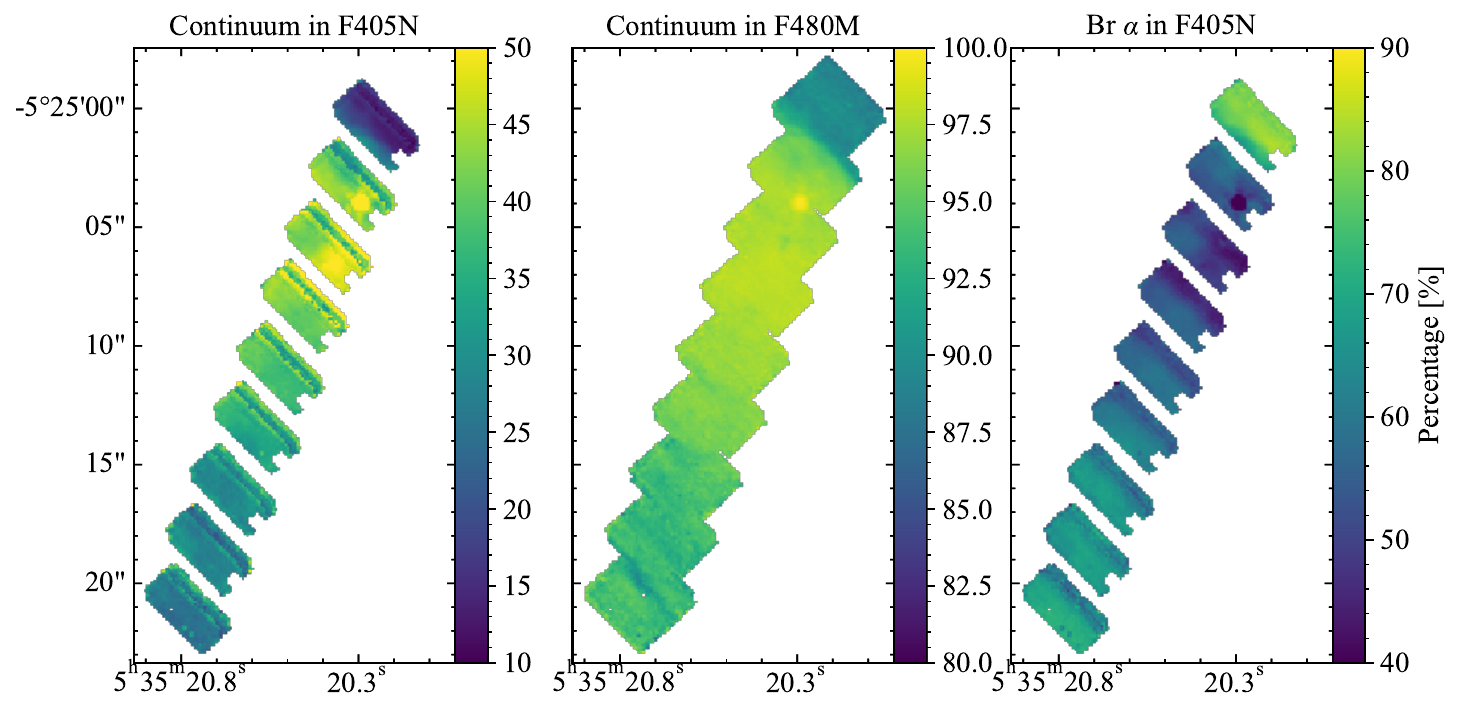}
\includegraphics[width=0.38\textwidth]{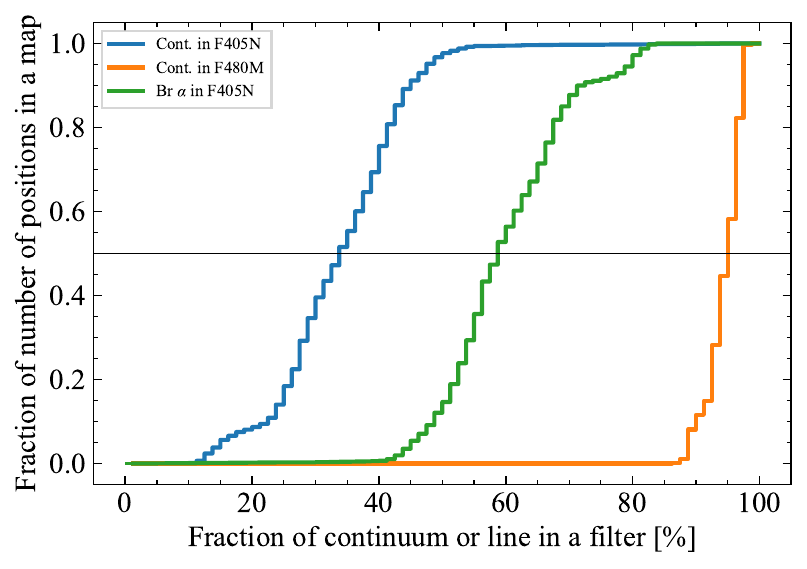}
\includegraphics[width=0.4\textwidth]{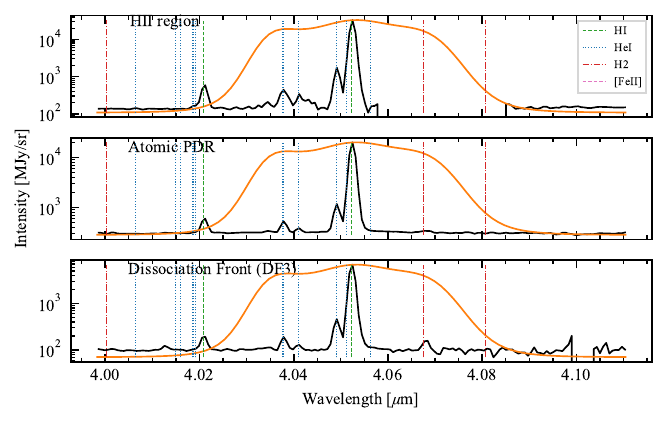}
\includegraphics[width=0.8\textwidth]{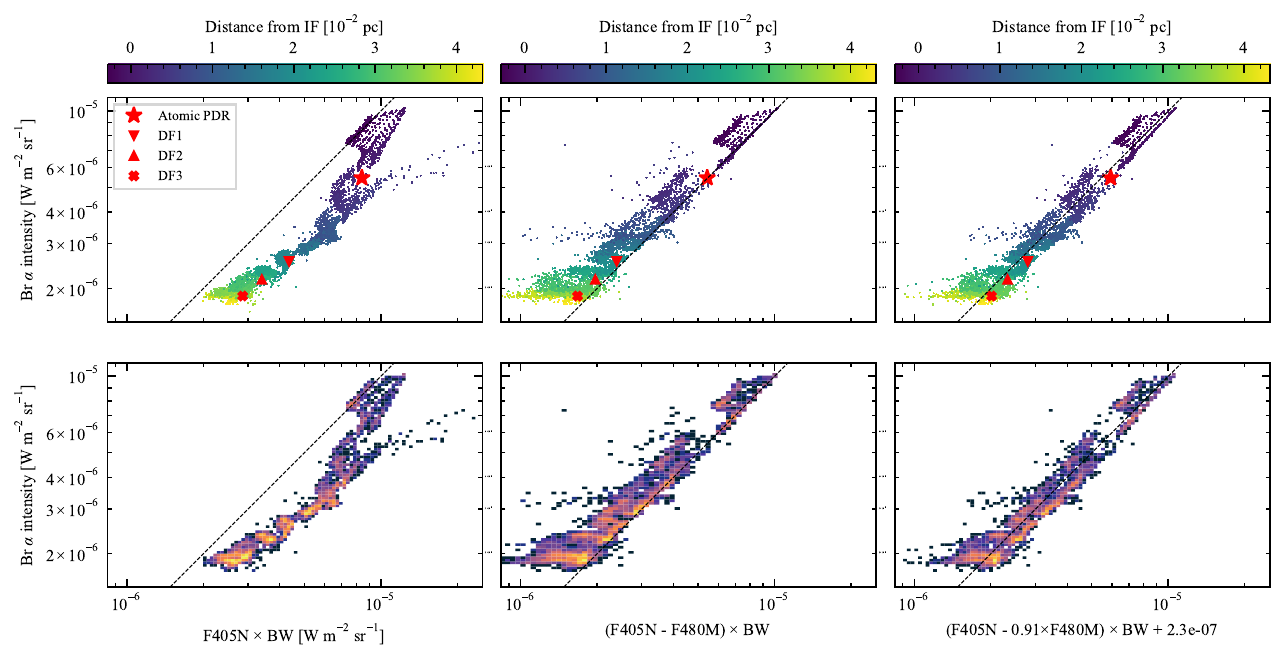}
\caption{Analysis of the continuum and Br~$\alpha$ in F405N and F480M calculated from the NIRSpec spectra. \textbf{Top: } Maps of continuum fraction in F405N and F480M (left two panels) and Br~$\alpha$ (right two panels). \textbf{Middle left: } Cumulative histograms of the continuum and line fractions in top panels. \textbf{Middle right: } Template spectra in the range covered by F405N. \textbf{Bottom: } Correlation between the synthetic images (left: F405N, middle: F405N$-$F480M, right: the best fit with Eq.~\ref{eq:correlation}) and the measured Br~$\alpha$ line intensities. We excluded the data point of the HII region template, because the wavelength of Br~$\alpha$ falls into the gap of NIRSpec in most of the pixels that consist of the template, and thus it does not represent well the template region. Symbols, lines, and plot types are the same as in  Fig.~\ref{fig:continuum_line_fractions_FeII}. See Sect.~\ref{subsec:brackett} for details.}
\label{fig:continuum_line_fractions_Br_a}
\end{center}
\end{figure*}

\begin{figure*}
\begin{center}
\includegraphics[width=0.83\textwidth]{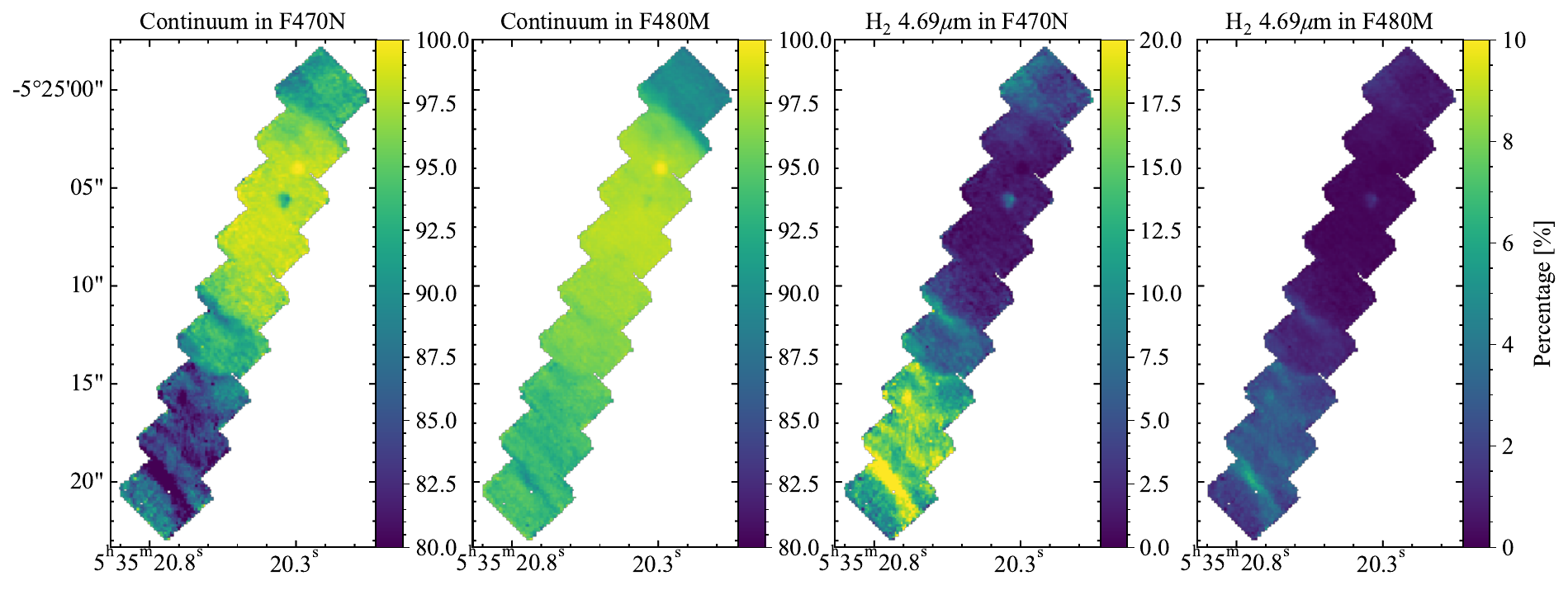}
\includegraphics[width=0.4\textwidth]{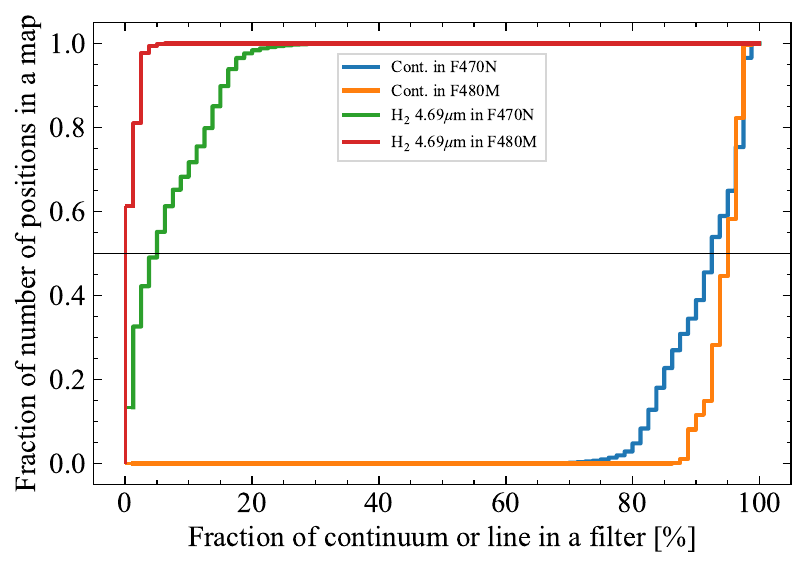}
\includegraphics[width=0.4\textwidth]{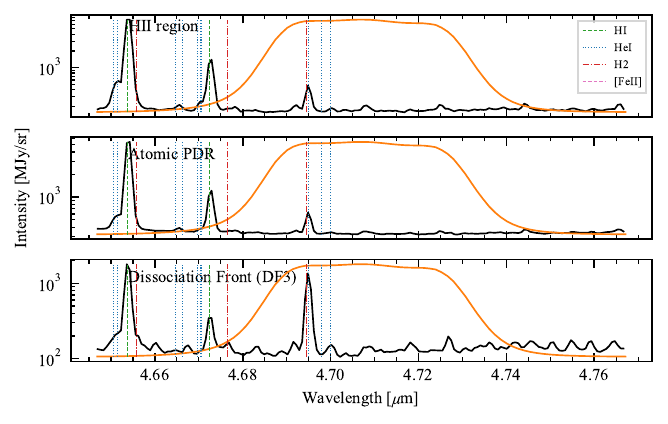}
\includegraphics[width=0.83\textwidth]{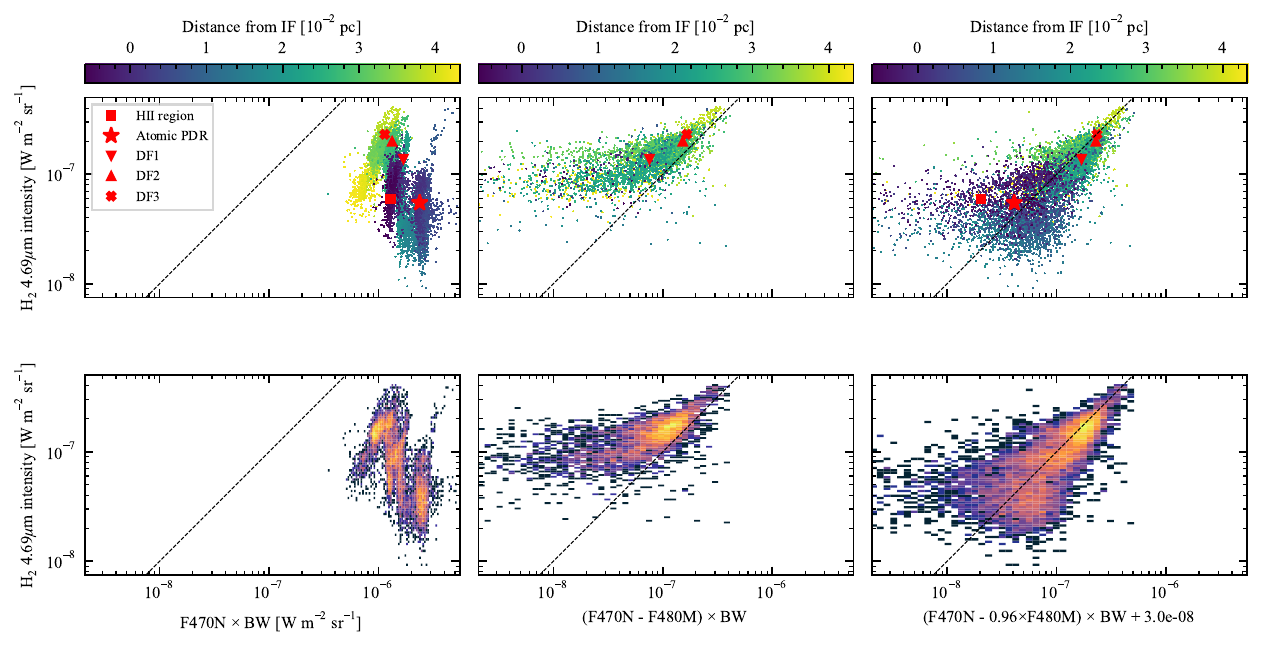}
\caption{Analysis of the continuum and H$_2$ 4.69~$\mu$m in F470N and F480M calculated from the NIRSpec spectra. \textbf{Top: } Maps of continuum fraction in F470N and F480M (left two panels) and H$_2$ 4.69~$\mu$m (right two panels). \textbf{Middle left: } Cumulative histograms of the continuum and line fractions in top panels. \textbf{Middle right: } Template spectra in the range covered by F470N. \textbf{Bottom: } Correlation between the synthetic images (left: F470N, middle: F470N$-$F480M, right: the best fit with Eq.~\ref{eq:correlation}) and the measured H$_2$ 4.69~$\mu$m line intensities. Symbols, lines, and plot types are the same as inFig.~\ref{fig:continuum_line_fractions_FeII}. See Sect.~\ref{subsec:h2_4p69} for details.}
\label{fig:continuum_line_fractions_H2_469}
\end{center}
\end{figure*}

\subsubsection{F323N--F300M as a tracer of H$_2$ 1--0 O(5) 3.23\um}
\label{subsec:h2_3p23}
The H$_2$ 3.23\um\ line is located on the blue shoulder of the 3.3\um\ PAH feature \citep[e.g., see][]{peeters2024}. The top row of Figure~\ref{fig:continuum_line_fractions_H2_323} shows an abrupt change in $\fcontrib{\mathrm{cont.},\mathrm{F323N}}$ at the Atomic PDR ($\approx 50$\% in the \ion{H}{ii} region, $\approx 30$\% elsewhere. The H$_2$ 3.23~\um\ line has the highest contribution to the F323N filter in the dissociation fronts ($\approx 10$\%), with a smaller but significant contribution in the diffuse molecular gas (similar to H$_2$ 2.12\um, Sec.~\ref{subsec:h2_2p12}). 

The contribution of this line to F323N is less than 15\% across the observed area, with a median of $\approx 5$\% (Fig.~\ref{fig:continuum_line_fractions_H2_323}). The correlation of F323N with the H$_2$ 3.23\um\ intensity is poor. F323N--F300M does not improve the correlation because F300M does not subtract the contribution of the 3.3~\um\ PAH feature to the F323N filter. In contrast, F323N--F335M results in negative flux because F323N contains a fraction of the PAH emission while F335M captures all of the PAH emission. The fit of Eq.~\ref{eq:correlation} with linear combinations of either (F323N, F300M) or (F323N, F335M) do not provide good estimates of the H$_2$ 3.23\um\ line intensity. Even when limiting the region included in the fit to the molecular PDR (i.e. further from the IF than DF2) and/or to the face-on PDR (distance from the IF being less than zero), linear combinations of these filters do not provide reliable estimators of H$_2$ 3.23\um\ line intensity. We do not recommend using F323N to estimate the intensity of the H$_2$ 3.23\um\ line.

\subsubsection{F335M--F300M as a tracer of the 3.3 \micron\ PAH}
\label{subsec:3p3_pah}
F335M is widely used as a tracer of the 3.3~\um\ PAH feature, which originates from small PAHs, and is used in combination with longer wavelength PAH bands to estimate the size of the PAHs in a given region. This filter also captures the weaker band at 3.4~\um\ attributed to aliphatic PAHs. The F300M filter is often used to subtract the underlying continuum \citep[e.g.][]{sandstrom2023}. Figure~\ref{fig:continuum_line_fractions_AIB} shows that $\fcontrib{\mathrm{cont.},\mathrm{F300M}}$ is close to 100\% in the atomic-to-molecular zone and slightly lower in the \hii\ region, while it is overall smooth. $\fcontrib{\mathrm{cont.},\mathrm{F335M}}$ is highest in the \hii\ region. 
The 3.3~\um\ PAH feature contributes 50--70\% to the F335M filter near the exciting sources, and 70--80\% at the remaining positions. %

The bottom rows of Figure~\ref{fig:continuum_line_fractions_AIB} show very tight correlations between the true 3.3~\um\ PAH intensity and that inferred from F335M or linear combinations of F335M and F300M. These panels indicate  that subtracting F300M from F335M clearly improves the correlation between F335M--F300M with the true 3.3~\um\ PAH intensity, and in general already gives the correct PAH intensity without any further corrections needed. The fit with Eq.~\ref{eq:correlation} (Table~\ref{table:N-W_fit}) does not significantly improve the correlation compared to the simple F335M--F300M. We recommend the simple F335M--F300M, but the two-parameter fit shown in Table~\ref{table:N-W_fit} gives also a good result and can be used as well.

\subsubsection{F405N--F480M as a tracer of Br$\alpha$}
\label{subsec:brackett}

Figure~\ref{fig:continuum_line_fractions_Br_a} shows that Br$\alpha$ contributes 50--80\% (median of $\approx 60$\%) to the F405N filter. As mentioned in Sect.~\ref{sec:data}, the F405N filter covers a NIRSpec wavelength gap, leading to stripes of missing data in the synthetic image. As described in Sect.~\ref{subsec:synthetic_images}, we allow a larger fraction of missing data compared to other filters in order to capture the Br$\alpha$ line, while not using the part of the map where Br$\alpha$ falls into the gap. This forces us to interpolate the gap when calculating the synthetic images and introduces more uncertainties in the baseline determination for the line fit. F480M is not affected by the gap, and the median continuum fraction is close to 100\%. Over the regions of the map where the synthetic image could be calculated, the top row of the figure shows that the continuum fraction in F405N is lowest in the \ion{H}{ii} region, increases to $\approx 40$\% in the Atomic PDR, and declines to $\approx 25$\% in the molecular zone. A similar trend is observed for F480M, but with continuum fractions ranging from 90--100\%. 

We use the F480M filter to trace the continuum under the Br~$\alpha$ line. The bottom rows show a strong overall correlation between Br~$\alpha$ and F405N, and with linear combinations of F405N and F480M. Overall, the intensity of the line increases as distance from the exciting sources decreases. The middle column in the bottom two rows show that F405N--F480M has a stronger, more linear correlation with the Br$\alpha$ line intensities compared to the F405N alone. F405N intensity alone slightly overestimates the line intensity. The right column of the bottom two rows shows that the best fit linear combination of F405N and F480M (Eq.~\ref{eq:correlation}, Table~\ref{table:N-W_fit}) improves the correlation with the line intensity, and is thus our recommended prescription for estimating this line intensity.

\subsubsection{F470N--F480M as a tracer of H$_2$ 0--0 S(9) 4.69\um}
\label{subsec:h2_4p69}

The top row of Figure~\ref{fig:continuum_line_fractions_H2_469} shows that the continuum fraction in F470N, which captures the H$_2$ 4.69\um\ line, is highest ($\approx 100$\%) in the atomic region and in less well-shielded regions of the molecular zone. The H$_2$ fraction of F470N rises significantly in the dissociation fronts, and that of continuum drops correspondingly.

The situation here is similar to F212N--F210M, tracing the H$_2$ 2.12\um\ (Sect.~\ref{subsec:h2_2p12}). The H$_2$ 4.69\um\ contributes up to 20\% (median of $\approx 5$\%) to the F470N filter. The bottom rows of Figure~\ref{fig:continuum_line_fractions_H2_469} show that  the correlation between F470N and H$_2$ 4.69\um\ line intensity is poor. Subtracting F480M from F470N leads to negative fluxes closer to the exciting sources due to \hi\ lines that are captured by the F480M filter but not F470N. In the brightest regions, F470N--F480M correlates relatively well with the H$_2$ 4.69\um\ intensity. We note an additional uncertainty here, namely that the underlying continuum increases with wavelength, and is larger in F480M than in F470N. This difference in continuum causes F470N--F480M to underestimate the integrated line intensity. The right column in the bottom two rows of the figure shows that fitting the  H$_2$ 4.69\um\ line intensity with a linear combination of F470N and F480M (Eq.~\ref{eq:correlation}, best-fit parameters in Table~\ref{table:N-W_fit}), leads to reduced scatter (Table~\ref{table:line_prescriptions}), and is therefore our recommended prescription for estimating H$_2$ 4.69\um\ line intensity.

\begin{figure*}
\begin{center}
\resizebox{.9\textwidth}{!}{%
\includegraphics[width=0.5\textwidth]{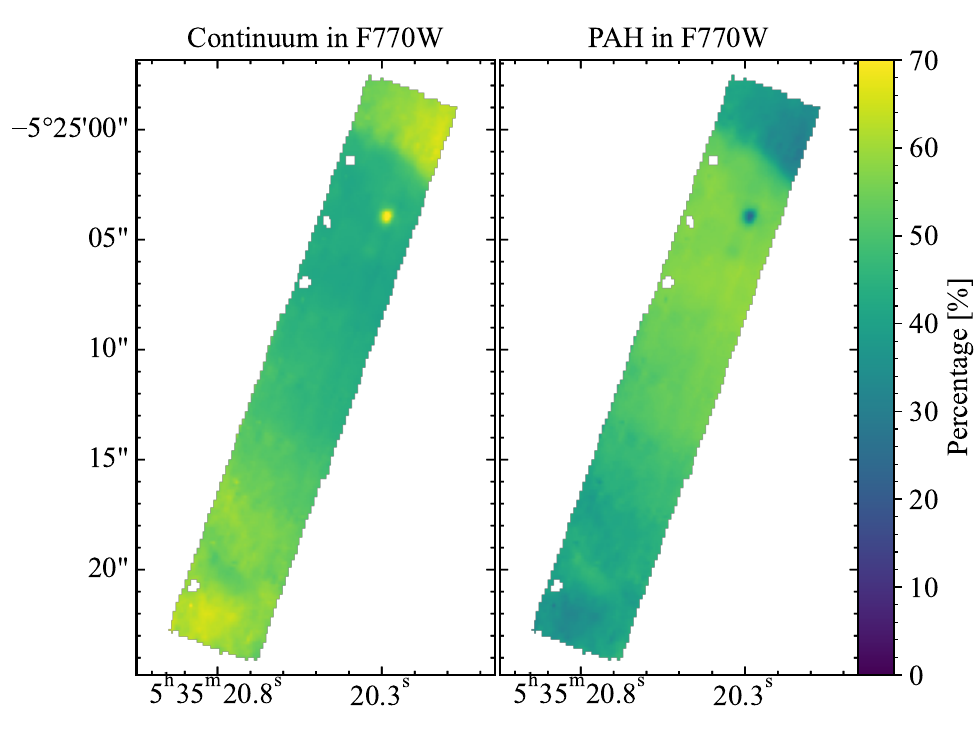}
\includegraphics[width=0.4\textwidth]{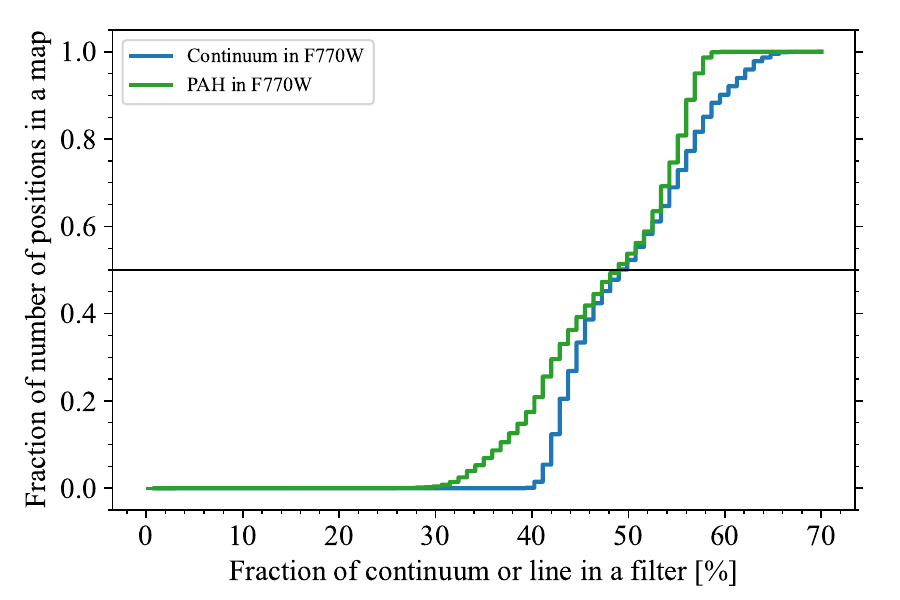}}
\includegraphics[width=0.9\textwidth]{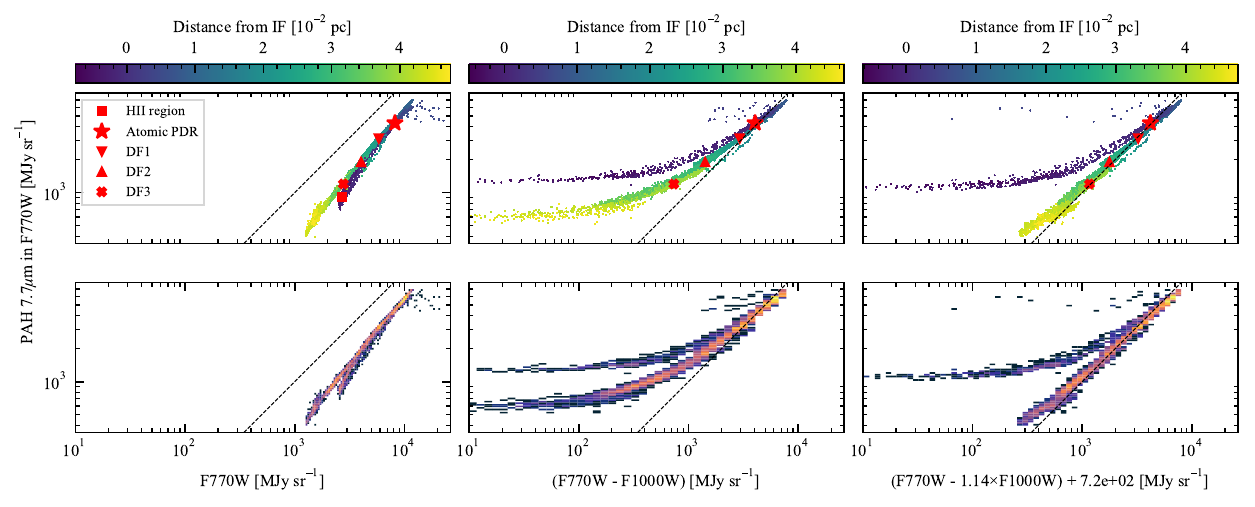}
\caption{\textbf{Top left:} Maps of continuum fraction in F770W (left) and PAH (right). \textbf{Top right:} Cumulative histograms of the continuum and PAH fractions in top panels. Horizontal black line indicates 50\%, i.e. the median value. \textbf{Bottom:} Correlation between the synthetic images (left: F770W, middle: F770W-F1000W, right: the best fit with Eq.~\ref{eq:correlation}) and the synthetic image of the PAH component in F770W. Colors indicate the distance from the ionization front as in Fig.~\ref{fig:continuum_line_fractions_FeII}. Red marks indicate the data points of the five template spectra.}
\label{fig:continuum_PAH_fractions_F770W}
\end{center}
\end{figure*}

\begin{figure*}
\begin{center}
\resizebox{.9\textwidth}{!}{%
\includegraphics[width=0.5\textwidth]{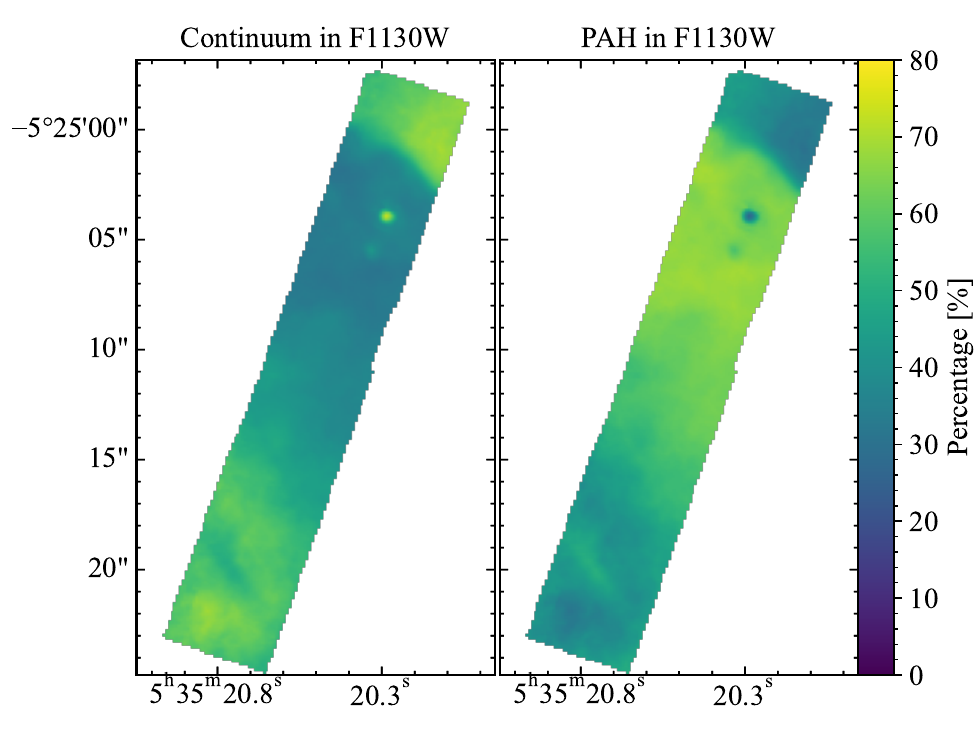}
\includegraphics[width=0.4\textwidth]{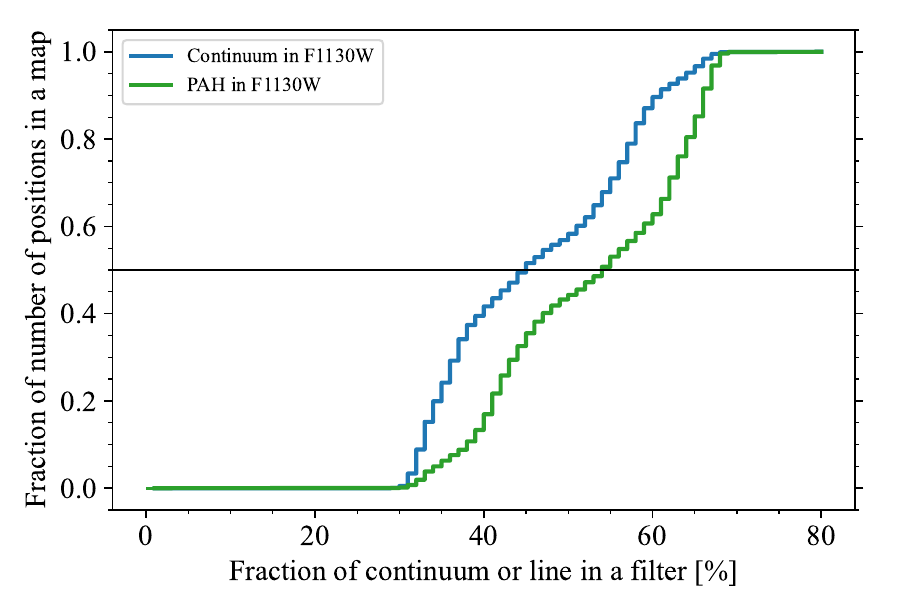}}
\includegraphics[width=0.9\textwidth]{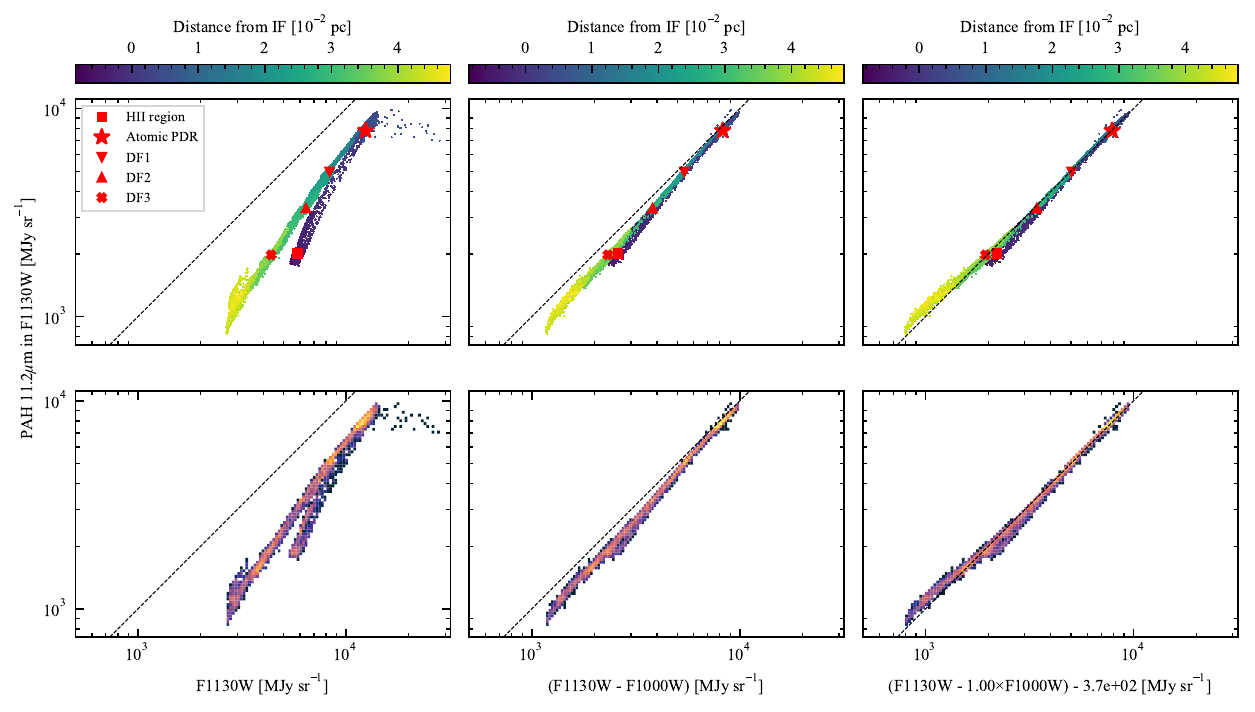}
\caption{\textbf{Top left:} Maps of continuum fraction in F1130W (left) and PAH (right). \textbf{Top right:} Cumulative histograms of the continuum and PAH fractions in top panels. Horizontal black line indicates 50\%, i.e. the median value. \textbf{Bottom:} Correlation between the synthetic images (left: F1130W, middle: F1130W-F1000W, right: the best fit with Eq.~\ref{eq:correlation}) and the synthetic image of the PAH component in F1130W. Colors indicate the distance from the ionization front as in Fig.~\ref{fig:continuum_line_fractions_FeII}. Red marks indicate the data points of the five template spectra.}
\label{fig:continuum_PAH_fractions_F1130W}
\end{center}
\end{figure*}

\begin{figure}
\begin{center}
\includegraphics[width=\columnwidth]{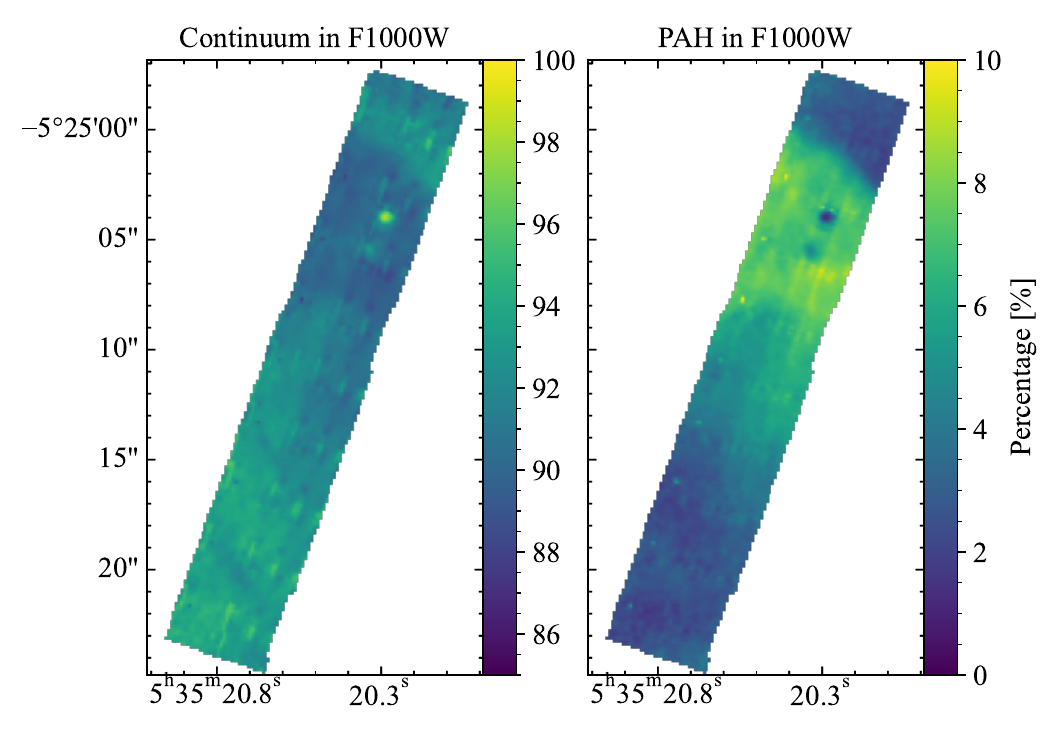}
\includegraphics[width=0.4\textwidth]{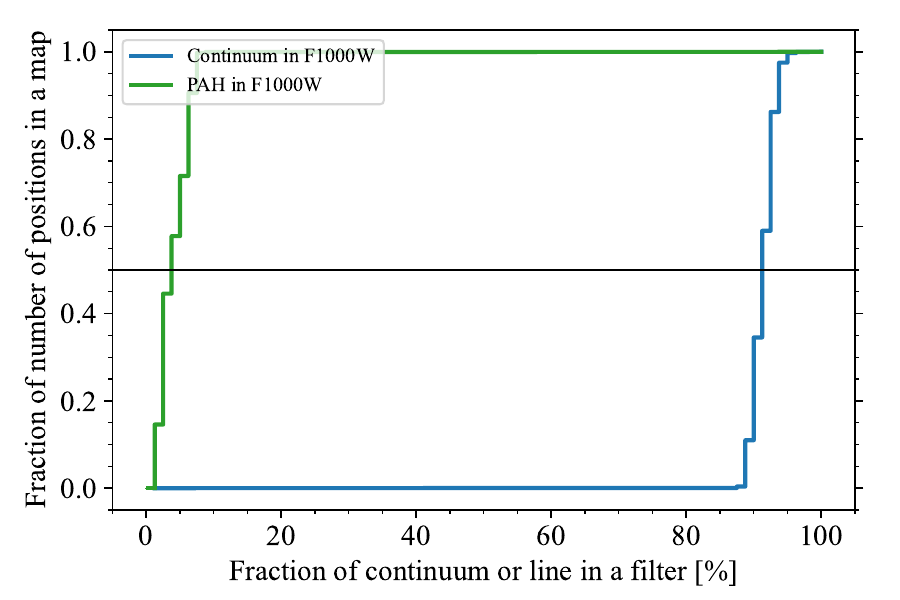}
\caption{Analysis of the continuum and PAH emission in F1000W calculated from the MRS spectra. \textbf{Top: } Maps of continuum fraction in F1000W (left) and PAH fraction (right). \textbf{Bottom: } Cumulative histograms of the continuum and PAH fractions in top panels. %
}
\label{fig:continuum_line_fractions_F1000W}
\end{center}
\end{figure}

\begin{figure}
\begin{center}
\includegraphics[width=\columnwidth]{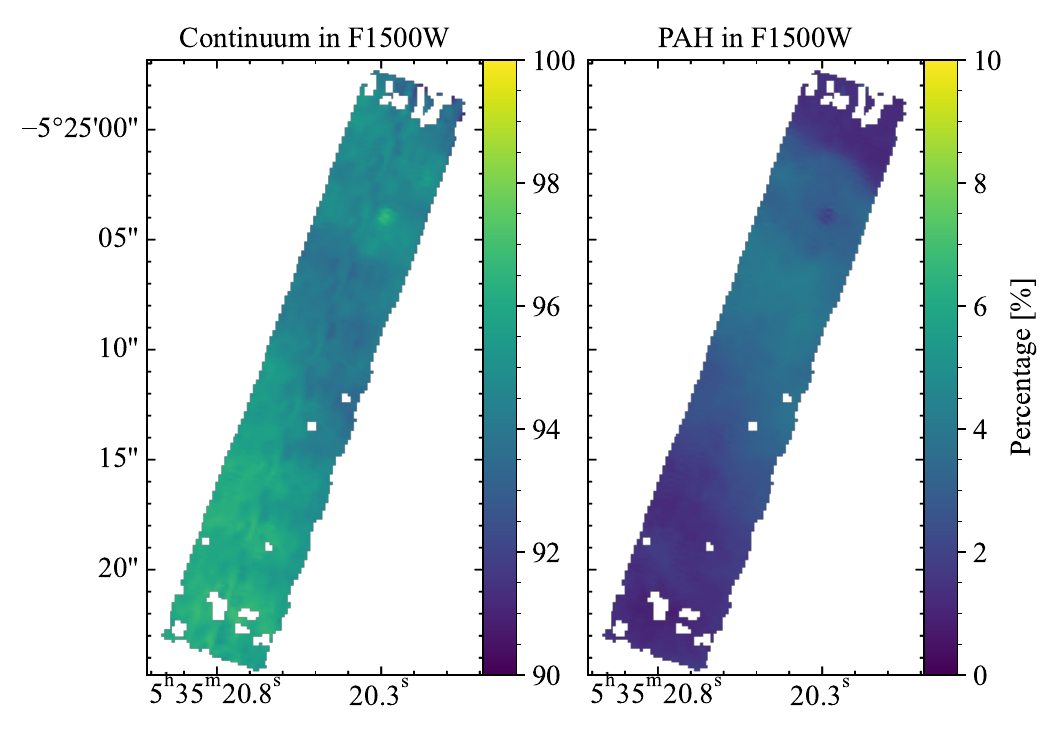}
\includegraphics[width=0.4\textwidth]{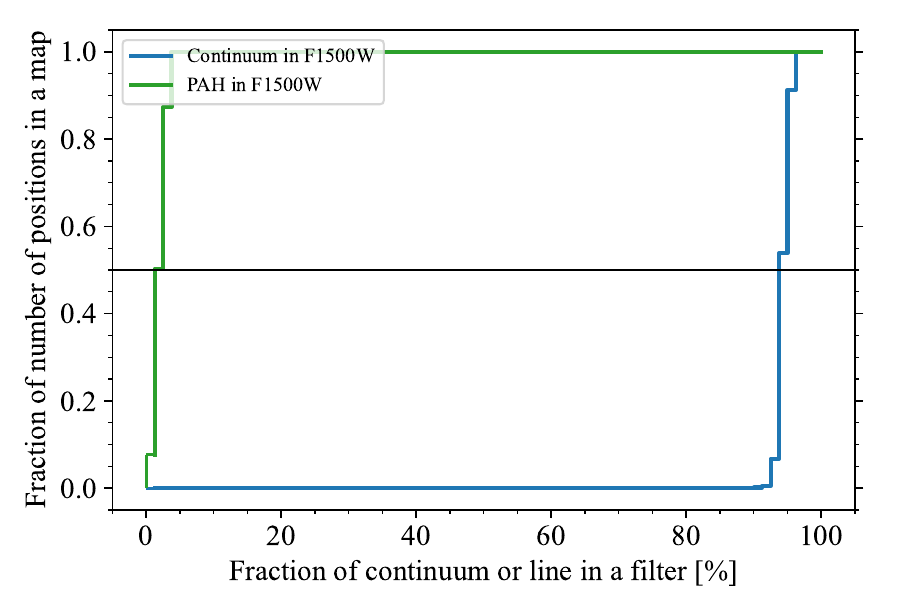}
\caption{Same as Fig.~\ref{fig:continuum_line_fractions_F1000W} but for F1500W. %
}
\label{fig:continuum_line_fractions_F1500W}
\end{center}
\end{figure}

\begin{figure}
\begin{center}
\includegraphics[width=0.9\columnwidth]{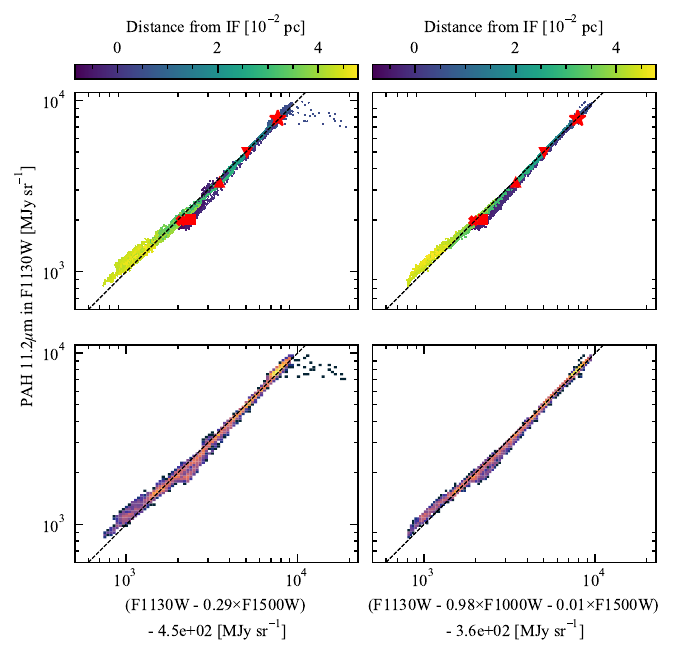}
\caption{Alternative correlation plots for the PAH emission in F1130W using the best fit with Eq.~\ref{eq:correlation} with F1500W (left) and the combination of F1000W and F1500W (right). Colors indicate the distance from the ionization front as in Fig.~\ref{fig:continuum_line_fractions_FeII}. Red marks indicate the data points of the five template spectra.}
\label{fig:scatterplots_PAH_F1130W_alt}
\end{center}
\end{figure}

\subsubsection{F770W--F1000W and F1130W--F1000W as tracers of 7.7~\um\ and 11.2~\um\ PAH emission}
\label{subsec:7p7_11p3_pah}

The F770W and F1130W filters capture the 7.7~\um\ and 11.2~\um\ PAH features (\S\ref{sec:intro}). Here we explore prescriptions for subtracting the underlying continuum from these filters to obtain the integrated intensities of the PAH features. The top row of both  Figure~\ref{fig:continuum_PAH_fractions_F770W} and~\ref{fig:continuum_PAH_fractions_F1130W} show that  the continuum fractions in these filters have similar spatial distributions, with 60--70\% in the \ion{H}{ii} region and molecular PDR, and $\approx 40$\% where the gas is mainly atomic. This is consistent with the physical picture where PAH emission arises primarily from warm, neutral regions excited by stellar FUV emission (i.e. PDRs). As mentioned above, the PAH emission is originating from the face-on background PDR directly beyond the \ion{H}{ii} region, and not from the ionized gas itself.

We explore using the F1000W filter to subtract the underlying continuum from F770W, and both F1000W and F1500W to subtract the continuum from F1130W. The top panels of Figures~\ref{fig:continuum_line_fractions_F1000W} and~\ref{fig:continuum_line_fractions_F1500W} show that these filters are indeed dominated by continuum emission, with $\mathrm{median}(\fcontrib{\mathrm{cont.},\mathrm{F1000W}})\approx 90$\% and $\mathrm{median}(\fcontrib{\mathrm{cont.},\mathrm{F1500W}})\approx 95$\%. The rest of the emission in these bands is mainly from PAHs (up to 10\%). 

The bottom two rows of Figure~\ref{fig:continuum_PAH_fractions_F770W} and~\ref{fig:continuum_PAH_fractions_F1130W} show very strong correlations between the true 7.7~\um\  (11.3~\um) PAH intensity and linear combinations of F1000W. We note that for these filters, we compute synthetic images of the extracted PAH feature profiles (in units of \mjysr), in contrast to the procedure for F335M where we measure the integrated intensity and multiply by the bandwidth. We opt for this method because the PAH features in the MIRI range are more continuous (broad) and there is no unique way to define each PAH feature. The F770W filter and F1130W filter on their own overestimate the underlying PAH intensity (as expected due to the underlying continuum), and the middle row of the figures shows that subtracting the appropriate continuum filter tightens the correlations. 

The resulting trend for the 11.2~\um\ feature is nearly linear, and fitting this relationship with Eq.~\ref{eq:correlation} (best-fit parameters in Table~\ref{table:N-W_fit}) leads to a tighter relationship (although mainly due to a vertical offset rather than a multiplicative factor). In contrast, subtracting F1000W from F770W results in the two distinct trends (middle row of Fig.~\ref{fig:continuum_PAH_fractions_F770W}), with pixels closest to the IF diverging from the main trend. The brightest pixels in the map show good agreement between the true 7.7~\um\ PAH intensity and F770W--F1000W.

Fitting the 7.7~\um\ PAH feature intensity as a function of F770W and F1000W yields a linear relationship for most pixels (bottom right corner of Fig.~\ref{fig:continuum_PAH_fractions_F770W}), while the pixels in the \ion{H}{ii} region remain separate. It seems possible that the nature of the continuum in the \ion{H}{ii} region is distinct (scattered starlight in addition to warm dust), and so while we are seeing PAH emission from the background PDR, the filter may be tracing continuum from the \ion{H}{ii} region. The fact that this effect is not seen for F1130W is consistent with this picture, because scattered starlight is primarily seen at shorter wavelengths. In nearby galaxies, stellar continuum contributes a significant amount to F770W but not to F1130W \citep{whitcomb2023a, sutter2024}.

Overall, Fig.~\ref{fig:continuum_PAH_fractions_F770W} shows that subtracting the underlying continuum using a linear fit (Eq.~\ref{eq:correlation} with F770W and F1000W) reproduces the 7.7\um\ PAH emission well for intermediate distances from the ionization front, namely the Atomic PDR, DF1, and DF2, while there is significant deviation in the \hii\ region and weaker deviation at DF3, i.e. deep in the molecular PDR. In contrast, the 11.2~\um\ PAH intensity is well described by a linear fit with Eq.~\ref{eq:correlation} either using F1000W to determine the continuum contribution (Fig.~\ref{fig:continuum_PAH_fractions_F1130W}) or using F1500W,  or both (Fig.~\ref{fig:scatterplots_PAH_F1130W_alt}). The goodness-of-fit and best-fit coefficients (Tables~\ref{table:cont_frac_temp_all} and \ref{table:N-W_fit}) indicate that F1000W alone works well to determine the continuum for F1130W, but a scaled F1500W can also be used to estimate the continuum contribution for F1130W if F1000W is not available.
For the 11.2~\um\ integrated intensity we recommend using the three-parameter fit (F1130W, F1000W, F1500W), or the fit using F1130W and F1500W (F1000W) alone if F1000W (F1500W) is unavailable (their RMS deviation from 1:1 in Table~\ref{table:N-W_fit} are small, and within 0.01~dex of each other). For the 7.7~\um\ integrated intensity we recommend using the two-parameter fit with F770W and F1000W, but limited to atomic regions and around dissociation fronts where the PAH emission is strong.

\section{Comparison to previous work}\label{sec:comparison_previous_work}

\subsection{3.3 \um\ PAH intensity}\label{subsec:aib3p3}

While the F335M filter is dominated by 3.3 \um\ PAH emission, under some circumstances it is challenging to separate the PAH emission from the underlying continuum emission without spectroscopic data. This continuum can be directly from stars or scattered starlight, stochastically-heated very small grains, blended overtone and combination bands from PAHs, as well as free-free and free-bound emission. In the NIRSpec FOV of the Orion Bar, the continuum in the 3~\mum region is dominated by emission from stochastically heated very small grains and/or blended overtone and combination bands from PAHs generating a `quasi-continuum' \citep[][]{boersma2023, peeters2024}. Here we compare approaches for subtracting continuum from the F335M filter and their ability to recover the true integrated intensity of the 3.3 \um\ PAH feature.

In the previous section, we found that a simple subtraction of F300M from F335M traces the actual intensity of the PAH quite well in the Orion Bar. However this does not necessarily hold for a broader range of environments. In the following, we follow the method proposed by \citet{sandstrom2023} using F300M and F360M to calibrate among different environments.

\begin{figure}
\begin{center}
\includegraphics[width=0.45\textwidth]{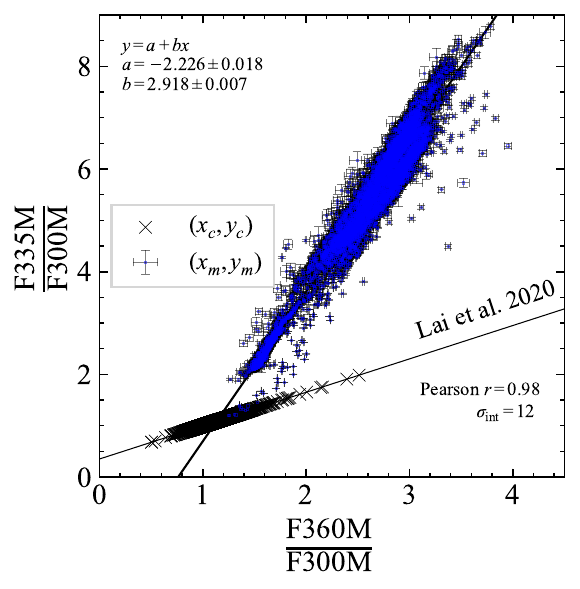}
\caption{The F360M/F300M vs. F335M/F300M relationship for the Orion Bar (blue points) and the galaxies of \citet{lai2020} where the continuum is dominated by stellar emission (black crosses). A linear fit is shown for both the Orion Bar and the galaxy sample by the solid lines. Fit parameters for the Orion Bar are given in the top left corner of the panel whereas fit parameters for the galaxy sample are taken from \citet{lai2020}. The Pearson correlation coefficient $r$ and intrinsic scatter $\sigma_{int}$ for the Orion Bar are given in the lower right corner of the panel.}
\label{fig:3p3_fit}
\end{center}
\end{figure}

\citet{sandstrom2023} expanded the method for subtracting continuum from F335M proposed by \citet{lai2020}, which quantifies the slope of the continuum using the two filters on either side of F335M, namely F300M and F360M, to PAH-dominated regions (defined as F1130W $> 10$ MJy sr$^{-1}$ in \citep{sandstrom2023}) of 
nearby galaxies from the Physics at High Resolution in Nearby GalaxieS (PHANGS) \jwst\ Treasury program \citep[see ][]{lee2023}. This approach works by first measuring the F360M/F300M and F335M/F300M colours ($x_m, y_m$ respectively) in PAH-dominated regions, and fitting a line to this relationship (see also Fig.~\ref{fig:3p3_fit}). We will refer this fit as PAH-fit. Previous work by \citet{lai2020} found that the fit to the colours $x_m, y_m$ in stellar continuum dominated regions is also linear but with a significantly shallower slope than for PAH-dominated regions. We will refer to this fit as the Lai-fit. For each measured colour ($x_m, y_m$), we determine a new line with the same slope as PAH-fit that goes through $x_m, y_m$, and we find its intersection with the Lai-fit. This intersection ($x_c, y_c$)  then gives an estimate of the colours if there were no PAH emission present, i.e. in stellar continuum dominated regions. These can then be converted into the continuum component in F335M surface brightness F335M$_\mathrm{cont}$. The PAH emission convolved with the filter function F335M$_\mathrm{PAH}$ is then given by 
\begin{align}
    \mathrm{F335M_{PAH}~[MJy~sr^{-1}]} &= \mathrm{F335M} - \mathrm{F335M_{cont}}\\
    & = \mathrm{F335M} - y_c \times 
    \mathrm{F300M}.
\end{align}

We apply this method of estimating F335M$_\mathrm{PAH}$ to synthetic images obtained from our NIRSpec data. This offers us the advantage of being able to directly compare the estimated F335M$_\mathrm{PAH}$ with the true value obtained by integrating the decomposed PAH spectra over the F335M filter -- which was not possible with the imaging-only dataset in \citet{sandstrom2023}. Fig.~\ref{fig:3p3_fit} shows the $F^\mathrm{synth.}_{\mathrm{F335M}}$/$F^\mathrm{synth.}_{\mathrm{F300M}}$ vs. $F^\mathrm{synth.}_{\mathrm{F360M}}$/$F^\mathrm{synth.}_{\mathrm{F300M}}$
relationship obtained from our NIRSpec cube. The best-fit to our data points (blue) has a slope of 2.9, which is steeper than the slope of 1.6 for the nearby galaxy data presented by \citet{sandstrom2023}. Following \citet{sandstrom2023}, each data point (blue) was shifted down and to the left, parallel to the best-fit line, to the corresponding position on the \citet{lai2020} line (black), yielding the continuum-dominated colours ($x_c, y_c$). 

\begin{figure}
\begin{center}
\includegraphics[width=0.45\textwidth]{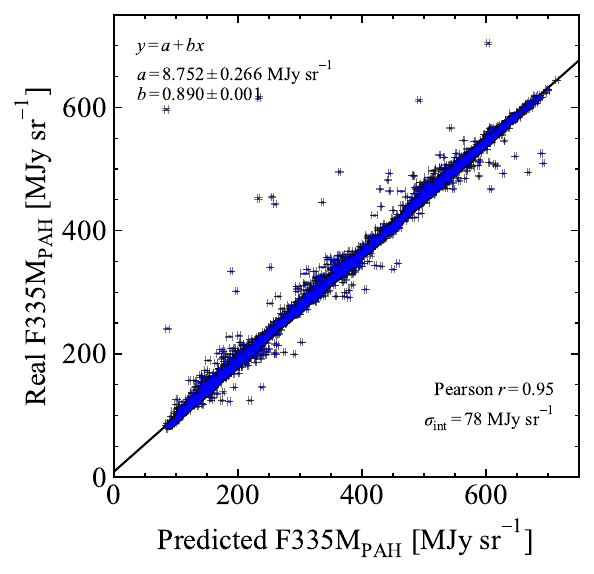}
\caption{Comparison of the 3.3 $\mu$m PAH emission obtained following the procedure of \citet{sandstrom2023} and Fig.~\ref{fig:3p3_fit} with the synthetic 3.3 $\mu$m PAH emission captured by the F335M filter using the NIRSpec data for the Orion Bar.  A linear fit is shown by the solid line and the fit parameters are given in the top left corner of the panel. The Pearson correlation coefficient $r$ and intrinsic scatter $\sigma_\mathrm{int}$ are given in the lower right corner of the panel.
}
\label{fig:3p3_fit_real_vs_predicted}
\end{center}
\end{figure}

Fig.~\ref{fig:3p3_fit_real_vs_predicted} shows the true F335M$_\mathrm{PAH}$ (synthetic image of the PAH component) vs. predicted F335M$_\mathrm{PAH}$ (using the method above). This plot shows that the predicted F335M$_\mathrm{PAH}$ is about 10 \% higher than the true value but the correlation is quite tight. Note that the true F335M$_\mathrm{PAH}$ is defined as the integrated PAH component in the F335M filter, excluding the PAH emission outside of this filter. The contribution of PAH emission outside of F335M is about 10\%, therefore the total PAH 3.3\um\ intensity matches well with F335M--F300M as shown in Sect.~\ref{subsec:continuum_fraction}. This means, at least for the Orion data, the method to estimate F335M$_\mathrm{PAH}$ proposed by \citet{sandstrom2023} traces the total PAH 3.3\um\ intensity rather than the contribution of PAH in the F335M filter. We confirm that the results from applying their method is consistent with our result for the Orion Bar.

\subsection{11.2 \um\ PAH intensity}\label{subsec:aib11p2}

\begin{figure}
\begin{center}
\includegraphics[width=0.45\textwidth]{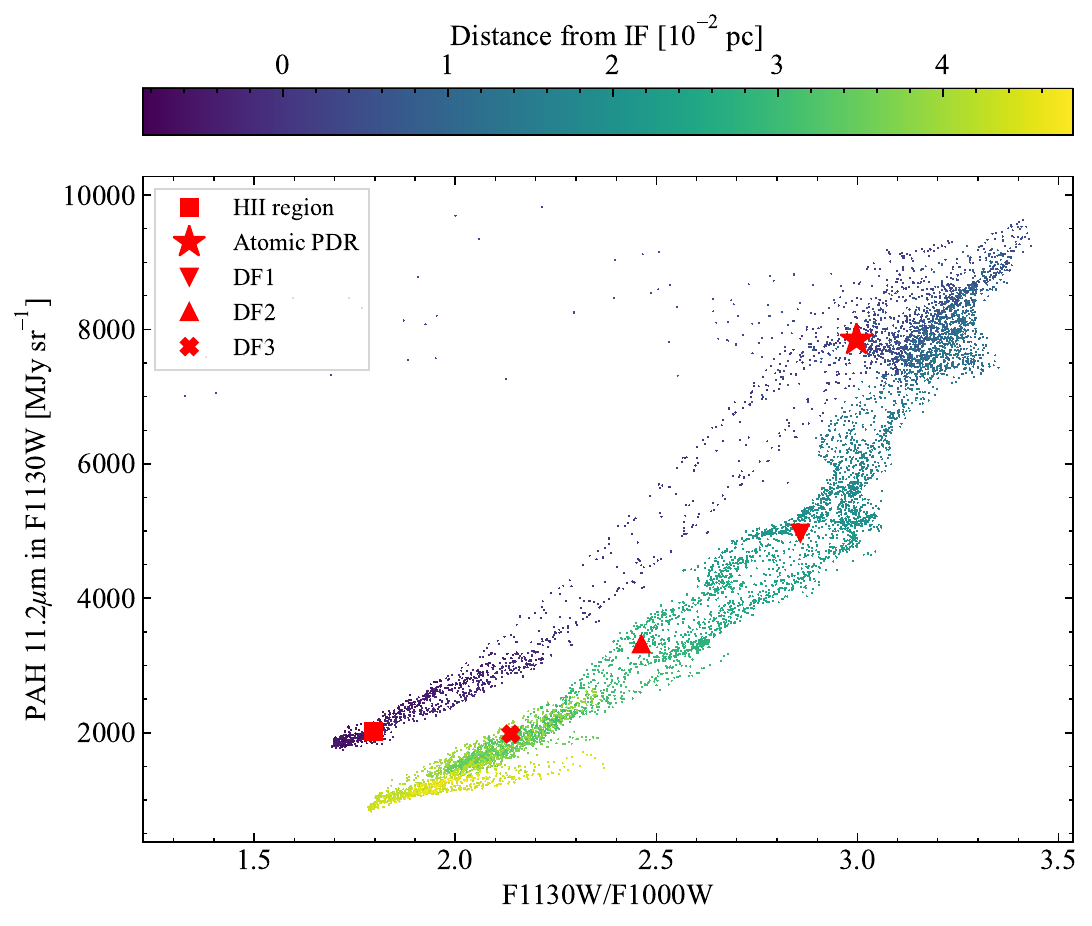}
\caption{The correlation between the synthetic image ratio of F1130W/F1000W and the PAH component in the F1130W band. Colors indicate the distance from the ionization front as in Fig.~\ref{fig:continuum_line_fractions_FeII}. Red marks indicate the data points of the five template spectra.
}
\label{fig:f1130wdivf1000w}
\end{center}
\end{figure}

As discussed in Sect.~\ref{subsec:continuum_fraction}, F1130W--F1000W is a good tracer of PAH 11.2~\um\ emission in the range of environments found in Orion. An alternative quantity to trace the PAH 11.2~\um\ is F1130W/F1000W, as used for example in \citet{wesson2024}. Fig.~\ref{fig:f1130wdivf1000w} shows the correlation between the synthetic image ratio of F1130W/F1000W and the PAH component in the F1130W (i.e. synthetic image of the PAH component, shown on the y-axis of the scatter plots in Fig.~\ref{fig:continuum_PAH_fractions_F1130W}). We confirm a general positive correlation between the PAH 11.2~\um\ strength and the F1130W/F1000W ratio. However, the area closer to the IF (along the line of sight of the \hii\ region) has a lower F1130W/F1000W ratio compared to the other regions with the same strength of the 11.2~\um\ PAH component. The off-branching of the \hii\ region is seen in the 11.2~\um\ vs. F1130W synthetic image scatter plot (Fig.~\ref{fig:continuum_PAH_fractions_F1130W}, left panels of the scatter plots), but not in the 11.2~\um\ vs. F1130W--F1000W plot (Fig.~\ref{fig:continuum_PAH_fractions_F1130W}, middle panels of the scatter plots). We interpret this off-branching as being due to higher continuum emission towards the ionized gas. The range of the F1130W/F1000W in our study is much higher than that in the planetary nebula NGC~6720 \citep{wesson2024}, indicating a stronger contribution of the PAH emission to the F1130W filter, or a different shape in the continuum emission, or stronger line emission in NGC~6720. 
We conclude that F1130W/F1000W can not be used as a universal prescription to quantitatively estimate the 11.2~\um\ PAH strength over a broad range of physical conditions. 

\section{Conclusions}

We investigate the contributions of line, PAH, and continuum emission in \jwst\ NIRCam and MIRI imaging filters using synthetic images created by the NIRSpec and MIRI-MRS data, observed towards the Orion Bar PDR from the \jwst\ Early Release Science program PDRs4All. We provide empirical prescriptions for tracing line and PAH emission with linear combinations of \jwst\ images, which we expect to be widely useful for the community to apply to their own JWST imaging observations. Our main results are as follows.

\begin{enumerate}
\item NIRSpec pixel-by-pixel cross-calibration with NIRCam with the pipeline version used in this study shows that the best-fit line has slopes of 0.8--1.0 (\S\ref{subsec:cross-cal}). The intrinsic scatter is of order 10 \mjysr, which is likely due to the not-yet-finalized NIRSpec flux calibration. The slope of best-fit lines for the MIRI imaging vs. MRS are significantly closer to 1.0.
\item For NIRSpec, some narrow band filters contain more than one strong emission line, and many medium filters have non-negligible contribution from line emission as a sum. Therefore, it is not straightforward to trace the intensity of some lines %
by a combination of the available imaging filters. 
\item For Pa~$\alpha$ (\S\ref{subsec:paschen}), Br~$\alpha$ (\S\ref{subsec:brackett}), H$_2$ 2.12\um\ (\S\ref{subsec:h2_2p12}) and 4.69\um (\S\ref{subsec:h2_4p69}), we recommend using a linear combination of narrow and medium filters as shown in Tables~\ref{table:N-W_fit} and~\ref{table:line_prescriptions}. 
\item For \FeII\ 1.644\um\ (\S\ref{subsec:feii}) and PAH 3.3\um\ (\S\ref{subsec:3p3_pah}), we recommend using a simple subtraction of the continuum filter, namely F164N--F162M and F335M--F300M, respectively, as introducing fitting parameters for a linear fit %
does not significantly improve the scatter.
\item For MIRI, all filters are dominated by either continuum or PAH emission, and the contribution of a given emission line is $< 5$\%. The strongest line is \NeIII\ 15.5\um\ towards the \hii\ region (Table~\ref{table:line_frac_temp_all}).
\item F1130W--F1000W is a reliable tracer of PAH 11.2\um, while a scaled version of F1500W can be also used to subtract the continuum from F1130W. A linear combination of F770W and F1000W traces the 7.7~\um\ PAH feature in the F770W filter for atomic regions and around dissociation fronts where the PAH emission is strong (\S\ref{subsec:7p7_11p3_pah}).
\item We applied the method to derive the PAH emission using F335M/F300M and F335M/F360M ratios proposed by \citet{sandstrom2023} based on nearby galaxies observed with \jwst. For the Orion Bar data, this method reproduces the total 3.3~\um\ PAH intensity, which is about 10\% smaller than the PAH contribution in F335M filter.
\end{enumerate}

\begin{acknowledgements}

We are very grateful to the \jwst\ Help Desk for their support with pipeline and calibration issues that were encountered while writing this paper. 
This work is based on observations made with the NASA/ESA/CSA James Webb Space Telescope. The data were obtained from the Mikulski Archive for Space Telescopes at the Space Telescope Science Institute, which is operated by the Association of Universities for Research in Astronomy, Inc., under NASA contract NAS 5-03127 for \jwst. These observations are associated with program \#1288. This research used the Canadian Advanced Network For Astronomy Research (CANFAR) operated in partnership by the Canadian Astronomy Data Centre and The Digital Research Alliance of Canada with support from the National Research Council of Canada the Canadian Space Agency, CANARIE and the Canadian Foundation for Innovation.

Support for program \#1288 was provided by NASA through a grant from the Space Telescope Science Institute, which is operated by the Association of Universities for Research in Astronomy, Inc., under NASA contract NAS 5-03127.
EP and JC acknowledge support from the University of Western Ontario, the Institute for Earth and Space Exploration, the Canadian Space Agency (CSA; 22JWGO1- 16), and the Natural Sciences and Engineering Research Council of Canada.
YO %
gratefully acknowledges the Collaborative Research Center 1601 (SFB 1601 sub-project A6) funded by the Deutsche Forschungsgemeinschaft (DFG, German Research Foundation) – 500700252. 
CB is grateful for an appointment at NASA Ames Research Center through the San Jos\'e State University Research Foundation (80NSSC22M0107).
TO acknowledges the support by the
271 Japan Society for the Promotion of Science (JSPS) KAKENHI Grant Number JP24K07087.
EH acknowledges supports of CNES.

We thank the anonymous referee for comments which improved the manuscript.

\end{acknowledgements}

\bibliographystyle{aa}
\bibliography{sep4}

\begin{appendix}

\FloatBarrier

\section{Additional NIRCam/NIRSpec cross-calibration results}
\label{sec:xcal_additional}

\begin{figure*}
\begin{center}
\includegraphics[width=0.83\textwidth]{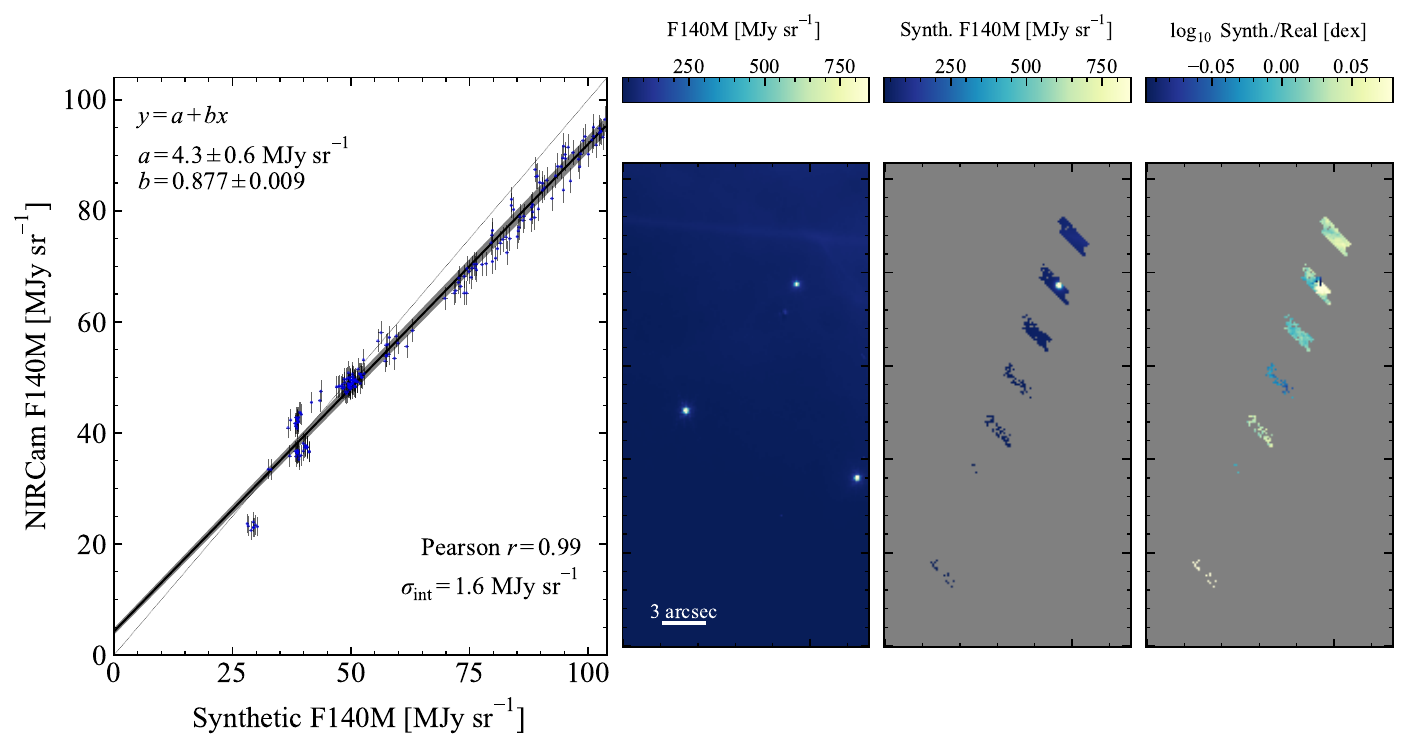}
\includegraphics[width=0.83\textwidth]{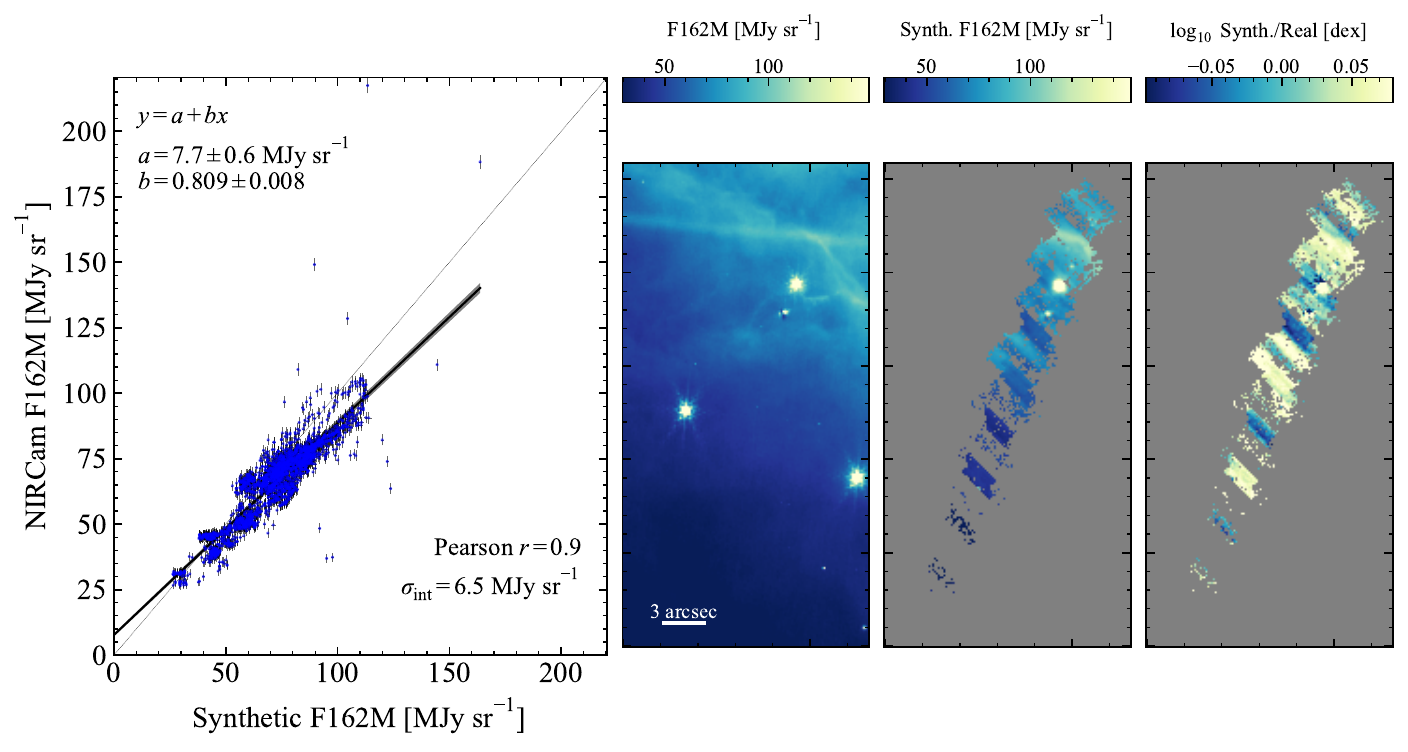}
\includegraphics[width=0.83\textwidth]{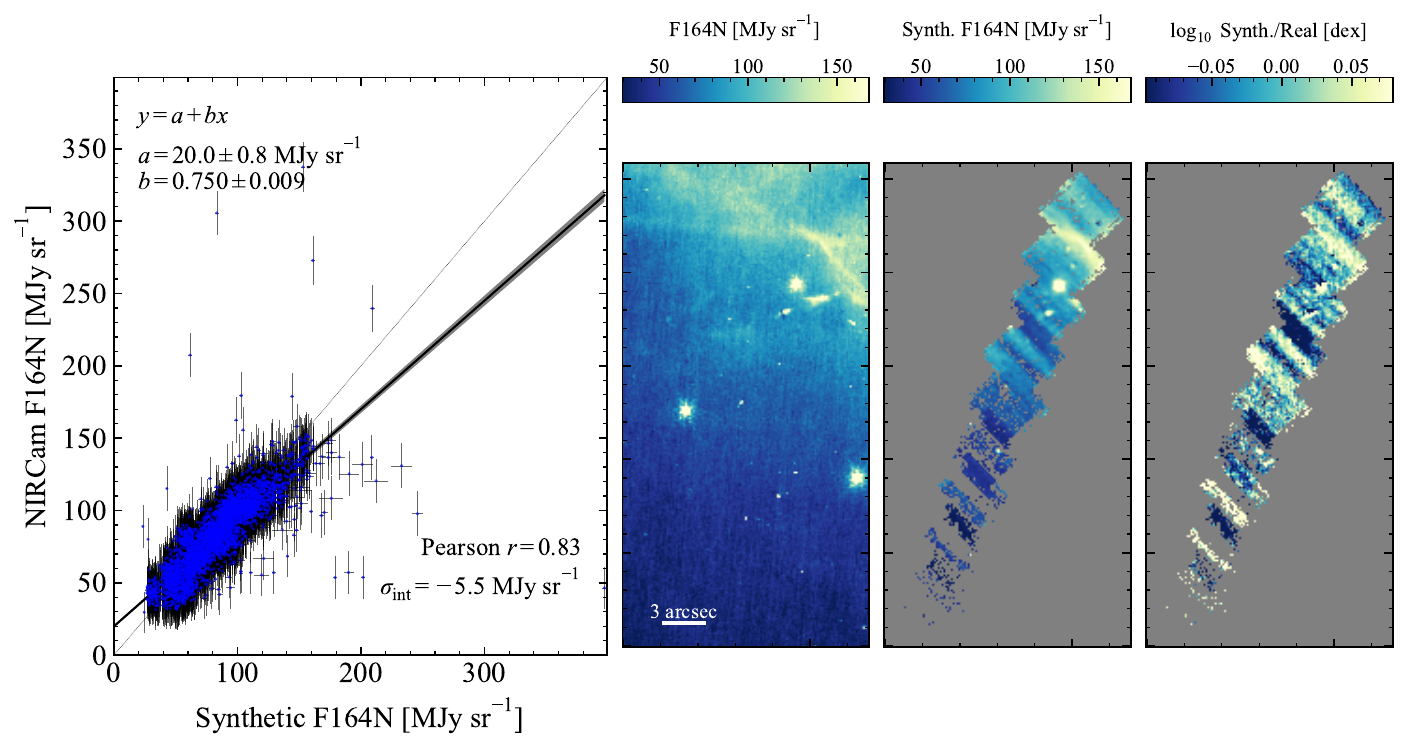}

\caption{NIRSpec/NIRCam cross-calibration results for  NIRCam filters that are not shown in Figure~\ref{fig:xcal_nirspec}: F140M, F162M, and F164N.}
\label{fig:xcal_nirspec2}
\end{center}
\end{figure*}

\begin{figure*}
\begin{center}
\includegraphics[width=0.83\textwidth]{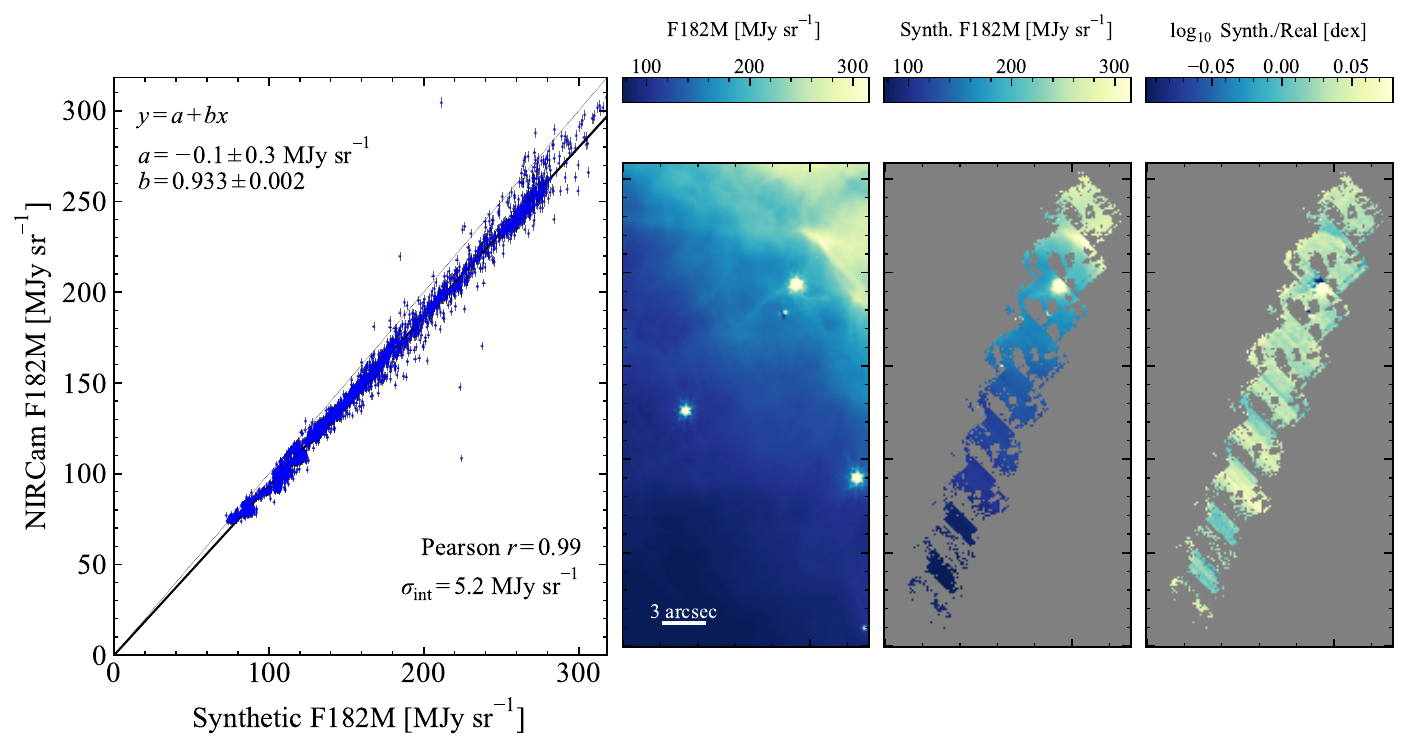}
\includegraphics[width=0.83\textwidth]{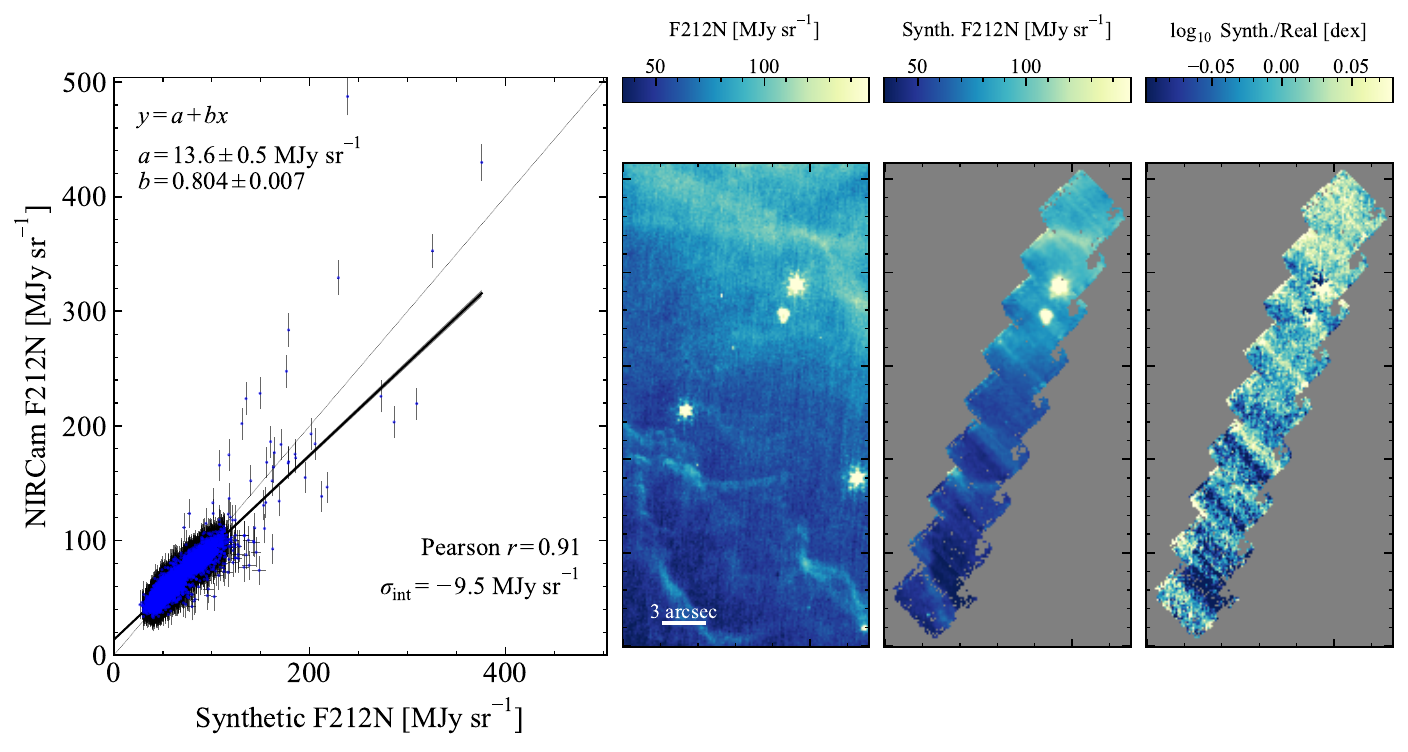}
\includegraphics[width=0.83\textwidth]{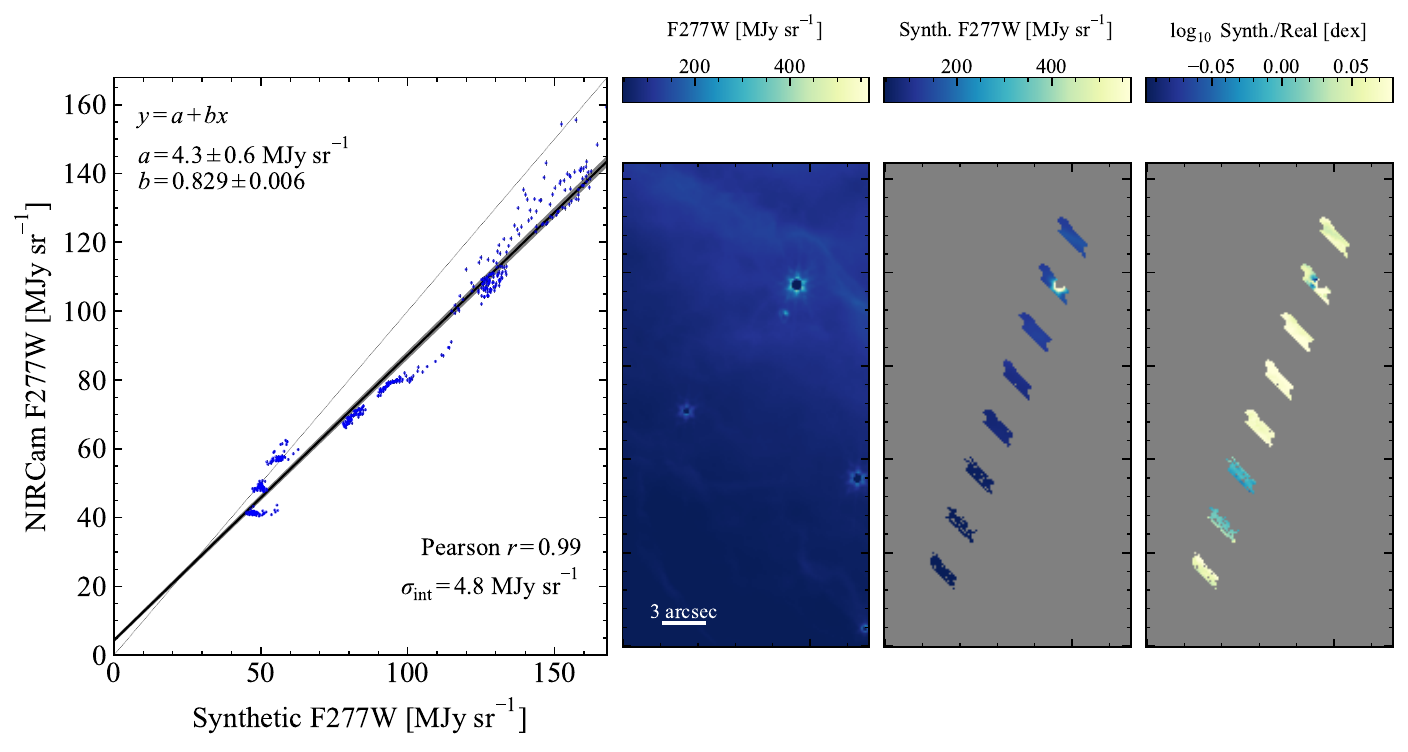}
\caption{NIRSpec/NIRCam cross-calibration results for  NIRCam filters that are not shown in Figure~\ref{fig:xcal_nirspec}: F182M, F212N, and F277W.
}
\label{fig:xcal_nirspec3}
\end{center}
\end{figure*}

\begin{figure*}
\begin{center}

\includegraphics[width=0.83\textwidth]{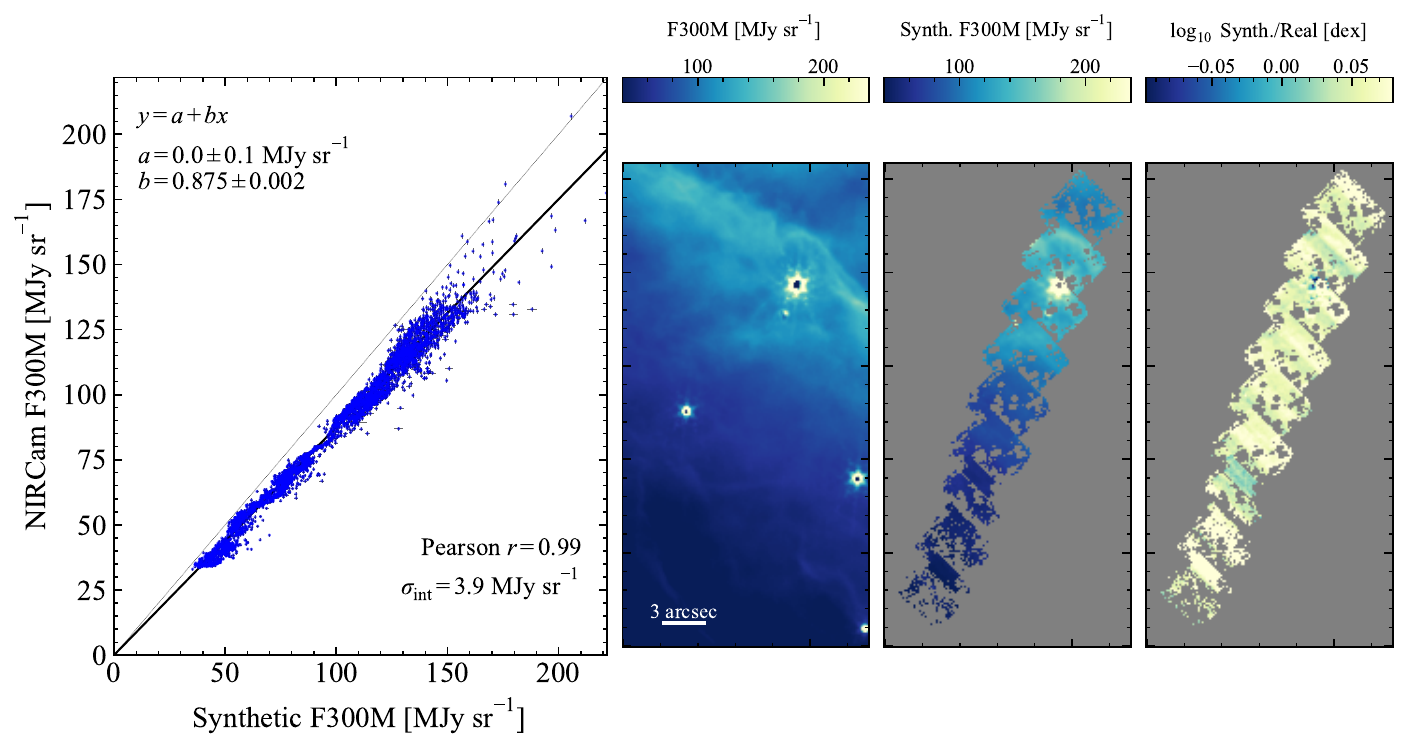}
\includegraphics[width=0.83\textwidth]{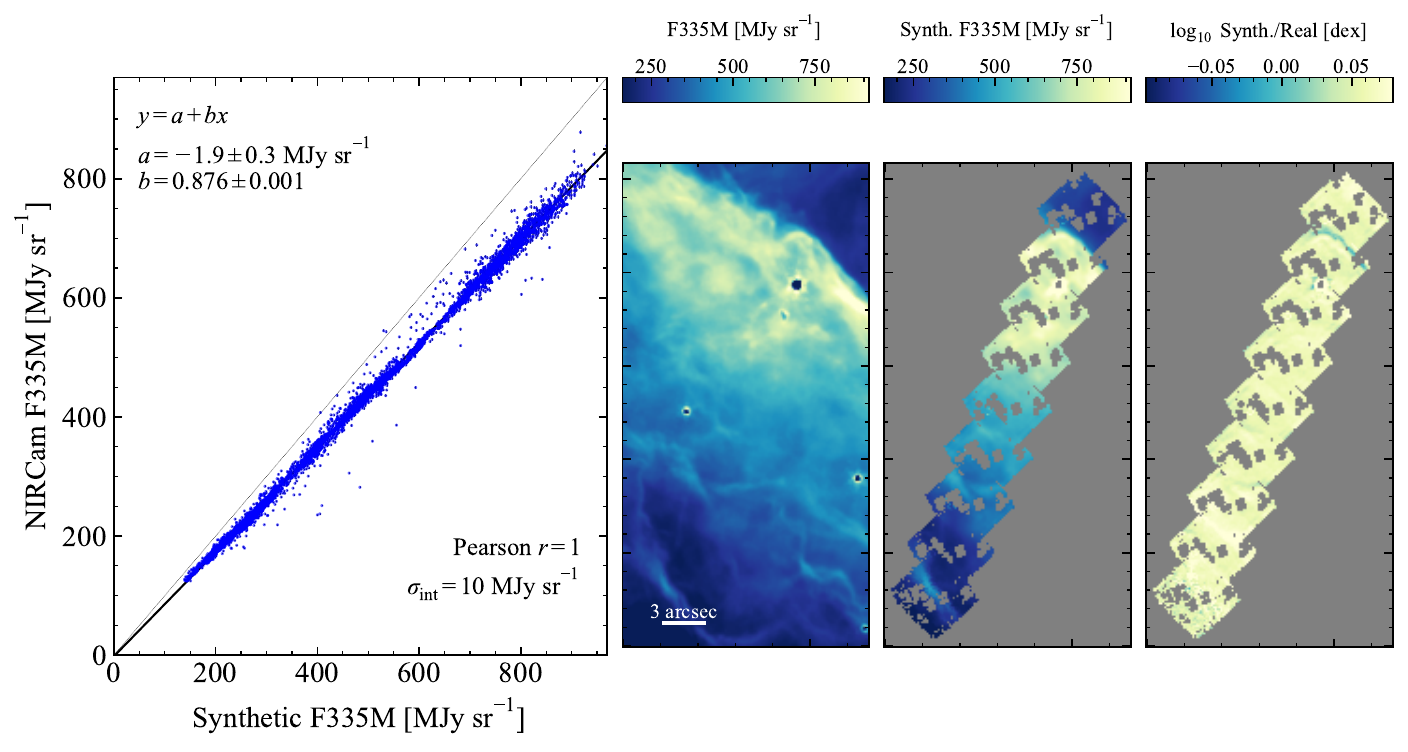}
\includegraphics[width=0.85\textwidth]{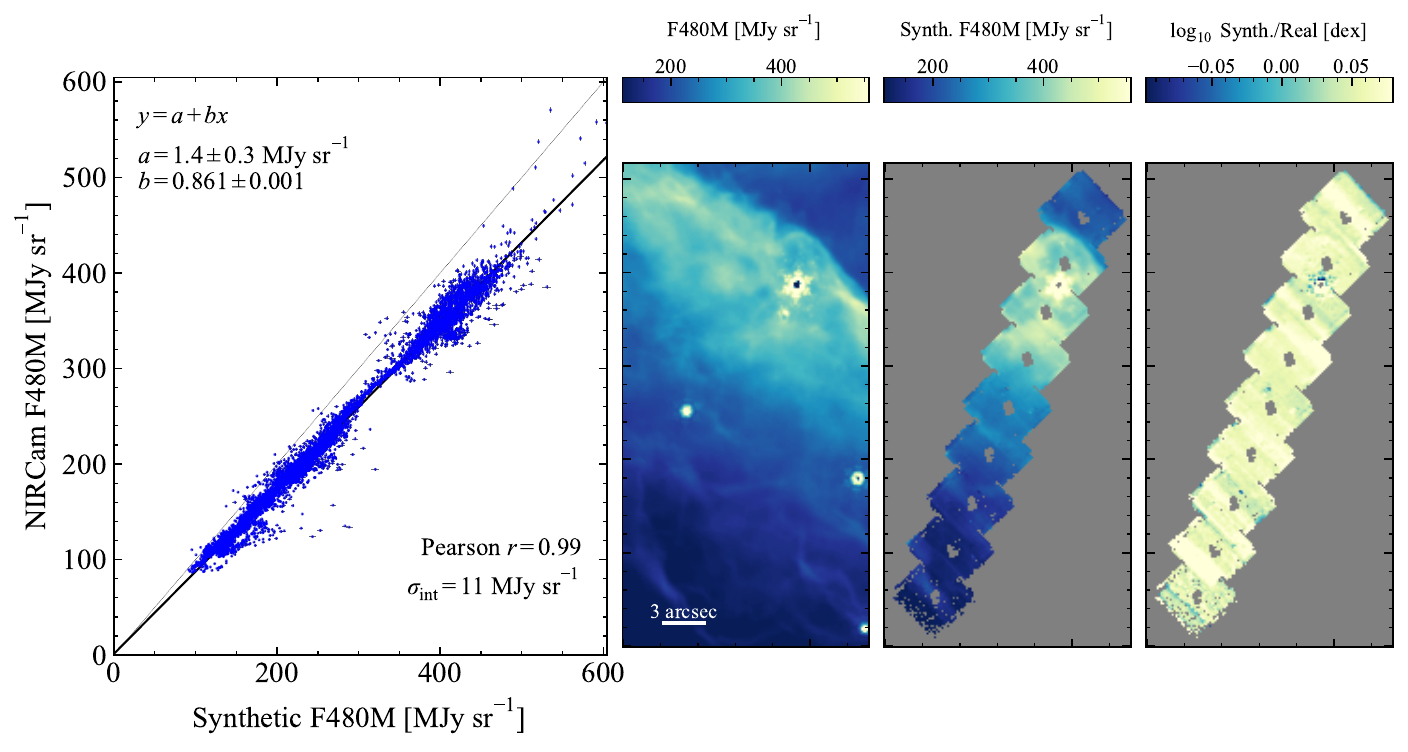}
\caption{NIRSpec/NIRCam cross-calibration results for  NIRCam filters that are not shown in Figure~\ref{fig:xcal_nirspec}: F300M,  F335M, and F480M.
}
\label{fig:xcal_nirspec4}
\end{center}
\end{figure*}

In Fig.~\ref{fig:xcal_nirspec}, the NIRSpec/NIRCam cross-calibration results for three selected NIRCam filters are shown. In this appendix, we show the same results for the remaining NIRCam filters (Figs.~\ref{fig:xcal_nirspec2}-~\ref{fig:xcal_nirspec4}).

\section{Correlation plots between the bright emission lines in MIRI and synthetic images}
\label{sec:MIRI_line_fractions}

\begin{figure*}
\begin{center}
\resizebox{.99\textwidth}{!}{%
\includegraphics[width=0.27\textwidth]{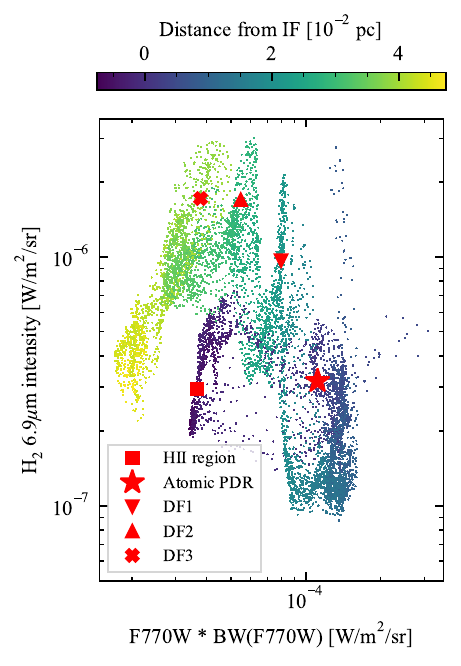}
\includegraphics[width=0.27\textwidth]{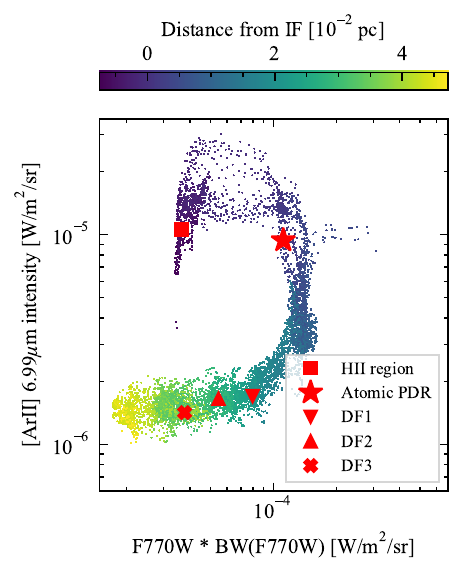}
\includegraphics[width=0.33\textwidth]{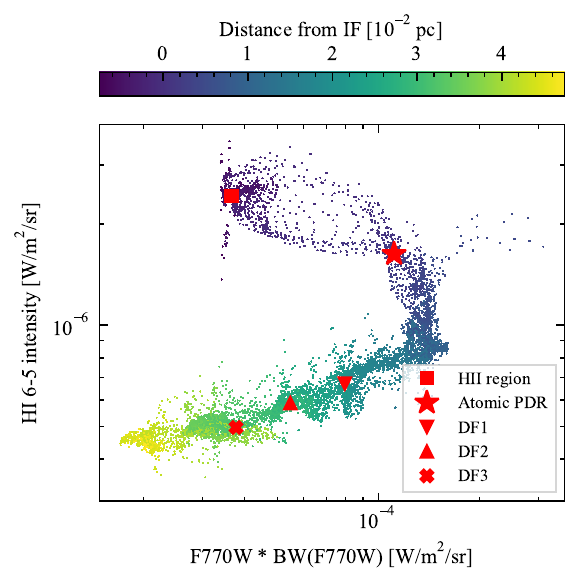}
\includegraphics[width=0.27\textwidth]{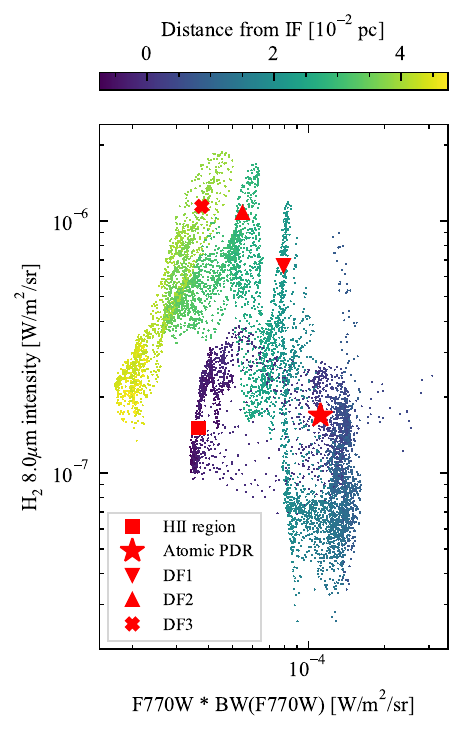}}
\resizebox{.99\textwidth}{!}{%
\includegraphics[width=0.3\textwidth]{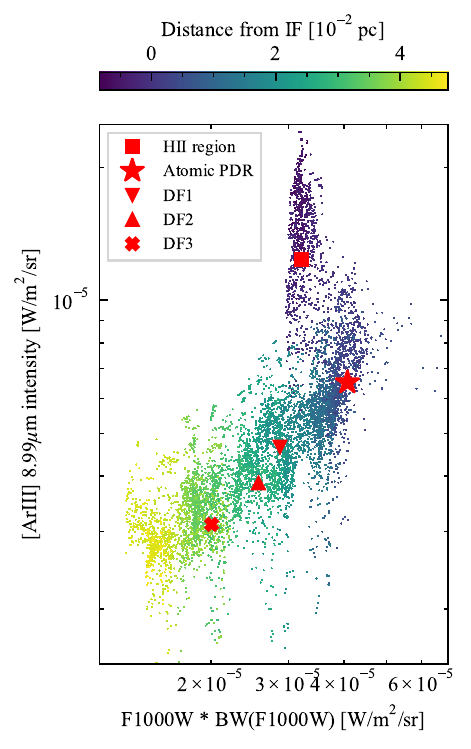}
\includegraphics[width=0.28\textwidth]{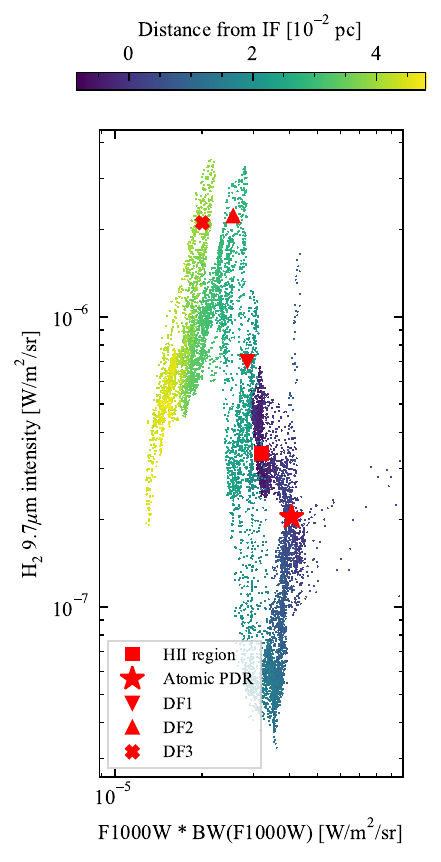}
\includegraphics[width=0.3\textwidth]{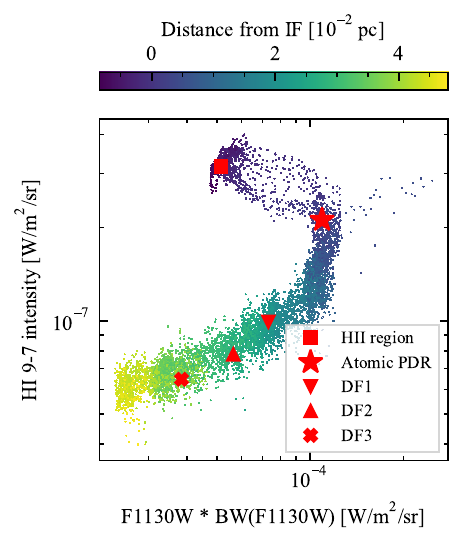}
\includegraphics[width=0.3\textwidth]{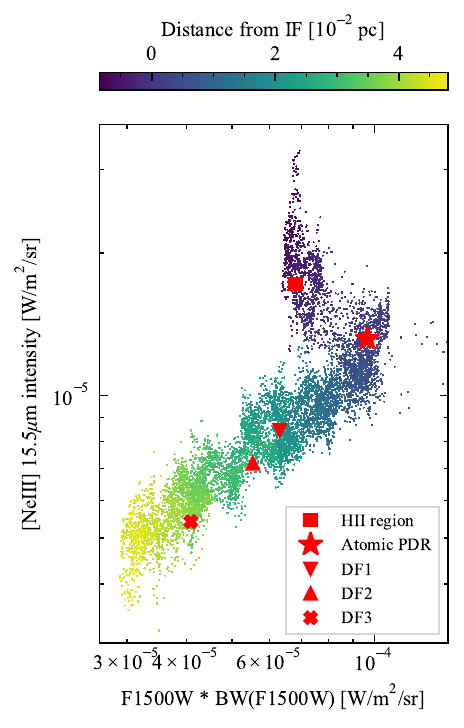}
}
\caption{Correlation between the measured line intensities and the synthetic images of the MIRI filters. \textbf{Top from left to right:} H$_2$ 6.9\,$\mu$m and F770W, \ArII\,6.99\,$\mu$m and F770W, HI 6--5 and F770W, and H$_2$ 8.0\,$\mu$m and F770W. \textbf{Bottom from left to right: } \ArIII\ 8.99\um\ and F770W, H$_2$ 9.7\,$\mu$m and F1000W, HI 9--7 and F1130W, and \NeIII\ 15.5\,$\mu$m and F1500W. We do not include the \SIV\ 10.5\um\ here due to the unsolved artifacts likely due to spurious jump detections in the ramps. Colors indicate the distance from the ionization front as in Fig.~\ref{fig:continuum_line_fractions_FeII}. Red marks indicate the data points of the five template spectra.%
}
\label{fig:MIRI_line_fractions}
\end{center}
\end{figure*}

Fig.~\ref{fig:MIRI_line_fractions} shows correlation plots of bright emission lines against the corresponding MIRI synthetic images.

\end{appendix}

\end{document}

%% file: affiliations.tex
\newcommand{\uwo}{1}
\newcommand{\wspace}{2}
\newcommand{\osu}{3}
\newcommand{\koln}{4}
\newcommand{\seti}{5}
\newcommand{\paris}{6}
\newcommand{\stsci}{7}
\newcommand{\toulouse}{8}
\newcommand{\umich}{9}
\newcommand{\ames}{10}
\newcommand{\cnrs}{11}
\newcommand{\madrid}{12}

\newcommand{\tokyo}{13}

\newcommand{\uwoname}{Department of Physics \& Astronomy, The University of Western Ontario, London ON N6A 3K7, Canada}
\newcommand{\wspacename}{Institute for Earth and Space Exploration, The University of Western Ontario, London ON N6A 3K7, Canada}
\newcommand{\osuname}{Department of Astronomy, The Ohio State University, 140 West 18th Avenue, Columbus, OH 43210, USA }
\newcommand{\kolnname}{I. Physikalisches Institut, Universit\"{a}t zu K\"{o}ln, Z\"{u}lpicher Stra{\ss}e 77, 50937 K\"{o}ln, Germany }
\newcommand{\setiname}{Carl Sagan Center, SETI Institute, 339 Bernardo Avenue, Suite 200, Mountain View, CA 94043, USA }
\newcommand{\parisname}{Institut d'Astrophysique Spatiale, Université Paris-Saclay, CNRS,  Bâtiment 121, 91405 Orsay Cedex, France}
\newcommand{\toulousename}{Institut de Recherche en Astrophysique et Planétologie, Université Toulouse III - Paul Sabatier, CNRS, CNES, 9 Av. du colonel Roche, 31028 Toulouse Cedex 04, France}
\newcommand{\stsciname}{Space Telescope Science Institute, 3700 San Martin Drive, Baltimore, MD 21218, USA}
\newcommand{\umichname}{Department of Astronomy, University of Michigan, 1085 South University Avenue, Ann Arbor, MI 48109, USA}
\newcommand{\amesname}{NASA Ames Research Center, MS 245-6, Moffett Field, CA 94035-1000, USA}
\newcommand{\tokyoname}{Department of Astronomy, Graduate School of Science, The University of Tokyo, 7-3-1 Bunkyo-ku, Tokyo 113-0033, Japan}
\newcommand{\cnrsname}{Institut des Sciences Mol\'{e}culaires d’Orsay, CNRS, Univ. Paris-Saclay, 91405 Orsay, France}
\newcommand{\madridname}{Instituto de Física Fundamental (CSIC), Calle Serrano 121-123, 28006 Madrid, Spain}

%% file: table_xcal_nirspec.tex
\begin{table*}
\caption{Best-fit parameters of the NIRCam vs. NIRSpec pixel-by-pixel cross-calibration.}
\label{tab:xcal_nirspec}
\centering
\begin{tabular}{llrrrr}
\hline
Filter & Overlapping IFU & Intercept $\calsymb{0, filter}$ & Slope $\calsymb{filter}$ &  Intrinsic   & Pearson's $r$ \\ 
 &   &  &  & scatter\tablefootmark{a} $\sigma_\mathrm{int}$   &  \\ 
 &   & MJy sr$^{-1}$ &  &  MJy sr$^{-1}$  &  \\ 
 \hline
F187N &  g235h-f170lp &  $-19\pm1$ &  $0.994\pm0.001$ &  $35$ &  $0.9955$ \\ 
F210M &  g235h-f170lp &  $5.3\pm0.3$ &  $0.897\pm0.003$ &  $7$ &  $0.9675$ \\ 
F212N &  g235h-f170lp &  $13.6\pm0.5$ &  $0.80\pm0.01$ &  $-9.5$ &  $0.9094$ \\ 
Average\tablefootmark{b} & g235h-f170lp & &  $0.980\pm0.001$ &  &  \\
\hline
F323N &  g395h-f290lp &  $1.7\pm0.3$ &  $0.919\pm0.001$ &  $7$ &  $0.9973$ \\ 
F335M &  g395h-f290lp &  $-1.9\pm0.3$ &  $0.876\pm0.001$ &  $10$ &  $0.9985$ \\ 
F405N &  g395h-f290lp &  $3\pm1$ &  $0.901\pm0.001$ &  $13$ &  $0.9986$ \\ 
F470N &  g395h-f290lp &  $9.6\pm0.4$ &  $0.870\pm0.001$ &  $13$ &  $0.9870$ \\ 
F480M &  g395h-f290lp &  $1.4\pm0.3$ &  $0.861\pm0.001$ &  $11$ &  $0.9929$ \\ 
Average\tablefootmark{b} & g395h-f290lp & &  $0.8901\pm0.0004$ &  &  \\
\hline
\end{tabular}
\tablefoot{
Statistical uncertainties in the slopes and intercepts are indicated.\\
\tablefoottext{a}{Intrinsic scatter $\sigma_\mathrm{int}$ is defined as $\sigma_\mathrm{int}^2\equiv \mathrm{mean}\left((y_i - f(x_i))^2\right) - \mathrm{mean}\left(\sigma_i^2\right)$, where $x_i,y_i$ are the individual measurements, $\sigma_i$ is the uncertainty on $y_i$, and $f$ is the best-fit line. Negative $\sigma_\mathrm{int}$ indicates that the uncertainties are likely overestimated.}
\tablefoottext{b}{The uncertainty in the inverse-variance-weighted mean $\calsymb{filter}$ for each grating is the formal uncertainty. The true uncertainties on these factors are likely closer to the scatter in $\calsymb{filter}$ for filters within a given grating.}
}
\end{table*}

%% file: table_xcal_miri.tex
\begin{table*}
\caption{Best-fit parameters of the MIRI MRS vs. MIRI Imager pixel-by-pixel cross-calibration (Eq.~\ref{eq:xcal}).}
\label{tab:xcal_miri}
\centering
\begin{tabular}{llrrrr}
\hline
Filter & Overlapping IFU & Intercept $\calsymb{0, filter}$ & Slope $\calsymb{filter}$ &  Intrinsic   & Pearson's $r$ \\ 
 &   &  &  & scatter $\sigma_\mathrm{int}$   &  \\ 
 &   & MJy sr$^{-1}$ &  &  MJy sr$^{-1}$  &  \\ 
\hline
F770W &  Channel 1 &  $95\pm4$ &  $1.020\pm0.001$ &  $165$ &  $0.9986$ \\ 
F1130W &  Channel 2 &  $1\pm5$ &  $1.018\pm0.001$ &  $166$ &  $0.9988$ \\ 
F1500W &  Channel 3 &  $191\pm9$ &  $0.935\pm0.001$ &  $246$ &  $0.9963$ \\ 
\hline
\end{tabular}
\end{table*}

%% file: table_continuum_fraction_all.tex
\begin{table}
\caption{Fraction of continuum emission ($\fcontrib{\mathrm{Cont.},~\mathrm{Filter}}$, Eq.~\ref{eq:fcont}), in units of \%, in the NIRCam and MIRI filters for the template spectra. }
\label{table:cont_frac_temp_all}
\centering
\begin{tabular}{lrrrrr}
\hline
Filter & \HII & Atomic & DF1 & DF2 & DF3 \\ 
\hline  \\[-7pt]
\multicolumn{6}{c}{\textbf{NIRCam}}\\[2pt]
F140M     & 99.5     & 99.4     & 99.6     & 99.7     & 99.8 \\ 
F162M     & 81.0     & 86.4     & 85.2     & 89.1     & 91.4 \\ 
F164N     & 66.7     & 73.5     & 72.1     & 82.1     & 84.8 \\ 
F182M     & 27.5     & 37.4     & 37.0     & 36.0     & 34.1 \\ 
F187N     & 4.9     & 7.3     & 7.3     & 7.1     & 6.4 \\ 
F210M     & 65.6     & 77.2     & 75.2     & 70.5     & 66.8 \\ 
F212N     & 88.4     & 92.8     & 81.7     & 62.0     & 52.3 \\ 
F277W     & 78.4     & 86.5     & 84.8     & 77.3     & 75.5 \\ 
F300M     & 90.7     & 95.1     & 94.0     & 90.6     & 89.8 \\ 
F323N     & 51.5     & 31.6     & 27.2     & 23.6     & 25.5 \\ 
F335M     & 43.8     & 25.6     & 22.5     & 19.9     & 21.8 \\ 
F405N     & 13.9     & 32.2     & 37.6     & 32.2     & 30.9 \\ 
F470N     & 92.6     & 96.6     & 91.4     & 84.6     & 79.4 \\ 
F480M     & 89.4     & 95.9     & 96.1     & 94.1     & 92.8 \\ [2pt]
\multicolumn{6}{c}{\textbf{MIRI}}\\[2pt]

F770W     & 62.5     & 46.3     & 46.1     & 51.7     & 55.5 \\ 
F1000W     & 92.5     & 90.5     & 91.8     & 92.6     & 93.0 \\ 
F1130W     & 65.3     & 36.7     & 40.2     & 48.1     & 54.5 \\ 
F1500W     & 94.6     & 94.8     & 94.7     & 95.7     & 96.3 \\ 

\hline
\end{tabular}
\end{table}

%% file: table_line_fraction_all.tex
\begin{table*}
\caption{Fraction of line or PAH emission ($\fcontrib{\mathrm{Line},~\mathrm{Filter}}$ and $\fcontrib{\mathrm{PAH},~\mathrm{Filter}}$, Eq.~\ref{eq:fcont}), in units of \%, in the NIRCam and MIRI filters for the template spectra.
}
\label{table:line_frac_temp_all}
\centering
\begin{tabular}{lSlSSSSS}
\hline
\multicolumn{2}{c}{Line ($\mu$m)} & Filter(s)  & \HII & Atomic & \multicolumn{1}{c}{DF1} & \multicolumn{1}{c}{DF2} & \multicolumn{1}{c}{DF3} \\ 
\hline\\[-7pt]
\multicolumn{8}{c}{\textbf{NIRCam}}\\[2pt]
\FeII & 1.64 & F164N & 12.4 & 12.4 & 9.0 & 4.5 & 4.9 \\ 
&& F162M & 1.8 & 1.8 & 1.2 & 0.6 & 0.7 \\  [5pt]
Pa $\alpha$ & 1.88 & F187N & 90.2 & 88.5 & 87.9 & 88.2 & 88.7 \\ 
&& F182M & 61.3 & 53.2 & 52.8 & 52.4 & 53.5 \\ [5pt]
H$_2$ 1-0 S(1) & 2.12 & F212N & 6.8 & 4.7 & 16.1 & 36.2 & 46.0 \\ 
&& F210M & 0.6 & 0.5 & 2.0  & 5.3 & 7.2 \\ [5pt] 
H$_2$ 1-0 O(5) & 3.24 & F323N & 2.0 & 0.8 & 3.8 & 7.6 & 9.0 \\  [5pt]
PAH & 3.3 & F335M & 51.2 & 73.1 & 76.2 & 78.4 & 76.5 \\  [5pt]
Br $\alpha$ & 4.05 & F405N & 82.8 & 65.3 & 59.2 & 64.3 & 65.9 \\  [5pt]
H$_2$ 0-0 S(9) & 4.69 & F470N & 4.5 & 2.3 & 8.0 & 14.7 & 19.9 \\ 
&& F480M & 0.7 & 0.4 & 1.4 & 2.7 & 3.8 \\  [2pt]
\multicolumn{8}{c}{\textbf{MIRI}}\\[2pt]
H$_2$ 0-0 S(5) & 6.9 & F770W & 0.07 & 0.02 & 0.10 & 0.26 & 0.38 \\ [2pt]
\ArII & 6.99 & F770W   & 2.58 & 0.75  & 0.19 & 0.27 & 0.33 \\ [2pt]
\HI\, 6-5 & 7.46 & F770W  & 0.79  & 0.17  & 0.10  & 0.13  & 0.16 \\ [2pt]
PAH & 7.7 & F770W & 34.0 & 52.7 & 53.4 & 47.3 & 43.2 \\ [2pt]
H$_2$ 0-0 S(4) & 8.0 & F770W & 0.06 & 0.02  & 0.13 & 0.30 & 0.46 \\ [2pt]
\ArIII &  8.99  & F1000W & 1.68  & 0.69  & 0.70  & 0.65  & 0.67 \\ [2pt]
H$_2$  0-0 S(3) & 9.7  & F1000W  & 0.17  & 0.08  & 0.38  & 1.36  & 1.65 \\ [2pt]
\SIV &  10.5     & F1000W  & 3.24 & 2.08 & 1.89  & 2.00  & 1.93 \\ [2pt]
PAH & 11.2 & F1130W  & 34.4  & 63.2  & 59.7  & 51.8  & 45.4  \\ [2pt]
\HI\, 9-7  & 11.3   & F1130W  & 0.31 & 0.10 & 0.07 & 0.07  & 0.09 \\ [2pt]
\NeIII & 15.5 & F1500W  & 4.21 & 2.27 & 2.23 & 2.16  & 2.20 \\ [2pt]
\hline
\end{tabular}
\end{table*}

%% file: table_N-W_fit_all_pointings_paper.tex
    \begin{table}
    \caption{Fit parameters of Eq.~\ref{eq:correlation} for selected line intensities and filter combinations.}
    \label{table:N-W_fit}
    \centering
    \begin{tabular}{llcccc}
    \hline
    \multicolumn{2}{c}{Line (\um)} & Filter A & Filter B & $\alpha$ & $\beta$ \\
    \hline  \\[-7pt]
\multicolumn{6}{c}{\textbf{NIRCam}}\\[2pt]
\FeII & 1.64 & F164N & F162M & 1.10 & 8.3e-08\\ 
Pa $\alpha$ & 1.88 & F187N & F182M & 0.72 & -1.3e-07\\ 
H$_2$ 1-0 S(1) & 2.12 & F212N & F210M & 0.78 & 1.5e-08\\ 
H$_2$ 1-0 O(5) & 3.24 & F323N & F300M & 0.69 & 1.4e-08\\
PAH & 3.3 & F335M & F300M & 1.13 & 2.0e-06\\ 
Br $\alpha$ & 4.05 & F405N & F480M & 0.91 & 2.3e-07\\ 
H$_2$ 0-0 S(9) & 4.69 &F470N & F480M & 0.96 & 3.0e-08\\  [2pt]
\multicolumn{6}{c}{\textbf{MIRI}}\\[2pt]
PAH & 7.7 & F770W & F1000W & 1.14 & 7.2e+02\\
PAH & 11.2 & F1130W & F1000W & 1.00 & -3.7e+02\\
PAH & 11.2 & F1130W & F1500W & 0.29 & -4.5e+02\\

    \hline
    \end{tabular}
    \end{table}

%% file: table_line_prescriptions.tex
\begin{table*}
\caption{Ratio of the calculated line or PAH intensities using various prescriptions over the actual line intensities for the template spectra. $\alpha$ and $\beta$ for each filter combination are shown in Table~\ref{table:N-W_fit}. Negative ratios indicate that the prescription gives a negative intensity by over-correcting the continuum. The ``Deviation'' column shows the mean deviation (in dex) from the 1:1 relation (the expected relation) between the actual line intensity and the calculated combination of the filters derived from all the data shown in scatter plots (Figs.~\ref{fig:continuum_line_fractions_FeII}--\ref{fig:continuum_PAH_fractions_F1130W}). Rows in bold indicate our recommended prescriptions for tracing line or PAH emission. }

\label{table:line_prescriptions}
\centering
\begin{tabular}{llrrrrrr}
\hline
Line ($\mu$m) & Filter(s)  & \HII & Atomic & DF1 & DF2 & DF3 & Deviation\\ 
\hline\\[-7pt]
\multicolumn{8}{c}{\textbf{NIRCam}}\\[2pt]
\FeII\ 1.64 & F164N  & 7.85         & 7.89         & 10.80         & 22.03         & 19.89         & 0.753 \\
& \textbf{F164N-F162M} & 1.42         & 1.38         & 0.66         & 2.08         & 1.99         & 0.279 \\
& (F164N-$\alpha$F162M)+$\beta$ & 1.08         & 1.00         & 0.61         & 1.86         & 1.91         & 0.285 \\ [5pt]
Pa $\alpha$ 1.87 & F187N & 1.12         & 1.14         & 1.15         & 1.15         & 1.14         & 0.043 \\ 
& F187N-F182M  & 0.95         & 0.95         & 0.96         & 0.95         & 0.95         & 0.014 \\ 
& \textbf{(F187N-$\alpha$F182M)+$\beta$} & 1.00         & 1.00         & 1.00         & 1.00         & 0.99         & 0.004 \\ [5pt]
H$_2$ 1-0 S(1) 2.12 & F212N & 14.54         & 20.72         & 6.09         & 2.72         & 2.14         & 0.692 \\ 
& F212N-F210M & -5.29         & -3.62         & -0.29         & 0.30         & 0.36         & 0.591 \\ 
& \textbf{(F212N-$\alpha$F210M)+$\beta$} & -0.79         & 1.92         & 1.20         & 0.87         & 0.78         & 0.194 \\ [5pt] 
H$_2$ 1-0 O(5) 3.24 & F323N & 48.73         & 123.66         & 26.08         & 13.04         & 10.92         & 1.216 \\ 
& F323N-F300M & 22.72         & 90.87         & 20.29         & 10.44         & 8.54         & 1.116 \\ 
& (F323N-$\alpha$F300M)+$\beta$ & 31.12         & 101.35         & 22.19         & 11.31         & 9.34         & 1.153 \\ [5pt]
PAH 3.3 & F335M  & 1.69         & 1.18         & 1.14         & 1.11         & 1.14         & 0.072 \\ 
& \textbf{F335M-F300M} & 0.97         & 0.96         & 0.96         & 0.95         & 0.96         & 0.017 \\
& (F335M-$\alpha$F300M)+$\beta$  & 1.03         & 0.97         & 0.99         & 0.99         & 1.03         & 0.015 \\ [5pt]
Br $\alpha$ 4.05 & F405N & 1.30         & 1.54         & 1.70         & 1.57         & 1.53         & 0.140 \\ 
& F405N-F480M & 1.10         & 1.00         & 0.94         & 0.91         & 0.90         & 0.072 \\ 
& \textbf{(F405N-$\alpha$F480M)+$\beta$} & 1.15         & 1.09         & 1.09         & 1.07         & 1.08         & 0.040 \\ [5pt]
H$_2$ 0-0 S(9) 4.69 & F470N  & 21.71         & 43.30         & 12.26         & 6.66         & 4.92         & 0.954 \\ 
& F470N-F480M  & -1.06         & -1.56         & 0.55         & 0.76         & 0.71         & 0.353 \\ 
& \textbf{(F470N-$\alpha$F480M)+$\beta$}  & 0.36         & 0.78         & 1.23         & 1.15         & 1.01         & 0.229 \\ [2pt]
\multicolumn{8}{c}{\textbf{MIRI}}\\[2pt]
PAH 7.7$\mu$m & F770W  & 2.94         & 1.90         & 1.87         & 2.11         & 2.31         & 0.234 \\
& F770W-F1000W  & -0.63         & 0.94         & 0.94         & 0.75         & 0.61         & 0.251 \\ 
& \textbf{(F770W-$\alpha$F1000W)+$\beta$}   & -0.34         & 0.97         & 1.04         & 0.93         & 0.98         & 0.157 \\ [5pt]
PAH 11.2$\mu$m & F1130W   & 2.91         & 1.58         & 1.67         & 1.93         & 2.20         & 0.213 \\ 
& F1130W-F1000W   & 1.29         & 1.05         & 1.09         & 1.15         & 1.17         & 0.045 \\ 
& \textbf{(F1130W-$\alpha$F1000W)+$\beta$}   & 1.10         & 1.01         & 1.01         & 1.04         & 0.98         & 0.015 \\ 
& F1130W-F1500W  & -2.29         & -0.31         & -0.28         & -0.63         & -0.98         & 1.531 \\ 
& \textbf{(F1130W-$\alpha$F1500W)+$\beta$} & 1.18         & 0.98         & 1.02         & 1.05         & 1.05         & 0.021 \\ 
& \textbf{(F1130W-$\alpha_1$F1000W-$\alpha_2$F1500W)+$\beta$}\tablefootmark{a} & 1.10         & 1.01         & 1.01         & 1.04         & 0.98         & 0.015 \\ 
\hline
\end{tabular}
\tablefoottext{a}{$\alpha_1 = 0.98$, $\alpha_2 = 0.01$, $\beta=-3.6$e+02}
\end{table*}